\documentclass[11pt]{article}
\usepackage{slashbox}
\usepackage{epsf}
\usepackage{smallgrf}
\usepackage{color}
\usepackage{amsmath,amsfonts,amsbsy,amssymb}

\begin{document}

\begin{titlepage}

\def\slash#1{{\rlap{$#1$} \thinspace/}}

\begin{flushright} 
\end{flushright} 

\vspace{0.1cm}

\begin{Large}
\begin{center}

{\bf  Non-Compact Hopf Maps and Fuzzy Ultra-Hyperboloids}
\end{center}
\end{Large}

\vspace{1cm}

\begin{center}
{\bf Kazuki Hasebe}   \\ 
\vspace{0.5cm} 
\it{Kagawa National College of Technology, Takuma, 
Mitoyo, Kagawa 769-1192, Japan} \\ 

\vspace{0.5cm} 
{\sf
hasebe@dg.kagawa-nct.ac.jp} 

\vspace{0.8cm} 

{\today} 

\end{center}

\vspace{1.5cm}

\begin{abstract}
\noindent

\baselineskip=18pt

Fuzzy hyperboloids naturally emerge in the geometries of branes, twistor theory, and higher spin theories. 
 In this work,  we perform a systematic study of higher dimensional fuzzy 
hyperboloids (ultra-hyperboloids) based on non-compact Hopf maps.  Two types of non-compact Hopf maps; split-type and hybrid-type, are introduced  from the  cousins of division algebras.  
We construct arbitrary even-dimensional fuzzy ultra-hyperboloids by applying the Schwinger operator formalism and indefinite Clifford algebras. It is shown that fuzzy hyperboloids, $H_F^{2p,2q}$, are represented by the coset, $H_F^{2p,2q}\simeq SO(2p,2q+1)/U(p,q)$, and  exhibit two types of  generalized  dimensional hierarchy; hyperbolic-type (for $q\neq 0$) and hybrid-type (for $q=0$). Fuzzy hyperboloids can be expressed  as fibre-bundle  of fuzzy fibre over  hyperbolic basemanifold.  
Such bundle structure of fuzzy hyperboloid gives rise to non-compact monopole gauge field.   
 Physical realization of fuzzy hyperboloids is argued in the context of lowest Landau level physics. 

\end{abstract}

\end{titlepage}

\newpage 

\tableofcontents

\section{Introduction}\label{sec:intro}

 Fuzzy geometry has been an active research field in the past decades. Fuzzy geometry naturally introduces a cut-off in small scale, which softens UV divergence of field theory.    
A most typical and well understood fuzzy manifold is fuzzy (two-)sphere \cite{berezin1975,Hoppe1982,madore1992}. Fuzzy two-sphere and its higher dimensional cousins naturally arise as non-trivial classical solutions of matrix model of string theory (as a review, see Ref.\cite{hep-th/9801182} for instance  and references therein).    
Interestingly, the original symmetry of classical manifolds is generally enhanced to be a larger  symmetry in their  corresponding fuzzy manifolds. Such symmetry enhancement is interpreted as quantum  geometrical effect particular to fuzzy manifolds \cite{balachandran2001,Hasebe2011}: quantum fluctuations on fuzzy manifold  ``smear'' the original classical geometry to generate a larger fuzzy geometry.   
Fuzzy spheres have mathematical structures closely related to Clifford algebra.   
The coordinates on fuzzy $2p$-sphere correspond to $SO(2p+1)$ gamma matrices of fully symmetric representation \cite{HoRamgoolam2002}.  
A most convenient way to handle the fully symmetric representation is to adopt the Schwinger operator formalism\footnote{The Schwinger operator formalism is sometimes referred to as oscillator formalism or oscillator realization in literature.}:  the fuzzy coordinates of fuzzy two-spheres \cite{watamuras1997,watamuras2000} and four-spheres  \cite{hep-th/9602115} are constructed by sandwiching the $SO(3)$  and $SO(5)$ gamma matrices by the Schwinger operators.  
Similarly, the coordinates of even higher dimensional fuzzy spheres can be constructed\cite{Kimura2002,Kimura2003,azuma2003,hep-th/0402010,hep-th/0406135}.  Furthermore, 
supersymmetric generalizations of fuzzy spheres \cite{Hasebe2011,math-ph/9804013,hep-th/0204170} and general fuzzy Riemann surfaces \cite{hep-th/0602290} have also been explored.  

The Schwinger construction of the fuzzy spheres is regarded as an operator version of the Hopf maps (see Ref.\cite{Hasebe2010} as a review, and Refs.\cite{hep-th/9510083,Sheikh-Jabbari2004, Sheikh-JabbarTorabian2005}). The Hopf maps are topological maps from sphere to sphere in different dimensions, introduced by Heinz Hopf  about eighty years ago \cite{Hopf1931,Hopf1935}:  
\begin{center}
\begin{tabular}{ccccccc}
\\ 
 & & $S^{3}$ &   $\overset{S^{1}}\longrightarrow $ & $S^{2}$ & & ~~~~~~~~~~(1st)\\
 &  $S^{7}$ & $\longrightarrow$ & $S^{4}$ &  & & ~~~~~~~~~~(2nd) \\
 $S^{15}$ & $\longrightarrow$ &   $S^{8}$ &&  & & ~~~~~~~~~~(3rd) \\  
\end{tabular}
\end{center}
The Hopf maps (which we designate as the compact Hopf maps) are related to the division algebras, $i.e.$ complex number (1st), quaternions (2nd) and octonions (3rd) \cite{Hasebe2010,Nakaharabook}.  
Quaternions and octonions are constructed by applying Cayley-Dickson construction to complex number \cite{Baez2002}.    
 The geometry of the Hopf maps reflects such particular construction and exhibits a hierarchical structure as shown in the above picture.  
 Meanwhile,  Clifford algebra is another generalization  of the complex number, and the coordinates of higher dimensional fuzzy spheres are essentially regarded as  gamma matrices of orthogonal groups. 
As is well known, higher dimensional Clifford algebra is constructed from lower dimensional one. This particular structure of construction of gamma matrices brings a hierarchical geometry to  fuzzy spheres \cite{Kimura2003,hep-th/0310274}. 

A main goal of this work is to explore a systematic construction of  higher dimensional fuzzy hyperboloids (fuzzy ultra-hyperboloids).    
Fuzzy hyperboloids are typical curved fuzzy manifolds that are regarded as  non-compact analogue of fuzzy spheres.  They arise as classical solutions of the matrix model on a pp-wave background geometry \cite{hep-th/0204033,sakaguchiyoshida2003}, and also play a crucial role in the discussion of UV/IR connection \cite{HoLi2001,HoLi2000}.   Furthermore, in the   context of twistor theory \cite{arXiv:0902.2523,cond-mat/0401224,hep-th/0203264,arXiv:1112.5210} and higher spin theory \cite{vasiliev2004,vasiliev2003}, fuzzy hyperboloids naturally emerge as their underlying geometry.   
Though low dimensional fuzzy hyperboloids are fairly well investigated \cite{berezin1975,grossepresnajder1993,Kishimoto2001,FakhriImaanpur2003,arXiv:0809.4885}, studies of fuzzy ultra-hyperboloids have a rather short history  \cite{Gazeauetal2006,Gazeau&Toppan,DeBellisetal2010}.   
 We show that the close relations between fuzzy spheres and  the Hopf maps can naturally be extended in the construction of fuzzy ultra-hyperboloids.     
 It is shown that there are two kinds of non-compact Hopf maps:  the first is the  split-type  (split Hopf maps) constructed by the split algebras\cite{arXiv:0905.2792,Baditoiu2010} 
\footnote{See also the preceding literatures \cite{Konderak1990,Blazic1996} for the 1st and 2nd split Hopf maps.}
:  
\begin{center}
\begin{tabular}{ccccccc}
\\ 
 & & $H^{2,1}$ &   $\overset{H^{1,0}}\longrightarrow $ & $H^{1,1}$ & & ~~~~~~~~~~(1st)\\
 &  $H^{4,3}$ & $\longrightarrow$ & $H^{2,2}$ &  & & ~~~~~~~~~~(2nd) \\
 $H^{8,7}$ & $\longrightarrow$ &   $H^{4,4}$ &&  & & ~~~~~~~~~~(3rd) \\  
\end{tabular}
\end{center}
The other is the hybrid type (hybrid Hopf maps\footnote{
The hybrid Hopf maps are a hybridization of the compact and split Hopf maps in the sense that the total manifolds are same as those of the split Hopf maps and the fibres are those of the compact Hopf maps. }):   
\begin{center}
\begin{tabular}{ccccccc}
\\  \vspace{0.3cm}
 & & $H^{2,1}$ &   $\overset{ S^{1}}\longrightarrow $ & $H^{2,0}$ & & ~~~~~~~~~~(1st)\\
  \vspace{0.3cm}
 &  $H^{4,3}$ &  $\overset{ S^{3}}\longrightarrow$ & $H^{4,0}$ &  & & ~~~~~~~~~~(2nd) \\
 $H^{8,7}$ &  $\overset{S^{7}}\longrightarrow$ &   $H^{8,0}$ &&  & & ~~~~~~~~~~(3rd) \\  
\end{tabular}
\end{center}
 With these non-compact Hopf maps, we argue geometrical structures of low dimensional fuzzy hyperboloids\footnote{ Interestingly, the compact \cite{MosseriDandoloff2001,BernevigChen2003} and split Hopf maps \cite{Rios2011} are also related to entangled qubit geometry of quantum information,   and black hole physics \cite{Borstenetal2009}. The quaternionic and split quaternionic gauge fields have also been used to construct a generalized Chern-Simons theory \cite{Kawamot-Watabiki1992I,Kawamot-Watabiki1992II}.  }. In general, the coordinates on fuzzy hyperboloids are given by gamma matrices of indefinite orthogonal groups.  Corresponding to  representations of the non-compact groups, 
two formulations have been proposed to realize  fuzzy hyperboloids  (see Ref.\cite{DeBellisetal2010} and references therein);  one of which is to adopt unitary infinite dimensional  representation of non-compact group, and the other is to adopt the non-unitary finite dimensional representation.  
We address their relations in the Schwinger operator formalism. 
As the hierarchical structure of the gamma matrices reflects the dimensional hierarchy of fuzzy spheres,  generalized dimensional hierarchies are observed in their non-compact counterparts.     
As coset, fuzzy spheres are given by \cite{HoRamgoolam2002} 
\begin{equation}
S_{F}^{2p}~\simeq ~SO(2p+1)/U(p),    
\end{equation}
which are locally expressed as  \cite{Kimura2002,Kimura2003}: 
\begin{equation}
S_F^{2p}\sim S^{2p}\otimes S_F^{2p-2}.  
\label{localeqfuzzysphere}
\end{equation}
Here, $\sim$ signifies local equivalence (throughout the paper, we adopt $\sim$ to express local equivalence): $S_F^{2p}$ is locally, but not globally, equivalent to  the trivial fibration of $S^{2p-2}_F$ over the basemanifold $S^{2p}$.  
Thus,  fuzzy sphere 
$S_F^{2p}$ can be regarded as a ``twisted'' fuzzy fibre-bundle of  fuzzy fibre $S^{2p-2}_F$ over  basemanifold $S^{2p}$.  
For fuzzy hyperboloids, we have 
\begin{equation}
H_{F}^{2p,2q}~\simeq~ SO(2p,2q+1)/U(p,q). 
\end{equation}
Corresponding to the split and hybrid Hopf maps, the fuzzy hyperboloids exhibit two types of the dimensional hierarchy, one of which is the hyperbolic type 
\begin{equation}
H_F^{2p,2q}\sim H^{2p,2q}\otimes H_F^{2p,2q-2} ~~(q\neq 0), 
\end{equation}
and the other is the hybrid type 
\begin{equation}
H_F^{2p,0}\sim H^{2p,0}\otimes S_F^{2p-2}. 
\end{equation}
That is,   $H_F^{2p,2q}$ is locally equivalent to fibre-bundle of  fibre $H_F^{2p,2q-2}$ over basemanifold $H^{2p,2q}$, and similarly $H_F^{2p,0}$ is equivalent to fibre-bundle of  $S_F^{2p-2}$ over $H^{2p,0}$.  
In either cases, the connection of fuzzy fibre gives rise to  (non-compact) monopole gauge field on hyperboloid\footnote{ Such  non-compact monopole and corresponding algebraic structure (split-quaternions) play important roles in pseudo-hermitian quantum mechanics \cite{BG11,BG11_3,2008nesterovcruz} and topological phase of  non-hermitian systems \cite{Sato1106,Esaki2011,HuHughes2011}.  }.   As the lowest Landau level physics in such monopole background,  a physical interpretation of  fuzzy hyperboloids is provided, too.    

This paper is organized as follows. 
In Section \ref{sec:cousins}, we review some basic mathematics of quaternions, symplectic groups, and indefinite gamma matrices.    
In Section \ref{sec:ind1}, we present close relations between the non-compact 1st Hopf map and the Schwinger operator construction of fuzzy two-hyperboloid.   
We extend the discussions to four-dimensional fuzzy hyperboloids with use of the non-compact 2nd Hopf maps in Section \ref{sec:ind2}.  
We further discuss  construction of even higher dimensional fuzzy hyperboloids based on gamma matrices of indefinite orthogonal groups in Section \ref{sec:fuzh}.   
In Section \ref{sec:fuz},  we demonstrate a physical realization of fuzzy hyperboloid in the context of lowest Landau level physics.  
Section \ref{sec:sum} is devoted to summary and discussions.

\section{Cousins of Quaternions and Indefinite Gamma Matrices}\label{sec:cousins}

In this section, we give a brief introduction of ultra-hyperboloids (Section \ref{subsec:ultrahy}), cousins of quaternions  (Section \ref{subsec;cousins}), and indefinite Clifford algebras (Section \ref{subse:indefi}). 

\subsection{Ultra-hyperboloids }\label{subsec:ultrahy}

The coordinates of ultra-hyperboloid $H^{p,q}$, $x^i$ $(i=1,2,\cdots,p)$ and $y^j$ $(j=1,2,\cdots,q+1)$, are defined so as to satisfy 
\begin{equation}
  \sum_{i=1}^{p} x^i x^i-\sum_{j=1}^{q+1} y^j y^j=-1. 
\label{defofultrahyper}
\end{equation}
Note $H^{p.q}\neq H^{q,p}$ $(p\neq q)$. 
Ultra-hyperboloids are expressed by the following cosets\footnote{ With $S^p\simeq H^{0,p}$, one may readily see that Eq.(\ref{costeshyperbolo}) reproduces the coset realizations of sphere:  
$S^{p}\simeq SO(p+1)/SO(p)$, $S^{2p+1}\simeq SU(p+1)/SU(p)$, and $S^{4p+3}\simeq Sp(p+1)/Sp(p)$. 
}:  
\begin{subequations}
\begin{align}
&H^{p,q}\simeq SO(p,q+1)/SO(p,q) \simeq SO(q+1,p)/SO(q,p),\label{cosetforhighhyper1}\\ 
&H^{2p,2q+1}\simeq SU(p,q+1)/SU(p,q)\simeq SU(q+1,p)/SU(q,p) 
\label{cosetforhighhyper2},\\
&H^{4p,4q+3}\simeq Sp(p,q+1)/Sp(p,q)\simeq  Sp(q+1,p)/Sp(q,p). 
\end{align}\label{costeshyperbolo}
\end{subequations}
Topology of $H^{p,q}$ is given by 
\begin{equation}
H^{p,q}\simeq R^p\otimes S^q, 
\end{equation}
and then, 
\begin{equation}
\pi_q (H^{p,q})=\pi_q(S^q)=\mathbb{Z}. 
\end{equation}
Several examples are described as 
\begin{subequations}
\begin{align}
&S^p\equiv H^{0,p},\\ 
&dS^p\equiv H^{1,p-1}\simeq R^1\otimes S^{p-1},\\
&AdS^p\equiv H^{p-1,1}\simeq R^{p-1}\otimes S^1,\\
&EAdS^p(= H^p)\equiv H^{p,0}\simeq R^{p}\otimes Z_2, \label{eucladsz2}
\end{align}
\end{subequations}
where $S^p$, $dS^p$, $AdS^p$ and $EAdS^p$ denote $p$-dimensional sphere, de Sitter, anti-de Sitter and Euclidean anti-de Sitter spaces, respectively. Note $dS^2= AdS^2(=H^{1,1})$. 
$S^p$ $(p> 2)$, $dS^p$ $(p> 2)$, and $EAdS^p$ are simply connected manifolds:  
\begin{align}
&\pi_1(S^{p>2})~\simeq  ~\pi_1(dS^{p>2})~\simeq ~  \pi_1(EAdS^p)~\simeq ~1. 
\end{align}
 $EAdS^p$ denotes $p$-dimensional two-leaf hyperboloid. ($Z_2$ in (\ref{eucladsz2}) corresponds to the two-leaves.) $dS^p$ is one-leaf hyperboloid. $AdS^p$ are connected but not simply connected: 
\begin{equation}
\pi_1(AdS^p)\simeq \mathbb{Z}. 
\end{equation}

As described in Introduction, 
the basemanifolds of the compact, non-compact, and hybrid Hopf maps are respectively given by  
\begin{center}
\begin{tabular}{ccccc}
 ~~~~~~~~~~~~~~~~~~ Compact  & ~~~~ & Split   &  ~~~~&  Hybrid  \\
\\  \vspace{0.3cm}
(1st)  ~~~~~~~~~~~~ $H^{0,2} =  S^2 $ &~~~~  &  $H^{1,1}$ & ~~~~ &  $H^{2,0}= EAdS^2$  \\
  \vspace{0.3cm}
  (2nd) ~~~~~~~~~~~ $H^{0,4} =  S^4$ &~~~~ &  $H^{2,2}$ & ~~~~& $ H^{4,0} = EAdS^4$   \\
  (3rd) ~~~~~~~~~~~~ $H^{0,8} = S^8$ & ~~~~&  $H^{4,4}$ & ~~~~ &  $H^{8,0}= EAdS^8 $  \\  
\end{tabular}
\end{center}
The hybrid Hopf maps represent trivial fibration, since the basemanifolds, $H^{2,0}$, $H^{4,0}$ and $H^{8,0}$, are two-leaf hyperboloids. The symmetry groups of the basemanifolds are  
\begin{center}
\begin{tabular}{ccccc}
 ~~~~~~~~~~~~~~~~~~ Compact  & ~~~~ & Split   &  ~~~~&  Hybrid  \\
\\  \vspace{0.3cm}
(1st)  ~~~~~~~~~~~~~ $SO(3) $ &~~~~  &  $SO(2,1)$ & ~~~~ &  $SO(1,2)$  \\
  \vspace{0.3cm}
  (2nd) ~~~~~~~~~~~~ $SO(5)$ &~~~~ &  $SO(3,2)$ & ~~~~& $ SO(1,4)$   \\
  ~(3rd) ~~~~~~~~~~~~ $SO(9)$ & ~~~~&  $SO(5,4)$ & ~~~~ &  $SO(1,8) $  \\  
\end{tabular}
\end{center}
The symmetry groups of 1st and 2nd Hopf maps are compactly restated by quaternion and split-quaternion groups: 
\begin{center}
\begin{tabular}{ccccc}
 ~~~~~~~~~~~~~~~~~~ Compact  & ~~~~ &  Split   &  ~~~~&  Hybrid  \\
\\  \vspace{0.3cm}
(1st)  ~~~~~~~~~~~~ $U(1;\mathbb{H}) $ &~~~~  & $U(1;\mathbb{H}') $ & ~~~~ &  $U(1;\mathbb{H}') $  \\
  \vspace{0.3cm}
  (2nd) ~~~~~~~~~~~ $U(2;\mathbb{H}) $ &~~~~ & $U(2;\mathbb{H}')\simeq U(1,1;\mathbb{H'}) $ & ~~~~& $ U(1,1;\mathbb{H}) $  \\ 
\end{tabular}
\end{center}
Here, $\mathbb{H}$ and $\mathbb{H}'$ respectively denote quaternion and split-quaternion. 
Note that $U(p,q; \mathbb{H})$ and $U(p,q; \mathbb{H}')$ $(p+q\le 2)$ all appear in this table.  
See also Eqs.(\ref{quaternionsgroupsone}) and (\ref{quaternionsgroupstwo}).

\subsection{Cousins of quaternions and symplectic groups}\label{subsec;cousins}

Quaternions and their cousins $(1,q_1,q_2,q_3)$ are defined so as to satisfy 
\begin{equation}
q_iq_j=-q_jq_i~(i\neq j),~~~~~q_1q_2q_3=-1, 
\label{propcouqua}
\end{equation}
and  
\begin{equation}
(q_1)^2=\epsilon_1,~~(q_2)^2=\epsilon_2,~~(q_3)^2=\epsilon_3,  
\label{squarequaternions}
\end{equation}
where each of $\epsilon_1$, $\epsilon_2$ and $\epsilon_3$ takes either $+1$ or $-1$.  
Then, there are four types of quaternions: 
\begin{itemize}
\item Quaternions \cite{Hamilton1844}: all of $\epsilon_{1},\epsilon_{2},\epsilon_{3}$ in (\ref{squarequaternions}) are $-1$. 
\item ``Hybrid'' quaternions \footnote{The author does not know how this type of quaternions is called in literature. In this paper, we call this type of quaternions  hybrid quaternions, since they are related to the hybrid Hopf maps. The hybrid quaternions do not respect associativity like hyperbolic quaternions. However, in both cases, $q_1q_2q_3$ in (\ref{propcouqua}) is defined to be independent on the associative order, $i.e.$ $(q_1q_2)q_3=q_1(q_2q_3)$.  } : two of $\epsilon_{1},\epsilon_{2},\epsilon_{3}$ are  $-1$ and the remaining is $+1$ . 
\item Split quaternions \cite{Cockle1848}:  two of $\epsilon_{1},\epsilon_{2},\epsilon_{3}$ are $+1$ and the remaining is $-1$. 
\item Hyperbolic quaternions \cite{Macfarlane1892}: all of $\epsilon_{1},\epsilon_{2},\epsilon_{3}$ are $+1$.  
\end{itemize}
$q_1$, $q_2$ and $q_3$ are generalization of imaginary unit and called imaginary quaternions, and their conjugation is given by    
\begin{equation}
{q_1}^*=-q_1,~~~{q_2}^*=-q_2,~~~{q_3}^*=-q_3. 
\end{equation}
With four real parameters, $r_0,r_1,r_2,r_3$, an arbitrary quaternion number is constructed as    
\begin{equation}
q=r_0+r_1q_1+r_2q_2+r_3q_3. 
\label{arbitquater}
\end{equation}
Similarly, arbitrary hybrid, split and hyperbolic quaternions are respectively constructed by replacing $q_i$ $(i=1,2,3)$ in (\ref{arbitquater}) with hybrid, split and hyperbolic imaginary quaternions. 
The conjugate of $q$ is given by   
\begin{equation}
q^*=r_0-r_1q_1-r_2q_2-r_3q_3,   
\end{equation}
and  $q^*q$ is derived as  
\begin{equation}
q^*q=qq^*={r_0}^2-\epsilon_1 {r_1}^2-\epsilon_2{r_2}^2-\epsilon_3 {r_3}^2, 
\label{qqstarinner}
\end{equation}
with $\epsilon_{1},\epsilon_{2},\epsilon_{3}$  (\ref{squarequaternions}). 
Thus, for quaternions, (\ref{qqstarinner}) provides the inner product in  Euclidean space.  
Similarly,  (\ref{qqstarinner}) respectively yields the inner product in split  (signature)  space  for split quaternions, and  Lorentzian (signature)  space for both hybrid and hyperbolic quaternions. It may be worthwhile to write down the algebra of four-types of quaternions explicitly:  
\begin{itemize}
\item Quaternions:
\begin{align}
&(q_1)^2=(q_2)^2=(q_3)^2=-1,\nonumber\\
&q_1q_2=-q_2q_1=q_3,~~ q_2q_3=-q_3q_2=q_1,~~q_3q_1=-q_1q_3=q_2. 
\label{propcouqua1}
\end{align}
\item Hybrid quaternions: 
\begin{align}
&(q_1)^2=+1, ~~(q_2)^2=(q_3)^2=-1,\nonumber\\
&q_1q_2=-q_2q_1=q_3,~~ q_2q_3=-q_3q_2=-q_1,~~q_3q_1=-q_1q_3=q_2. 
\label{propcouqua2}
\end{align}
\item Split quaternions:  
\begin{align}
&(q_1)^2=(q_2)^2=+1,~~(q_3)^2=-1,\nonumber\\
&q_1q_2=-q_2q_1=q_3,~~ q_2q_3=-q_3q_2=-q_1,~~q_3q_1=-q_1q_3=-q_2. 
\label{propcouqua3}
\end{align}
\item Hyperbolic quaternions:  
\begin{align}
&(q_1)^2=(q_2)^2=(q_3)^2=+1,\nonumber\\
&q_1q_2=-q_2q_1=-q_3,~~ q_2q_3=-q_3q_2=-q_1,~~q_3q_1=-q_1q_3=-q_2. 
\label{propcouqua4}
\end{align}
\end{itemize}
One may find that in either type of quaternions, $q_1q_2q_3=(q_1q_2)q_3=q_1(q_2q_3)=-1$ (\ref{propcouqua}) holds.    

Replacing the imaginary unit of the Pauli matrices with imaginary quaternions, we have 
\begin{align}
&
\sigma^2=\begin{pmatrix}
0 & -i \\
i & 0 
\end{pmatrix},~~\sigma^1=\begin{pmatrix}
0 & 1 \\
1 & 0 
\end{pmatrix},~~ \sigma^3=\begin{pmatrix}
1 & 0 \\
0 & -1
\end{pmatrix}\nonumber\\
\rightarrow ~~~&\gamma^1=\begin{pmatrix}
0 & -q_1\\
q_1 & 0 
\end{pmatrix},~~ 
\gamma^2=\begin{pmatrix}
0 & -q_2 \\
q_2 & 0 
\end{pmatrix},~~\gamma^3=\begin{pmatrix}
0 & -q_3 \\
q_3 & 0 
\end{pmatrix},~~ 
\gamma^4=\begin{pmatrix}
0 & 1 \\
1 & 0 
\end{pmatrix},~~\gamma^5=\begin{pmatrix}
1 &  0 \\
0 & -1
\end{pmatrix}. 
\label{substituteso5}
\end{align}
They satisfy the anti-commutation relations of $SO(5)$ gamma matrices, $\{\gamma^a,\gamma^b\}=2\eta^{ab}$, with $\eta^{ab}=diag(+,+,+,+,+)$. 
When we adopt the hybrid, split, and  hyperbolic quaternions instead of quaternions in (\ref{substituteso5}),  $\gamma^a$ satisfy the anti-commutation relations with  $\eta^{ab}=diag(-,+,+,+,+)$,  $\eta^{ab}=diag(-,-,+,+,+)$  and  $\eta^{ab}=diag(-,-,-,+,+)$, respectively.  Thus, we obtain $SO(4,1)$,  $SO(3,2)$  and $SO(2,3)$,  gamma matrices for hybrid, split and hyperbolic quaternions, respectively: 
\begin{itemize}
\item Quaternions ~~~~~~~~~~~~~~~~~~$\rightarrow $~~ $SO(5)$ gamma matrices 
\item Hybrid quaternions ~~~~~~~~$\rightarrow $~~~$SO(4,1)$ gamma matrices  
\item Split quaternions~~~~~~~~~~~~$\rightarrow $~~  $SO(3,2)$ gamma matrices  
\item Hyperbolic quaternions ~~~$\rightarrow $~~ $SO(2,3)$ gamma matrices  
\end{itemize}
 The $SO(5)$ and $SO(3,2)$ gamma matrices are crucial in constructing the 2nd compact \cite{zhanghu2001} and split Hopf maps \cite{arXiv:0905.2792}. Then, one may expect that  $SO(4,1)$ gamma matrices play a similar role in constructing the 2nd hybrid Hopf map \footnote{Meanwhile, since the $SO(2,3)$ gamma matrices of hyperbolic quaternions are equivalent to $SO(3,2)$ gamma matrices of split quaternions up to imaginary unit, we do not consider the $SO(2,3)$ case.}. This expectation turns out to be true in Section \ref{subse:hybrid2}.  
Note, however, there is a crucial difference: hybrid quaternions (and also hyperbolic quaternions) do not respect associativity unlike quaternions and split-quaternions, and hence  hybrid quaternions cannot be realized by matrices\footnote{It is well known that the quaternions are represented by  (imaginary unit times) $SU(2)$ Pauli matrices, and similarly the split-quaternions are by  (imaginary unit times) $SU(1,1)$ Pauli matrices (see Section \ref{subsec:non-compact1st}).}.    
 For instance, from (\ref{propcouqua2}) we find a non-associative relation:   
\begin{equation}
(q_1 q_1)q_2\neq q_1(q_1q_2).    
\end{equation}
The left-hand side is $(q_1 q_1 ) q_2 =(+1) q_2=q_2$, while the right-hand side is $q_1(q_1 q_2)=q_1 q_3=-q_2$.  
Thus, among the cousins of quaternions, only the original and split-quaternions satisfy associative algebras, and their groups are consistently defined.    
Low dimensional quaternion  and split-quaternions groups, $U(p,q;\mathbb{H})$ and $U(p,q;\mathbb{H}')$ ($p+q\le 2$),  are all exhausted as  
\begin{subequations}
\begin{align}
&U(1;\mathbb{H})\equiv Sp(1)\simeq USp(2)\simeq SU(2)\simeq SO(3), \\
&U(1;\mathbb{H}')\simeq Sp(2;\mathbb{R}) \simeq SU(1,1)\simeq SO(2,1)\simeq SO(1,2),  
\end{align}\label{quaternionsgroupsone}
\end{subequations}
and 
\begin{subequations}
\begin{align}
&U(2;\mathbb{H})\equiv Sp(2)\simeq  USp(4)\simeq SO(5),\\
&U(2;\mathbb{H}') \simeq U(1,1;\mathbb{H}')\simeq Sp(4;\mathbb{R}) \simeq  SO(3,2)\simeq SO(2,3),\\ 
&U(1,1;\mathbb{H})\equiv Sp(1,1)\simeq USp(2,2)\simeq  SO(4,1)\simeq SO(1,4). 
\end{align}\label{quaternionsgroupstwo}
\end{subequations}
Note that $SO(5)$, $SO(3,2)$ and $SO(4,1)$ structures naturally appear in  $U(p,q;\mathbb{H})$ and $U(p,q;\mathbb{H}')$ for $p+q= 2$ (\ref{quaternionsgroupstwo}).  
Low dimensional unitary groups of quaternion and split-quaternions provide  basic examples of indefinite orthogonal groups.

\subsection{Indefinite gamma matrices}\label{subse:indefi}

 Here, 
we introduce general indefinite orthogonal groups and their gamma matrices.    
 In indefinite orthogonal groups, finite dimensional representation of their gamma matrices is generally given by non-hermitian matrix. Multiplied by a suitable matrix,   non-hermitian gamma matrices  are transformed to hermitian matrices. We mainly discuss such ``hermitianization'' of gamma matrices of indefinite orthogonal groups, which will play a crucial role in explicit construction of the non-compact Hopf maps. 
For detail properties of indefinite gamma matrices, readers may consult Ref.\cite{KugoTownsend1983}. 

\subsubsection{$SO(p,q)$  gamma matrices $(p+q: \text{even})$ }

First, we consider even dimensional space-times:  
\begin{equation}
 p+q : \text{even},  
\end{equation}
 where $SO(p,q)$ gamma matrices,  $\gamma^{\mu}$ $(\mu=1,2,\cdots,p+q)$, satisfy 
\begin{equation}
\{\gamma^{\mu},\gamma^{\nu}\}=2\eta^{\mu\nu},\label{antigammaetaeven}
\end{equation}
with 
\begin{equation}
\eta^{\mu\nu}=diag(\overbrace{+,+,\cdots,+}^{p},\overbrace{-,-,\cdots,-}^{q}). 
\end{equation}
Since $(\gamma^{\mu})^2=+1$,  $\gamma^{\mu}$ $(\mu=1,2,\cdots,p)$ may be taken as hermitian matrices.  On the other hand, since $(\gamma^{\mu})^2=-1$, $\gamma^{\mu}$ $(\mu=p+1,p+2,\cdots,p+q)$ may be taken as anti-hermitian matrices \cite{KugoTownsend1983}. Thus, the $SO(p,q)$ gamma matrices are classified into hermitian and anti-hermitian matrices.  We have two different (hermitian and anti-hermitian) matrices that hermitianize  the gamma matrices. 
One hermitianizing matrix is constructed by multiplying all of the anti-hermitian gamma matrices 
\begin{equation}
k=(i)^{\frac{1}{2}q(q-1)}\gamma^{p+1}\gamma^{p+2}\cdots \gamma^{p+q}, 
\label{defofk}
\end{equation}
and the other is by multiplying the remaining all hermitian gamma matrices\footnote{$k'$ corresponds to $A$ matrix in Ref.\cite{KugoTownsend1983} up to a proportional factor. } 
\begin{equation}
k'=(i)^{\frac{1}{2}p(p+1)+1}\gamma^{1}\gamma^{2}\cdots \gamma^{p}.
\label{defofk'}
\end{equation}
With (\ref{antigammaetaeven}), 
it is straightforward to  show that $k\gamma^{\mu}$ and $k'\gamma^{\mu}$ are indeed hermitian    
\begin{subequations}
\begin{align}
&(k\gamma^{\mu})^{\dagger}=k\gamma^{\mu},\\
&(k'\gamma^{\mu})^{\dagger}=k'\gamma^{\mu}. 
\end{align}
\end{subequations}
Thus, all of the gamma matrices of $SO(p,q)$ are hermitianized multiplied by either of $k$ and $k'$. 
Hermitian conjugates of $k$ and $k'$ are respectively given by 
\begin{subequations}
\begin{align}
&k^{\dagger}=(-1)^q k,\\
&{k'}^{\dagger}=(-1)^{q+1} k'.
\end{align}
\end{subequations}
Therefore, in the case $(p,q)$=(even,even)\footnote{Remember $p+q$ is even.}, $k$ and $k'$ are hermitian and  anti-hermitian matrices, respectively. On the other hand, in the case $(p,q)$=(odd,odd), $k$ and $k'$ are anti-hermitian and hermitian matrices, respectively. 

 The $SO(p,q)$ generators are constructed by 
\begin{equation}
\sigma^{\mu\nu}=-i\frac{1}{4}[\gamma^{\mu},\gamma^{\nu}], 
\end{equation}
and they satisfy 
\begin{subequations}
\begin{align}
&(k\sigma^{\mu\nu})^{\dagger}=(-1)^q k\sigma^{\mu\nu},\\
&(k'\sigma^{\mu\nu})^{\dagger}=(-1)^{q+1} k'\sigma^{\mu\nu}. 
\end{align}
\end{subequations}
Therefore, when $(p,q)$=(even,even), $\gamma^{\mu}$ and $\sigma^{\mu\nu}$ are simultaneously hermitianized only by $k$. Meanwhile, when $(p,q)$=(odd,odd),  $\gamma^{\mu}$ and $\sigma^{\mu\nu}$ are simultaneously hermitianized only by $k'$.

\subsubsection{$SO(l,m)$ gamma matrices ($l+m$~:~odd) from $SO(p,q)$ ($p+q$~:~even) }

Next, we consider odd dimensional space-times:  
\begin{equation}
l+m: \text{odd}.
\end{equation}
As is well known,  $SO(l,m)$ ($l+m$~:~odd) gamma matrices can be ``constructed'' by  $SO(p,q)$ ($p+q$ : even) gamma matrices. 

\begin{itemize}
\item In the case of $SO(l,m)=SO(p+1,q)$
\end{itemize}

The $SO(p+1,q)$ gamma matrices $\gamma^{a}$ ($a=1,2,\cdots,p+q+1$) satisfy 
\begin{equation}
\{\gamma^{a},\gamma^{b}\}=2\eta^{ab},\label{antigammaetaodd1}
\end{equation}
with 
\begin{equation}
\eta^{ab}=diag(\overbrace{+,+,\cdots,+}^{p},\overbrace{-,-,\cdots,-}^{q},+). 
\end{equation}
$\gamma^{a}$ consist of the $SO(p,q)$ gamma matrices 
 $\gamma^{\mu}$ $(\mu=1,2,\cdots,p+q)$ and 
 \begin{equation}
 \gamma^{p+q+1}\equiv (i)^{\frac{1}{2}(p-q)}\gamma^1\gamma^2\cdots\gamma^{p+q}. 
 \label{onegammalast}
 \end{equation}
 $\gamma^{p+q+1}$ is a hermitian matrix that satisfies $(\gamma^{p+q+1})^2=1$. 
With $k$ (\ref{defofk}), $\gamma^{a}=(\gamma^{\mu},\gamma^{p+q+1})$ are hermitianized as 
\begin{equation}
(k\gamma^{a})^{\dagger}=k\gamma^{a}. 
\label{kgammahermieq}
\end{equation}
Notice that unlike $k$, $k'$ does not hermitianize all of the $SO(p+1,q)$ gamma matrices: $k'\gamma^{\mu}$ are hermitian  (as stated above) but $k'\gamma^{p+q+1}$ is anti-hermitian.   
Then, all of the gamma matrices of $SO(p+1,q)$ can be  hermitian multiplied  only by $k$.  Similarly, it can be shown that the $SO(p+1,q)$ generators 
\begin{equation}
\sigma^{ab}=-i\frac{1}{4}[\gamma^a,\gamma^b]
\end{equation}
satisfy 
\begin{equation}
(k\sigma^{ab})^{\dagger}=(-1)^{q}k\sigma^{ab}. 
\end{equation}

\begin{itemize}
\item In the case of $SO(l,m)=SO(p,q+1)$
\end{itemize}

The $SO(p,q+1)$ gamma matrices $\gamma^{a}$ ($a=1,2,\cdots,p+q+1$) satisfy 
\begin{equation}
\{\gamma^{a},\gamma^{b}\}=2\eta^{ab},\label{antigammaetaodd1}
\end{equation}
with 
\begin{equation}
\eta^{ab}=diag(\overbrace{+,+,\cdots,+}^{p},\overbrace{-,-,\cdots,-}^{q},-). 
\end{equation}
$\gamma^{a}$ ($a=1,2,\cdots,p+q+1$) are explicitly given by the $SO(p,q)$ gamma matrices  
 $\gamma^{\mu}$ $(\mu=1,2,\cdots,p+q)$ and 
 \begin{equation}
 \tilde{\gamma}^{p+q+1}\equiv (i)^{\frac{1}{2}(p-q)+1}\gamma^1\gamma^2\cdots\gamma^{p+q}. 
 \label{anothgammalast}
 \end{equation}
 $\tilde{\gamma}^{p+q+1}$ is an anti-hermitian matrix that satisfies $(\tilde{\gamma}^{p+q+1})^2=-1$.  
With $k'$ (\ref{defofk'}), all of  $\gamma^{a}=(\gamma^{\mu},\tilde{\gamma}^{p+q+1})$ are hermitianized :   
\begin{equation}
(k'\gamma^{a})^{\dagger}=k'\gamma^{a}. 
\label{kdgammahermieq}
\end{equation}
Notice that $k$ does not hermitianize all of the $SO(p,q+1)$ gamma matrices: $k\gamma^{\mu}$ are hermitian (as stated above) but $k\tilde{\gamma}^{p+q+1}$ is  anti-hermitian. 
Consequently, the gamma matrices of $SO(p,q+1)$ are hermitianized only by $k'$.  
 Similarly,  $SO(p,q+1)$ generators 
\begin{equation}
\sigma^{ab}=-i\frac{1}{4}[\gamma^a,\gamma^b]
\end{equation}
satisfy 
\begin{equation}
(k'\sigma^{ab})^{\dagger}= (-1)^{q+1}k'\sigma^{ab}. 
\end{equation}

\subsubsection{Consistency }

The $SO(l,m)$ gamma matrices ($l+m$: odd) are constructed either by  $SO(l-1,m)$ or $SO(l,m-1)$ gamma matrices by the methods stated above.  
 Then, there  exist  two superficially different ways for the construction of  $SO(l,m)$ ($l+m$: odd) gamma matrices. We discuss consistency of such two methods.  
 
\begin{itemize}
\item $SO(l,m)=SO$(odd, even) 
\end{itemize}

The gamma matrices of $SO(l,m)$ $(l,m)=$(odd, even) constructed from those of $SO(l-1,m)$ become hermitian (\ref{kgammahermieq}) multiplied by $k$.  Since $(p,q)=(l-1,m)$=(even,  even), $k$ is hermitian in the present case.  
Meanwhile,  the gamma matrices of  $SO(l,m)$ $(l,m)=$(odd, even) constructed from those of $SO(l,m-1)$ are hermitianized (\ref{kdgammahermieq}) by  $k'$.  $k'$ is also hermitian, since $(p,q)=(l,m-1)$=(odd, odd). 
Thus, in either case, we reach the same conclusion: hermitianization of $SO$(odd, even) gamma matrices is performed  by the hermitian matrix \footnote{ The corresponding $SO$(odd,even) generators also become hermitian matrices multiplied by the same hermitian matrix.}.  
Consequently, for instance, $SO(1,2)$, $SO(1,4)$, $SO(3,2)$ gamma matrices respectively become hermitian matrices multiplied by their corresponding hermitian matrices. 

\begin{itemize}
\item $SO(m,l)=SO$(even, odd) 
\end{itemize}

Since the overall signs of metrics of $SO(m,l)$ and $SO(l,m)$ are interchanged,   
their gamma matrices are equal up to the overall imaginary unit.   
Therefore, their hermitianizing matrices are also equal up to the overall  imaginary unit.  As discussed above, the $SO$(odd,even) gamma matrices are hermitianized by a hermitian matrix. Therefore,  
the hermitianizing matrix of   $SO$(even,odd) gamma matrices is given by an  anti-hermitian matrix\footnote{The corresponding  $SO$(even,odd) generators also become anti-hermitian matrices multiplied by the same anti-hermitian matrix}. For instance, $SO(2,1)$, $SO(4,1)$, $SO(2,3)$ gamma matrices respectively become hermitian multiplied by their corresponding anti-hermitian matrices.

\subsubsection{Relations between $SO(1,2p)$ and $SO(2p+1)$ gamma matrices}\label{subse:relso2p+1}

The $SO(1,2p)$ gamma matrices consist of  $SO(0,2p)$ gamma matrices, $\gamma^1,\gamma^2,\cdots,\gamma^{2p}$, and $\gamma^{2p+1}$:   
\begin{equation}
\gamma^{2p+1}\equiv (-i)^{p}\gamma^1\gamma^2\cdots\gamma^{2p}.  
\label{gamma2p+1}
\end{equation}
Here, we substituted $(p,q)$ with $(0,2p)$ in (\ref{onegammalast}) to derive (\ref{gamma2p+1}).   
Their hermitianizing  matrix is given by  (\ref{defofk}) 
\begin{equation}
k=i^{p(2p-1)}\gamma^1\gamma^2\cdots\gamma^{2p}=(-1)^{p}{\gamma}^{2p+1},  
\label{so12nhermiti}
\end{equation}
which is hermitian. 
The hermitianized $SO(1,2p)$ gamma matrices are    $k\gamma^1,k\gamma^2,\cdots,k\gamma^{2p},k\gamma^{2p+1}$ where the last one is proportional to the unit matrix: 
$k\gamma^{2p+1}=(-1)^p \bold{1}$. From anti-commutativity between $k$ and $\gamma^a$, $i.e.$ $k\gamma^a=-\gamma^a k$,  
we can see that the hermitianized $SO(0,2p)$ gamma  matrices $k\gamma^1,k\gamma^2,\cdots,k\gamma^{2p}$ and the matrix $k$ (\ref{so12nhermiti}) amount to  $SO(2p+1)$ gamma matrices.  
This unique relation between $SO(1,2p)$ and $SO(2p+1)$ underlies relations between  the compact and  hybrid Hopf maps as we shall discuss in Section \ref{subsec:non-compact1st}.

\section{Non-compact 1st Hopf Map and Fuzzy Two-Hyperboloid}\label{sec:ind1}

In this section, we introduce the non-compact 1st Hopf map (Section \ref{subsec:non-compact1st}) and the corresponding fuzzy two-hyperboloid $H_F^{2,0}$ (Section \ref{subsec:twofuzzyhyper}), mainly based on Refs.\cite{FakhriImaanpur2003,arXiv:0905.2792}  
\footnote{ The split 1st Hopf map, $H^{2,1}\overset{H^{1,0}}\longrightarrow H^{1,1}$, cannot be realized by using the usual imaginary unit \cite{arXiv:0905.2792}. To realize the split 1st Hopf map, we need to utilize the split-imaginary unit, and such construction will be discussed in Appendix \ref{sec:split-al}.}. 

\subsection{Non-compact 1st Hopf map: $H^{2,0}\simeq H^{2,1}/S^1$}\label{subsec:non-compact1st}

The 1st non-compact Hopf map  is given by 
\begin{equation}
H^{2,1}\overset{S^{1}}\longrightarrow H^{2,0},
\label{differentnoncompacthopfmap}
\end{equation}
which can be rewritten as the following form more familiar to physicists,   
\begin{equation}
AdS^3 \overset{{U}(1)}\longrightarrow EAdS^2. 
\label{1stmapsrewritten}
\end{equation}
Here, $EAdS^2=H^{2,0}$ is a two-leaf hyperboloid whose  
 symmetry group  is $SO(1,2)$. The gamma matrices of the $SO(1,2)$ group,  $\tau^i$ ($i=1,2,3$),  
 satisfy the following anti-commutation relations  
\begin{equation}
\{\tau^i,\tau^j\}=2\eta^{ij}, 
\end{equation}
with $\eta_{ij}=diag(-,-,+)$. Note $SO(1,2)\simeq SU(1,1)$, and $\tau^i$ are given by the $SU(1,1)$ ``Pauli matrices'': 
\begin{equation}
\tau^1=i\sigma^2,~~\tau^2=-i\sigma^1,~~\tau^3=\sigma^3.  
\label{defoftaus}
\end{equation}
They satisfy the commutation relations 
\begin{equation}
[\tau^i,\tau^j]=-2i\epsilon^{ijk}\tau_k, 
\end{equation}
where $\epsilon^{ijk}$  is 3 rank antisymmetric tensor with $\epsilon^{123}=1$.  
With use of $SO(0,2)$ gamma matrices $\tau^1,\tau^2$, 
the hermitianizing matrix (\ref{so12nhermiti}) is constructed as 
\begin{equation}
\kappa=i\tau^1\tau^2=\sigma^3, 
\end{equation}
and the hermitianized gamma matrices are 
\begin{equation}
\kappa^i\equiv \sigma^3\tau^i=(\sigma^1,\sigma^2,1). 
\label{defkappa}
\end{equation}

With an $SO(1,2)$ Dirac spinor $\phi$ subject to a ``normalization'' constraint $\phi^{\dagger}\sigma^3\phi=1$,  the 1st non-compact Hopf map is given by 
\begin{equation}
\phi 
\rightarrow 
x^i=\phi^{\dagger}  \kappa^i\phi, 
\label{1stHopfmapexplicit}
\end{equation}
and $x^i$ in (\ref{1stHopfmapexplicit}) automatically satisfy the condition of $H^{2,0}$:  
\begin{equation}
\eta_{ij}x^i x^j=-x^2-y^2+z^2=(\phi^{\dagger}\sigma^3 \phi)^2=1,
\label{1sthybrideq}
\end{equation}
where $x=x^1$, $y=x^2$, $z=x^3$. 
Thus, $x^i$ (\ref{1stHopfmapexplicit}) are regarded as coordinates on two-leaf hyperboloid. In particular, since $z=\phi^{\dagger}\phi \ge 0$,  the present construction corresponds to the upper leaf of two-leaf hyperboloid.   

Here, we also mention the ``derivation'' of 
the hybrid Hopf map from the compact Hopf map.   
The original 1st Hopf map is given by 
\begin{equation}
\phi~~\rightarrow ~~x^i =\phi^{\dagger}\sigma^i\phi,
\label{1stnoncompex}
\end{equation}
where $\phi$ is subject to $\phi^{\dagger}\phi=1$. 
$x^i$ automatically satisfy the condition of $S^2$: 
\begin{equation}
x^2+y^2+z^2=(\phi^{\dagger}\phi)^2=1. 
\label{s2relation1st}
\end{equation}
As mentioned in Section \ref{subse:relso2p+1}, there is a unique relation between $SO(2p+1)$  and  $SO(1,2p)$ gamma matrices. In the present case $p=1$,    $\kappa=\sigma^3$ and  $(\kappa^1,\kappa^2)=(\sigma^1,\sigma^2)$  amount to the $SO(3)$ gamma matrices. With (\ref{1stnoncompex}),  the compact Hopf map (\ref{s2relation1st}) can be restated as 
\begin{equation}
( \phi^{\dagger}\kappa^1\phi)^2+( \phi^{\dagger}\kappa^2\phi)^2+( \phi^{\dagger}\kappa\phi)^2=( \phi^{\dagger}\phi)^2.
\label{restatedhopfeq}
\end{equation}
Interchanging the right-hand side and the last term on the left-hand side in (\ref{restatedhopfeq}), we have 
\begin{equation}
-( \phi^{\dagger}\kappa^1\phi)^2-( \phi^{\dagger}\kappa^2\phi)^2+( \phi^{\dagger}\phi)^2= ( \phi^{\dagger}\kappa\phi)^2, 
\end{equation}
which is exactly equal to the relation (\ref{1sthybrideq}). Thus, one can ``derive''  the non-compact 1st Hopf map from the compact Hopf map. Similarly,   the 2nd and 3rd hybrid Hopf maps can also be obtained by their corresponding  compact 2nd and 3rd Hopf maps from the relations of gamma matrices  discussed in Section \ref{subse:relso2p+1}.

\subsection{Fuzzy two-hyperboloid: $H^{2,0}_F$ }\label{subsec:twofuzzyhyper}

We argue the Schwinger operator formulation of fuzzy two-hyperboloid with emphasis on its relation to the non-compact 1st Hopf map. 

\subsubsection{Finite dimensional non-unitary representation}

By replacing $\phi$ with the Schwinger operator $\Phi$ in (\ref{1stnoncompex}),  the  coordinates on fuzzy two-hyperboloid are constructed as \cite{FakhriImaanpur2003}   
\begin{equation}
{X}^i=\Phi^{\dagger}\kappa^i\Phi. 
\label{defnofsu11coordinates}
\end{equation}
$X^i$ are  hermitian operators.   For $X^i$ to satisfy the $SU(1,1)$ algebra
\begin{equation}
[X^i,X^j]=-2i\epsilon^{ijk}X_k,  
\end{equation}
the components of Schwinger operator should satisfy  generalized creation and annihilation relations: 
\begin{equation}
[\Phi_\alpha,\Phi_{\beta}^{\dagger}]=(\sigma^3)_{\alpha\beta},~~~~[\Phi_{\alpha},\Phi_{\beta}]=[\Phi_{\alpha}^{\dagger},\Phi_{\beta}^{\dagger}]=0. 
\label{commutationrelation}
\end{equation}
Note that  $\Phi_2$ and $\Phi^{\dagger}_2$ obey an unusual commutation relation, $[\Phi_2,\Phi_2^{\dagger}]=-1$. 
From (\ref{defnofsu11coordinates}) and (\ref{commutationrelation}),  square of the radius of fuzzy two-hyperboloid is derived as 
\begin{equation}
\eta_{ij} {X}^i {X}^j=-X^2-Y^2+Z^2=(\Phi^{\dagger}\sigma^3\Phi) ( \Phi^{\dagger}\sigma^3\Phi   +2). 
\label{covariantsu11}
\end{equation}
 The eigenvalues of (\ref{covariantsu11}) are  given by 
\begin{equation}
\eta_{ij} {X}^i {X}^j=n (n+2), 
\end{equation}
with 
\begin{equation}
\Phi^{\dagger}\sigma^3\Phi=n_1+n_2\equiv n, 
\end{equation}
($n_1$ and $n_2$ are non-negative integers) and the eigenstates are 
\begin{equation}
|n_1,n_2\rangle=\frac{1}{\sqrt{n_1!n_2!}} (\Phi_1^{\dagger})^{n_1}(\Phi_2^{\dagger})^{n_2}|0\rangle, 
\label{n1n2ket}
\end{equation}
with $|0\rangle$ that satisfies $\Phi_1|0\rangle =\Phi_2|0\rangle=0$. 
Here, we used that  
\begin{subequations}
\begin{align}
&\Phi_1^{\dagger}\Phi_1|n_1,n_2\rangle =n_1 |n_1,n_2\rangle,\label{secondn1}\\
&\Phi_2^{\dagger}\Phi_2|n_1,n_2\rangle =-n_2 |n_1,n_2\rangle.  \label{secondn2} 
\end{align}\label{secondn1and2} 
\end{subequations}
Note the minus sign in front of the right-hand side of (\ref{secondn2}):  Physically, such minus sign indicates that  $|0\rangle$ is an ``unstable'' vacuum for the 2nd oscillator mode, since $|0\rangle$ corresponds to the highest ``energy'' state. 
From (\ref{secondn1and2}), one finds that $X^3$ takes the eigenvalues 
\begin{equation}
X^3=\Phi^{\dagger}_1\Phi_1+\Phi_2^{\dagger}\Phi_2=n_1-n_2=n,n-2,n-4,\cdots, -n.
\label{spectrax5} 
\end{equation}
These spectra are same as those of fuzzy two-sphere. 

Dual state of (\ref{n1n2ket}) 
\begin{equation}
\langle n_1,n_2|=\frac{1}{\sqrt{n_1!n_2!}} \langle 0|\Phi_1^{n_1}\Phi_2^{n_2},   
\end{equation}
yields  negative norm  for odd $n_2$:   
$\langle n_1,n_2|n'_1,n'_2\rangle =(-1)^{n_2}\delta_{n_1n_1'}\delta_{n_2n_2'}.$  
More suitable dual state may be given by    
\begin{equation}
\langle\!\langle n_1,n_2|=\frac{1}{\sqrt{n_1!n_2!}} \langle 0|{(\Phi^1)}^{n_1}{(\Phi^2)}^{n_2} =(-1)^{n_2}\langle n_1,n_2|,  
\label{dualapprosu11}
\end{equation}
where $\Phi^{\alpha}\equiv (\sigma^3)^{\alpha\beta}\Phi_{\beta}$, $i.e.$ $(\Phi^1,\Phi^2)=(\Phi_1,-\Phi_2)$.   (\ref{dualapprosu11}) satisfies $\langle\!\langle n_1,n_2|n'_1,n'_2\rangle=\delta_{n_1n_1'}\delta_{n_2n_2'}$. 

 Such non-unitary construction of fuzzy hyperboloid is a straightforward generalization of that of fuzzy sphere, and readily applicable to represent states on fuzzy hyperboloids in arbitrary higher dimensions \cite{DeBellisetal2010}. However, such non-hermitian construction  is rather counterintuitive, since  in the large $n$ limit,  the spectra (\ref{spectrax5}) do not reduce to the values of the 3rd coordinate of the classical upper leaf (or lower-leaf) hyperboloid, $i.e.$ $x^3\ge n$ (or $x^3\le -n$).

\subsubsection{Infinite dimensional unitary representation }\label{subsec:twohyperinf}

The above mentioned ``problem'' of the non-unitary construction of fuzzy hyperboloid 
is amended by adopting infinite dimensional unitary representation. 
Unitary representation can readily be obtained by a redefinition of the creation and annihilation Schwinger operators.  We  interchange $\Phi_2$ and  $\Phi_2^{\dagger}$ for the components of the Schwinger operator to satisfy the usual commutation relations:  
\begin{equation}
[\Phi_{\alpha},\Phi_{\beta}^{\dagger}]=\delta_{\alpha\beta},~~[\Phi_{\alpha},\Phi_{\beta}]=[\Phi_{\alpha}^{\dagger},\Phi_{\beta}^{\dagger}]=0. 
\label{usuopephire}
\end{equation}
With the new Schwinger operators, $X^i$ (\ref{defnofsu11coordinates}) are represented as 
\begin{subequations}
\begin{align}
&X^1=\Phi_1^{\dagger}\Phi_2^{\dagger}+\Phi_2\Phi_1,\\
&X^2=-i\Phi_1^{\dagger}\Phi_2^{\dagger}+i\Phi_2\Phi_1,\\
&X^3= \Phi_1^{\dagger}\Phi_1+\Phi_2\Phi_2^{\dagger}  =\Phi_1^{\dagger}\Phi_1+\Phi_2^{\dagger}\Phi_2+1. 
\label{newlysu11coor}
\end{align}\label{newlysu11coorall}
\end{subequations}
These are still hermitian operators, but  
$X^1$ and $X^2$ (that are originally constructed by anti-hermitian gamma matrices) become particle-number non-conserving operators.  
One may readily check that (\ref{newlysu11coorall}) indeed satisfy the $SU(1,1)$ commutation relations under  (\ref{usuopephire}). 
The radius of the fuzzy hyperboloid (\ref{covariantsu11}) is now written as 
\begin{equation}
\eta_{ij}{X}^i{X}^j=(\Phi^{\dagger}\sigma^3\Phi-1) ( \Phi^{\dagger}\sigma^3\Phi   +1),  
\label{2covariantsu11}
\end{equation}
and the simultaneous eigenstates of (\ref{2covariantsu11}) and (\ref{newlysu11coor}) are 
\begin{equation}
|n_1,n_2)=\frac{1}{\sqrt{n_1!n_2!}} (\Phi_1^{\dagger})^{n_1}(\Phi_2^{\dagger})^{n_2}|\text{vac}\rangle, 
\label{inftwostateupper}
\end{equation}
with non-negative integers, $n_1,n_2$, and  eigenvalues    
\begin{subequations}
\begin{align}
&\eta_{ij}{X}^i{X}^j=
(n_1-n_2)^2-1,    
\label{2covariantsu112}\\
&{X}^3=n_1+n_2+1.  
\label{eigenX3}
\end{align}\label{spectraradix3}
\end{subequations}
Here, $|\text{vac}\rangle$ is the true vacuum of the newly defined Schwinger operator $\Phi_{1,2}$: 
\begin{equation}
\Phi_1|\text{vac}\rangle =\Phi_2|\text{vac}\rangle  =0,  
\end{equation}
and the dual state of (\ref{inftwostateupper}) is obtained as  
\begin{equation}
(n_1,n_2|=\frac{1}{\sqrt{n_1!n_2!}}\langle \text{vac}|(\Phi_1)^{n_1}(\Phi_2)^{n_2},  
\end{equation}
which always yields  positive norm:  $(n_1,n_2|n'_1,n'_2)=\delta_{n_1n_1'}\delta_{n_2n_2'}$.

Since $\eta_{ij}X^iX^j$ indicates square of the radius of fuzzy hyperboloid, it should have  positive eigenvalues. By the interchange symmetry between $\Phi_1$ and $\Phi_2$ in (\ref{newlysu11coorall}), we can  take $n_1 \ge n_2$ in 
(\ref{inftwostateupper}) without loss of generality.  
Instead of $n_1$ and $n_2$, we introduce new parameters $n$ and $l$,     
\begin{subequations}
\begin{align}
&n\equiv n_1-n_2 \ge  0,\label{defn2}\\
&l\equiv n_2 \ge 0.    \label{defl2}
\end{align}\label{defnl2}
\end{subequations}
The spectra of (\ref{spectraradix3}) are rewritten as 
\begin{subequations}
\begin{align}
&\eta_{ij}{X}^i{X}^j=
n^2-1, \label{sqpcx3fuzzyhyper2} \\
&{X}^3=n+1+2l=n+1,n+3,n+5,n+7,\cdots.  
\label{2covariantsu112}
\end{align}\label{twocovariantsu11}
\end{subequations}
Here, $n$ and $l$ (\ref{defnl2}) indicate the radius of hyperboloid and  the 3rd coordinate on the upper hyperboloid, respectively.  
In the large $n$ limit, the radius of fuzzy hyperboloid behaves as $ \sim n$ and $X^3$ is $\ge n$.  
Thus in unitary representation, the spectra of fuzzy hyperboloid naturally reduce to (the upper leaf of) the classical two-leaf hyperboloid.   
For the (semi-)positivity of square of the radius in (\ref{sqpcx3fuzzyhyper2}), $n$  should be taken as $n\ge 1$. With $n$ and $l$, 
 (\ref{inftwostateupper}) is represented as  
\begin{equation}
|n,l)=\frac{1}{\sqrt{(n+l)!l!}}
(\Phi_1^{\dagger})^{n+l}(\Phi_2^{\dagger})^{l}|\text{vac}\rangle, 
\end{equation}
which describes a  state at  $X^3=n+1+2l$ $(l=0,1,2,3,\cdots)$ on the upper leaf  of the fuzzy-hyperboloid.  
In the language of the $SU(1,1)$ representation theory,   
the present unitary representation corresponds to the discrete series \cite{so(21)references,bookPerelomov} \footnote{Detail correspondences to the $SU(1,1)$ representation theory in Refs.\cite{so(21)references,bookPerelomov} are as follows. In the discrete series, the eigenvalues of the $SU(1,1)$ Casimir are given by  $\eta_{ij} {X}^i{X}^j=4j(j-1)$ with $j=1,{3}/{2},2,\cdots$ ($j$ corresponds to $k$ in Refs.\cite{so(21)references,bookPerelomov}),  and those of $\frac{1}{2}{X}^3$ are $\mu=j,j+1,j+2,j+3,\cdots$. The identification to the notation of the present paper is  
$j=\frac{1}{2}(n+1)$ and $\mu=\frac{1}{2}(n+1)+l$ $(l=0,1,2,\cdots)$.  Such identification exactly reproduces (\ref{twocovariantsu11}) from the known formulas of the $SU(1,1)$ representation theory.}.  

We have two kinds of two-dimensional hyperboloids, two-leaf hyperboloid $H^{2,0}$  and one-leaf hyperboloid $H^{1,1}$ ($AdS^2$).  
Either  coordinates of their corresponding fuzzy manifolds  satisfy the $SU(1,1)$ algebra. 
Their difference is specified by choosing different unitary irreducible representations of $SU(1,1)$.   
For fuzzy  $H^{1,1}$, one has to adopt the principal series \cite{HoLi2000,Gazeauetal2006}, while for fuzzy $H^{2,0}$ one has to adopt the discrete series.  Thus, 
choice of  unitary irreducible representation is crucial in description of  fuzzy manifold. 
It should also be mentioned that such distinction is lacking in the  non-unitary construction of fuzzy hyperboloid.    

The relevant mathematical structures of 2D fuzzy sphere and fuzzy hyperboloids are summarized in Table \ref{3Dspacetable}.  
\begin{table}
\renewcommand{\arraystretch}{1}
\hspace{-0.5cm}
\begin{tabular}{|c||c|c|c|c|c|
}
\hline   Symmetry    & Original manifold  &   Hopf Map & $k$  matrix  
& Enhanced Algebra & Fuzzy Manifold \\ 
\hline
\hline  $SO(3)$ 
  & $S^{2}= H^{0,2}$   & Compact  & Hermite 
  & $SO(3)\simeq SU(2)$   & $S_F^2\simeq \mathbb{C}P^{1,0}$     \\ 
\hline $SO(2,1)$ & $dS^2= AdS^2= H^{1,1}$  & Split & Anti-hermite 
   & $SO(1,2)\simeq SU(1,1)$  &  $H_F^{1,1}\simeq \mathbb{C'}P^{0,1}$ \\ 
\hline $SO(1,2)$ & $EAdS^2=H^{2,0}$ & Hybrid & Hermite 
  & $SO(2,1)\simeq SU(1,1)$    &$H_F^{2,0}\simeq \mathbb{C}P^{0,1}$ \\
\hline
\end{tabular}
 \caption{ Fuzzyfication of two-hyperboloids and related properties.  See Appendix \ref{sec:appeninfcom} for  $\mathbb{C}P^{p,q}$ and  $\mathbb{C}'P^{p,q}$.  } 
 \label{3Dspacetable}
\end{table} 

\section{Non-compact 2nd Hopf Maps and Fuzzy Four-Hyperboloids}\label{sec:ind2}

 In this section, we argue a construction of four-dimensional fuzzy 
 hyperboloids based on non-compact 2nd Hopf maps. First, we introduce  two types of non-compact 2nd Hopf maps, the split and hybrid types (Section \ref{subsec:split2ndhopf} and \ref{subse:hybrid2}). Corresponding to two types of non-compact Hopf maps, we construct fuzzy $H^{2,2}$ and fuzzy $H^{4,0}$ (Section \ref{subsec:fuzzysplit} and \ref{subsect:fyzzyfroutwoleaf}). 
Geometrical structures of the fuzzy four-hyperboloids are also discussed (Section \ref{subsec:quantumf}).  

\subsection{Split 2nd Hopf map: $H^{2,2}\simeq H^{4,3}/H^{2,1}$}\label{subsec:split2ndhopf}

In the explicit  realization of the split Hopf map \cite{arXiv:0905.2792},  the hermitianizing matrix for  $SO(3,2)$ gamma matrices plays a crucial role.      
The $SO(3,2)$  gamma matrices $\gamma^a$ ($a=1,2,3,4,5$)  satisfy    
\begin{equation}
\{\gamma^a,\gamma^b\}=2\eta^{ab},
\end{equation}
with $\eta^{ab}=\eta_{ab}=(\eta_{ij},+,+)=\text{diag}(-,-,+,+,+)$.
They are explicitly given by 
\begin{equation}
\gamma^i=-\tau^i\otimes \sigma^2,~~\gamma^4=1\otimes  \sigma^1,~~\gamma^5=\gamma^1\gamma^2\gamma^3\gamma^4 =1\otimes \sigma^3,  
\label{explicitso32constgam}
\end{equation}
or 
\begin{align}
&\gamma^i=
\begin{pmatrix}
0 & i\tau^i\\
-i\tau^i & 0
\end{pmatrix},~~\gamma^4=
\begin{pmatrix}
0 & 1\\
1 & 0
\end{pmatrix},~~\gamma^5=
\begin{pmatrix}
1 & 0\\
0 & -1
\end{pmatrix}, 
\label{gammasso(3,2)}
\end{align}
where $\tau^i$ ($i=1,2,3$) are the $SU(1,1)$ Pauli matrices (\ref{defoftaus}).  
$\gamma^1$ and $\gamma^2$ are anti-hermitian  while $\gamma^3$, $\gamma^4$ and $\gamma^5$ are hermitian.  
By the formula (\ref{defofk}), $SO(3,2)$ hermitianizing matrix is constructed as   
\begin{equation}
k=i\gamma^4\gamma^3=
\begin{pmatrix}
\sigma^3 & 0 \\
0 & \sigma^3
\end{pmatrix}, 
\label{hermitiso32}
\end{equation}
and  hermitianized ``gamma'' matrices are given by   
\begin{equation}
k^a=k\gamma^a, 
\end{equation}
or  
\begin{align}
&k^1=
\begin{pmatrix}
0 & i\sigma^1\\
-i\sigma^1 & 0
\end{pmatrix},~~
k^2=
\begin{pmatrix}
0 & i\sigma^2 \\
-i\sigma^2 & 0
\end{pmatrix},~~
k^3 =
\begin{pmatrix}
0 & i1_2 \\
-i1_2 & 0
\end{pmatrix},\nonumber\\
&k^4=
\begin{pmatrix}
0 & \sigma^3 \\
\sigma^3 & 0
\end{pmatrix},~~~~~~~
k^5=
\begin{pmatrix}
\sigma^3 & 0 \\
0 & -\sigma^3
\end{pmatrix}.
\end{align}

Coordinates on $H^{4,3}$ are represented by a ``normalized''  $SO(3,2)$ Dirac spinor $\psi$ (the $SO(3,2)$ Hopf spinor) subject to the constraint 
\begin{equation}
\psi^{\dagger} k\psi=\psi_1^*\psi_1-\psi_2^*\psi_2+\psi_3^*\psi_3-\psi_4^*\psi_4=1. 
\label{normaso32hopfh22}
\end{equation}
With such $\psi$, the 2nd split Hopf map, $H^{4,3}\overset{H^{2,1}}\longrightarrow H^{2,2}$, is realized as 
\begin{equation}
\psi~~\rightarrow~~ x^a=\psi^{\dagger}k^a\psi.
\label{2ndnoncompactHopf}
\end{equation}
It is easily checked that $x^a$ satisfy the condition of $H^{2,2}$: 
\begin{align}
\eta_{ab}x^a x^b=-x^1x^1-x^2x^2+x^3x^3+x^4x^4+x^5x^5=(\psi^{\dagger}k\psi)^2=1.
\label{constraintsofh22}
\end{align}

Commutators of $\gamma^a$ provide the $SO(3,2)$ generators   
\begin{equation}
\gamma^{ab}=-i\frac{1}{4}[\gamma^a,\gamma^b],
\end{equation}
or 
\begin{align}
&\gamma^{ij}=-\frac{1}{2}\epsilon^{ijk}
\begin{pmatrix}
\tau_k & 0 \\
0 & \tau_k
\end{pmatrix},~~
\gamma^{i4}=\frac{1}{2}
\begin{pmatrix}
\tau^i & 0 \\
0 & -\tau^i
\end{pmatrix},\nonumber\\
&\gamma^{i5}=-\frac{1}{2}
\begin{pmatrix}
0 & \tau^i \\
\tau^i & 0 
\end{pmatrix},~~
\gamma^{45}=
\frac{i}{2}
\begin{pmatrix}
0 & 1 \\
-1 & 0
\end{pmatrix},
\end{align}
and they satisfy   
\begin{equation}
[\gamma_{ab},\gamma_{cd}]=i(\eta_{ac}\gamma_{bd}-\eta_{ad}\gamma_{bc}+\eta_{bd}\gamma_{ac}-\eta_{bc}\gamma_{ad}).
\end{equation}
As easily checked, $k\gamma^{ab}$ are hermitian as well. 

\subsection{Fuzzy split four-hyperboloid: $H_F^{2,2}$}\label{subsec:fuzzysplit}

By replacing the $SO(3,2)$ Hopf spinor with a four-component Schwinger operator, we construct fuzzy four-hyperboloid  $H_F^{2,2}$. 

\subsubsection{Finite dimensional unitary representation }\label{so32fuzzy4}

As a natural generalization of the $SU(1,1)$ Schwinger operator, we 
introduce the $SO(3,2)$ Schwinger operator  $\Phi=(\Phi_1,\Phi_2,\Phi_3,\Phi_4)^t$ whose components satisfy  
\begin{equation}
[\Phi_\alpha,\Phi_{\beta}^{\dagger}]=(k)_{\alpha\beta},~~~~[\Phi_{\alpha},\Phi_{\beta}]=[\Phi_{\alpha}^{\dagger},\Phi_{\beta}^{\dagger}]=0,  
\label{commuwithk}
\end{equation}
where $k$ is the $SO(3,2)$ hermitianizing matrix (\ref{hermitiso32}). 
With $\Phi$,  coordinates on $H_F^{2,2}$ are constructed as   
\begin{equation}
{X}^a=\Phi^{\dagger}k^a\Phi.  
\label{defanothX2}
\end{equation}
Square of the radius and the 5th coordinate of $H_F^{2,2}$ are explicitly given by  
\begin{subequations}
\begin{align}
&\eta_{ab}{X}^a{X}^b=-X^1X^1-X^2X^2+X^3X^3+X^4X^4+X^5X^5=(\Phi^{\dagger}k\Phi)(\Phi^{\dagger}k\Phi+4),\label{squarexxso32} \\
&X^5=\Phi_1^{\dagger}\Phi_1-\Phi_2^{\dagger}\Phi_2-\Phi_3^{\dagger}\Phi_3+\Phi_4^{\dagger}\Phi_4.\label{squareradix5}
\end{align} \label{twoequationsaboutfuzzyh22}
\end{subequations}
 Their simultaneous eigenstates are 
\begin{equation}
|n_1,n_2,n_3,n_4\rangle=\frac{1}{\sqrt{n_1!n_2!n_3!n_4!}} (\Phi^{\dagger}_1)^{n_1} (\Phi^{\dagger}_2)^{n_2}(\Phi^{\dagger}_3)^{n_3}(\Phi^{\dagger}_4)^{n_4}|0\rangle,
\label{so32eigen}
\end{equation}
where $n_1,n_2,n_3,n_4$ are non-negative integers. Furthermore,  
$|n_1,n_2,n_3,n_4\rangle$ are simultaneous eigenstates of four number operators made of the Schwinger operator:  
\begin{align}
&\Phi_1^{\dagger}\Phi_1|n_1,n_2,n_3,n_4\rangle =n_1 |n_1,n_2,n_3,n_4\rangle,\nonumber\\
&\Phi_2^{\dagger}\Phi_2|n_1,n_2,n_3,n_4\rangle =-n_2 |n_1,n_2,n_3,n_4\rangle,  \nonumber\\
&\Phi_3^{\dagger}\Phi_3|n_1,n_2,n_3,n_4\rangle =n_3 |n_1,n_2,n_3,n_4\rangle,\nonumber\\
&\Phi_4^{\dagger}\Phi_4|n_1,n_2,n_3,n_4\rangle =-n_4 |n_1,n_2,n_3,n_4\rangle.  
\label{schwingeropeeigenso32nonuni}
\end{align}
Hence, the eigenvalues of (\ref{twoequationsaboutfuzzyh22}) are derived as 
\begin{subequations}
\begin{align}
&\eta_{ab}X^a X^b=n(n+4), \label{squareradih22}\\
&X^5=n_1+n_2-n_3-n_4, 
\end{align}\label{spectrah22x}
\end{subequations}
where 
\begin{equation}
n\equiv n_1+n_2+n_3+n_4.
\end{equation}
With a given $n$, square of the radius of $H_F^{2,2}$ is fixed, and $X^5$ takes the following values:  
\begin{equation}
X^5=n,n-2,n-4,n-6,\cdots,-n. \label{spextx5h22}
\end{equation}
The spectra (\ref{spectrah22x})  coincide with those of the fuzzy four-sphere.  
The eigenstates (\ref{so32eigen}) are still degenerate for  given $n$ and $X^5$, and such ``internal structure'' gives rise to symmetry enhancement of fuzzy ultra-hyperboloids as we shall see later.   

The dual state of (\ref{so32eigen}) for positive inner product is  given by 
\begin{align}
\langle \!\langle n_1,n_2,n_3,n_4|&=\frac{1}{\sqrt{n_1!n_2!n_3!n_4!}} \langle 0|(\Phi^4)^{n_4} (\Phi^3)^{n_3}(\Phi^2)^{n_2}(\Phi^1)^{n_4}\nonumber\\
&=(-1)^{n_2+n_4}\frac{1}{\sqrt{n_1!n_2!n_3!n_4!}} \langle 0|(\Phi_4)^{n_4} (\Phi_3)^{n_3}(\Phi_2)^{n_2}(\Phi_1)^{n_4}, 
\end{align}
where $\Phi^{\alpha}=(k)^{\alpha\beta}\Phi_{\beta}$, $i.e.$ $(\Phi^1,\Phi^2,\Phi^3,\Phi^4)=( \Phi_1,-\Phi_2,\Phi_3,-\Phi_4)$.

\subsubsection{Infinite dimensional unitary representation }

We redefine the Schwinger operator by the interchange, $\Phi_2\leftrightarrow \Phi_2^{\dagger}$ and $\Phi_4\leftrightarrow \Phi_4^{\dagger}$,  for its components to satisfy the usual commutation relations:  
\begin{equation}
[\Phi_{\alpha},\Phi_{\beta}^{\dagger}]=\delta_{\alpha\beta},~~[\Phi_{\alpha},\Phi_{\beta}]=[\Phi_{\alpha}^{\dagger},\Phi_{\beta}^{\dagger}]=0.   
\end{equation}
With such newly defined Schwinger operators,  $X^a$ (\ref{defanothX2}) are rewritten as  
\begin{align}
&X^1=i\Phi_1^{\dagger}\Phi_4^{\dagger}+i\Phi_2\Phi_3-i\Phi_1\Phi_4-i\Phi_2^{\dagger}\Phi_3^{\dagger} ,\nonumber\\
&X^2=\Phi_1^{\dagger}\Phi_4^{\dagger}-\Phi_2\Phi_3+\Phi_1\Phi_4-\Phi_2^{\dagger}\Phi_3^{\dagger},\nonumber\\
&X^3=i\Phi_1^{\dagger}\Phi_3+i\Phi_4^{\dagger}\Phi_2-i\Phi_3^{\dagger}\Phi_1-i\Phi_2^{\dagger}\Phi_4 ,\nonumber\\
&X^4=\Phi_1^{\dagger}\Phi_3-\Phi_4^{\dagger}\Phi_2+\Phi_3^{\dagger}\Phi_1-\Phi_2^{\dagger}\Phi_4,\nonumber\\
&X^5= \Phi_1^{\dagger}\Phi_1-\Phi_2\Phi_2^{\dagger}-\Phi_3^{\dagger}\Phi_3+\Phi_4\Phi_4^{\dagger} =\Phi_1^{\dagger}\Phi_1-\Phi_2^{\dagger}\Phi_2-\Phi_3^{\dagger}\Phi_3+\Phi_4^{\dagger}\Phi_4. 
\label{expressx1tox5}
\end{align}
 Notice that both $X^1$ and $X^2$ originally constructed  by anti-hermitian gamma matrices become  particle-number non-conserving operators, and the others, $X^3$, $X^4$ and $X^5$, are particle-number conserving operators.  
 From (\ref{expressx1tox5}), a straightforward calculation shows 
\begin{equation}
\eta_{ab}{X}^a{X}^b=(\Phi^{\dagger}k\Phi-2)(\Phi^{\dagger}k\Phi+2). 
\label{schsqarah22}
\end{equation}
The simultaneous eigenstates of $X^5$  (\ref{expressx1tox5}) and (\ref{schsqarah22}) are given by\footnote{Details about the irreducible representation of the $SO(3,2)$ groups are found in Refs.\cite{so32references, Evans1967}.} 
\begin{equation}
|n_1,n_2,n_3,n_4)=\frac{1}{\sqrt{n_1!n_2!n_3!n_4!}} (\Phi^{\dagger}_1)^{n_1} (\Phi^{\dagger}_2)^{n_2}(\Phi^{\dagger}_3)^{n_3}(\Phi^{\dagger}_4)^{n_4}|\text{vac}\rangle,
\label{roundstate}
\end{equation}
with the eigenvalues 
\begin{subequations}
\begin{align}
&\eta_{ab}{X}^a{X}^b=(n-2)(n+2)=n^2-4, 
\label{spectraxxfour2} \\
&X^5=n_1-n_2+n_3-n_4,
\end{align}  
\end{subequations}
where $n$ denotes the eigenvalues of $\Phi^{\dagger}k\Phi$:  
\begin{equation}
n\equiv n_1-n_2-n_3+n_4.
\end{equation}
For the semi-positive definiteness of  square of the radius, $n$ in (\ref{spectraxxfour2}) should be taken as  $n\ge 2$.  
With fixed $n$, $X^5$ is given by 
\begin{equation}
X^5=n+2(n_3-n_4)=n+2\Delta n, 
\label{rangex5fuzzyh22}
\end{equation}
where $\Delta n\equiv n_3-n_4=0,\pm 1,\pm 2, \pm 3, \cdots.$
Since $\Delta n$ takes an arbitrary integer, the spectra of $X^5$ range from $-\infty$ to $+\infty$ with interval 2. 

 The dual state of (\ref{roundstate}) is given by 
\begin{equation}
(n_1,n_2,n_3,n_4|=\frac{1}{\sqrt{n_1!n_2!n_3!n_4!}} \langle \text{vac}|(\Phi_4)^{n_4} (\Phi_3)^{n_3}(\Phi_2)^{n_2}(\Phi_1)^{n_1},
\end{equation}
which gives rise to positive norm: 
\begin{equation}
(n_1,n_2,n_3,n_4|n'_1,n'_2,n'_3,n'_4)=\delta_{n_1 n'_1} \delta_{n_2 n'_2}\delta_{n_3 n'_3}\delta_{n_4 n'_4}.
\end{equation}

\subsection{Hybrid 2nd Hopf map: $H^{4,0}\simeq H^{4,3}/S^3$}\label{subse:hybrid2}

Next, we construct the hybrid 2nd Hopf map. As suggested in Section \ref{subsec;cousins}, the $SO(1,4)$ gamma matrices are crucial in constructing the hybrid 2nd Hopf map.  The $SO(1,4)$ gamma matrices are introduced so as to satisfy 
\begin{equation}
\{\gamma^a,\gamma^b\}=2{\eta}^{ab},
\end{equation}
with ${\eta}^{ab}={\eta}_{ab}=diag(-,-,-,-,+)$.  Explicitly, the $SO(1,4)$ gamma matrices are  given by 
\begin{align}
&\gamma^1= 
\begin{pmatrix}
0  & i\sigma^1  \\
i\sigma^1 & 0 
\end{pmatrix},~~\gamma^2=
\begin{pmatrix}
0 & i\sigma^2 \\
i\sigma^2 & 0 
\end{pmatrix},~~\gamma^3=
\begin{pmatrix}
0 & i\sigma^3 \\
i\sigma^3 & 0 
\end{pmatrix},\nonumber\\
&\gamma^4= 
\begin{pmatrix}
 0& 1_2  \\
 -1_2 & 0
\end{pmatrix},~~\gamma^5= 
\begin{pmatrix}
1_2 & 0 \\
0 & -1_2  
\end{pmatrix},
\label{so14gammaexp}
\end{align}
From (\ref{so12nhermiti}),  the $SO(1,4)$ hermitianizing matrix  is constructed as 
\begin{equation}
k=-\gamma^1\gamma^2\gamma^3\gamma^4=\begin{pmatrix}
 1_2 & 0  \\
0 & -1_2 
\end{pmatrix}=\gamma^5. 
\label{hermitiso14}
\end{equation}
Notice that $SO(1,4)$ and  $SO(3,2)$ hermitianizing matrices,  (\ref{hermitiso14})  and (\ref{hermitiso32}),  are unitary equivalent. 
The hermitianized $SO(1,4)$ gamma matrices $k^a=k\gamma^a$ are derived as 
\begin{align}
&k^1=
\begin{pmatrix}
0 & i\sigma^1 \\
-i\sigma^1 & 0  
\end{pmatrix},~~k^2=
\begin{pmatrix}
0 &  i\sigma^2 \\
-i\sigma^2 & 0   
\end{pmatrix},~~k^3=
\begin{pmatrix}
 0 & i\sigma^3 \\
-i\sigma^3 & 0 
\end{pmatrix},\nonumber\\
&k^4= 
\begin{pmatrix}
0 & 1_2 \\
1_2 & 0 
\end{pmatrix},~~k^5=  
\begin{pmatrix}
1_2 & 0 \\
0 & 1_2  
\end{pmatrix}. 
\end{align}
With $k$, we introduce an $SO(1,4)$ Dirac spinor $\psi$ subject to the ``normalization'' condition  
\begin{equation}
\psi^{\dagger}k\psi=\psi_1^*\psi_1+\psi_2^*\psi_2-\psi_3^*\psi_3-\psi_4^*\psi_4=1, 
\end{equation}
which geometrically represents $H^{4,3}$. Since the hermitianizing matrices of $SO(1,4)$ and $SO(3,2)$ are equivalent, the total manifolds of the split and hybrid  2nd Hopf maps are identically given by $H^{4,3}$.  
With $\psi$, the hybrid Hopf map is realized as 
\begin{equation}
x^a=\psi^{\dagger}k^a\psi, 
\label{so41map2nd}
\end{equation}
or 
\begin{align}
&x^1=i\psi_1^*\psi_4+i\psi_2^*\psi_3-i\psi_3^*\psi_2-i\psi_4^*\psi_1,\nonumber\\
&x^2=\psi_1^*\psi_4-\psi_2^*\psi_3-\psi_3^*\psi_2+\psi_4^*\psi_1,\nonumber\\
&x^3=i\psi_1^*\psi_3-i\psi_2^*\psi_4-i\psi_3^*\psi_1+i\psi_4^*\psi_2,\nonumber\\
&x^4=\psi_1^*\psi_3+\psi_2^*\psi_4+\psi_3^*\psi_1+\psi_4^*\psi_2,\nonumber\\
&x^5=\psi_1^*\psi_1+\psi_2^*\psi_2+\psi_3^*\psi_3+\psi_4^*\psi_4,  
\label{coodxso14}
\end{align}
which automatically satisfy the condition of two-leaf four-hyperboloid $H^{4,0}$: 
\begin{equation}
{\eta}_{ab}x^ax^b =-x^1x^1-x^2x^2-x^3x^3-x^4x^4+x^5x^5=(\psi^{\dagger}k\psi)^2=1.  
\end{equation}
From (\ref{coodxso14}), we find that $x^5\ge 0$ and  
 $x^a$ (\ref{coodxso14}) are coordinates on the upper leaf of $H^{4,0}$. 

\subsection{Fuzzy two-leaf four-hyperboloid: $H_F^{4,0}=EAdS_F^4$}\label{subsect:fyzzyfroutwoleaf}

We explore the fuzzy version of $H^{4,0}$.   

\subsubsection{Finite dimensional non-unitary representation}

We introduce  $SO(1,4)$ Schwinger operator $\Phi=(\Phi_1,\Phi_2,\Phi_3,\Phi_4)^t$ whose components satisfy 
\begin{equation}
[\Phi_\alpha,\Phi_{\beta}^{\dagger}]=(k)_{\alpha\beta},~~~~[\Phi_{\alpha},\Phi_{\beta}]=[\Phi_{\alpha}^{\dagger},\Phi_{\beta}^{\dagger}]=0, 
\label{commuwithk2}
\end{equation}
where $k$ is the $SO(1,4)$ hermitianizing matrix (\ref{hermitiso14}).  
 Coordinates on $H_F^{4,0}$ are constructed as 
\begin{equation}
{X}^a=\Phi^{\dagger}k^a\Phi, 
\end{equation}
which provide 
\begin{subequations}
\begin{align}
&{\eta}_{ab}{X}^a{X}^b=-X^1X^1-X^2X^2-X^3X^3-X^4X^4+X^5X^5=(\Phi^{\dagger}k\Phi)(\Phi^{\dagger}k\Phi+4),
\label{etaxxso14}\\
&X^5=\Phi_1^{\dagger}\Phi_1+\Phi_2^{\dagger}\Phi_2+\Phi_3^{\dagger}\Phi_3+\Phi_4^{\dagger}\Phi_4. 
\label{twosquarex514}
\end{align}\label{twosquarex514total}
\end{subequations}
Their simultaneous eigenstates are given by 
\begin{equation}
|n_1,n_2,n_3,n_4\rangle=\frac{1}{\sqrt{n_1!n_2!n_3!n_4!}} (\Phi^{\dagger}_1)^{n_1} (\Phi^{\dagger}_2)^{n_2}(\Phi^{\dagger}_3)^{n_3}(\Phi^{\dagger}_4)^{n_4}|0\rangle,    
\label{irrso41re}
\end{equation}
with the eigenvalues    
\begin{subequations}
\begin{align}
&{\eta}_{ab}{X}^a{X}^b=n(n+4),\\
&X^5=n_1+n_2-n_3-n_4.   \label{x5spectraso14}
\end{align}
\end{subequations}
Here, $n\equiv n_1+n_2+n_3+n_4$. 
(\ref{irrso41re}) is also a simultaneous eigenstate of  the four number operators made of the Schwinger operator:  
\begin{align}
&\Phi_1^{\dagger}\Phi_1|n_1,n_2,n_3,n_4\rangle =n_1 |n_1,n_2,n_3,n_4\rangle,\nonumber\\
&\Phi_2^{\dagger}\Phi_2|n_1,n_2,n_3,n_4\rangle =n_2 |n_1,n_2,n_3,n_4\rangle,  \nonumber\\
&\Phi_3^{\dagger}\Phi_3|n_1,n_2,n_3,n_4\rangle =-n_3 |n_1,n_2,n_3,n_4\rangle,\nonumber\\
&\Phi_4^{\dagger}\Phi_4|n_1,n_2,n_3,n_4\rangle =-n_4 |n_1,n_2,n_3,n_4\rangle.    
\label{schwingerso14eigenonuni}
\end{align}
Comparing to the non-unitary representation of $H_{F}^{2,2}$ (Section \ref{so32fuzzy4}), one may find that the spectra of $SO(1,4)$  (\ref{schwingerso14eigenonuni}) and  $SO(3,2)$ Schwinger operators (\ref{schwingeropeeigenso32nonuni}) are identical by the interchange, $n_2\leftrightarrow n_3$.  Thus, 
$H_F^{2,2}$ and $H^{4,0}_F$  are not ``distinguished'' only by their non-unitary representation\footnote{
The non-unitary representation (\ref{irrso41re}) is  naturally regarded as fully symmetric representation of $SU(2,2)$. Also, $k$ (\ref{hermitiso14}) is the $SU(2,2)$ invariant matrix. This suggests that the enhanced symmetry of
  $H^{4,0}_F$ is  $SU(2,2)$. We will revisit this in Section \ref{subsec:quantumf}.}.

\subsubsection{Infinite dimensional unitary representation}

By the replacement $\Phi_3\leftrightarrow \Phi_3^{\dagger}$ and $\Phi_4 \leftrightarrow \Phi_4^{\dagger}$, we can define new creation and annihilation operators that satisfy the usual commutation relations:  
\begin{equation}
[\Phi_{\alpha},\Phi_{\beta}^{\dagger}]=\delta_{\alpha\beta},~~[\Phi_{\alpha},\Phi_{\beta}]=[\Phi_{\alpha}^{\dagger},\Phi_{\beta}^{\dagger}]=0. 
\end{equation}
With the newly defined Schwinger operator, the fuzzy coordinates on $H_{F}^{4,0}$ are represented as 
\begin{align}
&X^1=i\Phi_1^{\dagger}\Phi_4^{\dagger}+i\Phi_2^{\dagger}\Phi_3^{\dagger}-i\Phi_1\Phi_4-i\Phi_2\Phi_3,\nonumber\\
&X^2=\Phi_1^{\dagger}\Phi_4^{\dagger}-\Phi_2^{\dagger}\Phi_3^{\dagger}+\Phi_1\Phi_4-\Phi_2\Phi_3,\nonumber\\
&X^3=i\Phi_1^{\dagger}\Phi_3^{\dagger}-i\Phi_2^{\dagger}\Phi_4^{\dagger}-i\Phi_1\Phi_3+i\Phi_2\Phi_4,\nonumber\\
&X^4=\Phi_1^{\dagger}\Phi_3^{\dagger}+\Phi_2^{\dagger}\Phi_4^{\dagger}+\Phi_1\Phi_3+\Phi_2\Phi_4,\nonumber\\
&X^5=\Phi_1^{\dagger}\Phi_1+\Phi_2^{\dagger}\Phi_2+\Phi_3\Phi_3^{\dagger}+\Phi_4\Phi_4^{\dagger}. 
\label{xanewso14}
\end{align}
Again, the fuzzy coordinates $X^1$, $X^2$, $X^3$ and $X^4$  originally constructed by anti-hermitian gamma matrices become particle-number non-conserving operators.  
From (\ref{xanewso14}), we have  
\begin{subequations}
\begin{align}
&{\eta}_{ab}{X}^a{X}^b=-X^1X^1-X^2X^2-X^3X^3+X^4X^4+X^5X^5=(\Phi^{\dagger}k\Phi-2)(\Phi^{\dagger}k\Phi +2),\\
&X^5=\Phi_1^{\dagger}\Phi_1+\Phi_2^{\dagger}\Phi_2+\Phi_3^{\dagger}\Phi_3+\Phi_4^{\dagger}\Phi_4+2. 
\end{align}\label{radiussquarex2}
\end{subequations}
Their simultaneous eigenstates are 
\begin{equation}
|n_1,n_2,n_3,n_4)=\frac{1}{\sqrt{n_1!n_2!n_3!n_4!}} (\Phi^{\dagger}_1)^{n_1} (\Phi^{\dagger}_2)^{n_2}(\Phi^{\dagger}_3)^{n_3}(\Phi^{\dagger}_4)^{n_4}|\text{vac}\rangle,
\label{siminfstateo14}
\end{equation}
where $n_1,n_2,n_3,n_4$ are non-negative integers and $|\text{vac}\rangle$ is defined as $\Phi_1 |\text{vac}\rangle=\Phi_2 |\text{vac}\rangle=\Phi_3 |\text{vac}\rangle=\Phi_4 |\text{vac}\rangle=0$. 
The eigenvalues  are given by  
\begin{subequations}
\begin{align}
&{\eta}_{ab}{X}^a{X}^b=(n-2)(n+2)=n^2-4, \label{squareradi2014unit}\\
&X^5=n_1+n_2+n_3+n_4+2, 
\label{x54dhyperboloidfuzzy2}
\end{align}
\end{subequations}
with  
\begin{equation}
n\equiv n_1+n_2-n_3-n_4.
\end{equation}
For (semi-)positive definiteness of square of the radius, $n$ in (\ref{squareradi2014unit}) should be taken as  $n\ge 2$. With a given $n$, the spectra of $X^5$ read as  
\begin{equation}
X^5=n+2+2(n_3+n_4)=n+2,~ n+4,~ n+6,~ \cdots.
\label{rangefourhyper} 
\end{equation}
Note that the range of the spectra of  $X^5$ (\ref{rangefourhyper}) is different from that of  $H_F^{2,2}$ (\ref{rangex5fuzzyh22}).  This is consistent with the fact that  $H_F^{4,0}$ corresponds to fuzzyfication (of the upper leaf) of two-leaf four-hyperboloid and $H_F^{2,2}$ corresponds to another four-fuzzy hyperboloid.  
In the language of $SO(1,4)$ representation theory, the present unitary representation of $H_F^{4,0}$ corresponds to the discrete series of $SO(1,4)$ 
 \footnote{
With use of $X^a$ (\ref{xanewso14}), the $SO(1,4)$ generators are constructed as 
\begin{equation}
X^{ab}\equiv -i\frac{1}{4}[X^a,X^b], 
\end{equation}
and the $SO(1,4)$ quadratic Casimir is given by 
\begin{equation}
\sum_{a<b} X_{ab}X^{ab}=\frac{1}{2}(\Phi^{\dagger}k\Phi+2)(\Phi^{\dagger}k\Phi-2)=\frac{1}{2}(n+2)(n-2).
\label{so41casimirspe}
\end{equation}
In the Dixmier notation\cite{Dixmier1961}, with $SO(1,4)$ generators $L_{ab}$,  the quadratic Casimir and its eigenvalues are generally given by 
\begin{equation}
C_2=\sum_{a<b}L_{ab}L^{ab}=p(p+1)+(q+1)(q-2).
\label{2ndcasimirso41ge} 
\end{equation}
The eigenvalues  of the unitary representation of (\ref{so41casimirspe}) are realized as the discrete series by the choice of $p=q=\frac{n}{2}$ in (\ref{2ndcasimirso41ge}). See Refs.\cite{Gazeau&Toppan,Dixmier1961,Bohm1973,newton,thomas} for more details about the representation theory of $SO(1,4)$.} .  

Similar to $H^{4,0}$,  the isometry of 
 $H^{1,3}(=dS^4)$  is given by  $SO(1,4)$.    Fuzzyfication of $H^{4,0}$ is realized by adopting the discrete series of $SO(1,4)$,  while  
fuzzyfication of $H^{1,3}(=dS^4)$  is by the principal series \cite{Gazeauetal2006,Gazeau&Toppan}  \footnote{Different from the present work, in Refs.\cite{Gazeauetal2006,Gazeau&Toppan}  $SO(1,4)$  quartic Casimir  made of Pauli-Lubanski vectors was adopted to define fuzzy $dS^4$. 
Such fuzzy $dS^4$ construction provides another natural generalization of fuzzy hyperboloid. }.  (This is a higher dimensional analogue of the relations between $H_F^{2,0}$ and $H_F^{1,1}$ mentioned in the last paragraph of Section \ref{subsec:twohyperinf}.)

\subsection{Enhanced Symmetry}\label{subsec:quantumf}

So far, everything is parallel between fuzzy two- and four-hyperboloids, except for the extra degeneracy of states in 4D case. Such ``extra'' degrees of freedom reflect symmetry enhancement particular to higher dimensional fuzzy hyperboloid.  

\subsubsection{Enhanced algebra and internal structure}

Unlike fuzzy two-hyperboloid, the coordinates  $X^a$ on fuzzy four-hyperboloid  do not satisfy a closed algebra by themselves.  
(In the following, we treat $H_{F}^{2,2}$ and $H_F^{4,0}$ in a unified way : For $H_F^{2,2}$,    $\eta^{ab}$ are taken as the $SO(3,2)$ metric, while for $H^{4,0}$,  the $SO(1,4)$ metric.)  
The commutators of $X^a$ yield ``new'' operators $X^{ab}$; 
\begin{equation}
[X^a,X^b]=4iX^{ab}.  
\label{commufirstimso32}
\end{equation}
Together with $X^{ab}$, $X^a$ satisfy the closed algebra 
\begin{align} 
&[X^{ab},X^c]=i\frac{1}{2}(\eta^{ac}{X}^b-\eta^{bc}{X}^a),\nonumber\\
&[X^{ab},X^{cd}]=i(\eta^{ac}{X}^{bd}-\eta^{ad}{X}^{bc}+\eta^{bd}{X}^{ac}-\eta^{bc}{X}^{ad}),  
\label{so24pre}
\end{align}
where 
\begin{equation}
X^{ab}=-i\frac{1}{4}[X^a,X^b]=\Phi^{\dagger}k\gamma^{ab}\Phi, 
\end{equation}
with
\footnote{ $\gamma^{ab}$ (\ref{defofgammab}) satisfy 
\begin{align}
&[\gamma^{ab},\gamma^{c}]=i({\eta}^{ac}\gamma^b-{\eta}^{bc}\gamma^a),\nonumber\\
&[\gamma^{ab},\gamma^{cd}]=i({\eta}^{ac}\gamma^{bd}-{\eta}^{ad}\gamma^{bc}+{\eta}^{bd}\gamma^{ac}-{\eta}^{bc}\gamma^{ad}). 
\label{relationsgammagamma}
\end{align}
These relations  are readily derived  by using $\{\gamma^a,\gamma^b\}=2\eta^{ab}$  and  the formula 
\begin{equation}
[AB,CD]=A\{B,C\}D-AC\{B,D\}+\{A,C\}DB-C\{A,D\}B.   
\end{equation}
} 
\begin{equation}
\gamma^{ab}=-i\frac{1}{4}[\gamma^a,\gamma^b]. 
\label{defofgammab}
\end{equation}
Define $X^{AB}$ $(A,B=1,2,\cdots,6)$  as  
\begin{align}
&X^{AB}\equiv X^a ~~~~~~~~\text{for}~~~~~ (A,B)=(6,a),\nonumber\\
&X^{AB}\equiv X^{ab}~~~~~~~\text{for}~~~~~ (A,B)=(a,b). 
\label{su22definition}
\end{align}
The above algebras, (\ref{commufirstimso32}) and (\ref{so24pre}), are concisely rewritten as 
\begin{equation}
[X^{AB},X^{CD}]=i(\eta^{AC}{X}^{BD}-\eta^{AD}{X}^{BC}+\eta^{BD}{X}^{AC}-\eta^{BC}{X}^{AD}), 
\label{so24algebraandso42}
\end{equation}
where $\eta^{AB}$ denotes  $SO(4,2)$  or $SO(2,4)$ metric corresponding to $H_F^{2,2}$ or $H_{F}^{4,0}$  
 \footnote{If we defined $X^{AB}$ as 
$X^{AB}\equiv iX^a$ for $(A,B)=(6,a)$ and 
$X^{AB}\equiv X^{ab}$ for $(A,B)=(a,b)$,  
 $X^{AB}$ would become  $SO(3,3)$ generators. However, in this case, hermiticity of $X^{AB}$ is not coherent: $X^{ab}$ are hermitian but $X^{6a}$ are anti-hermitian. Then, we adopt the definition (\ref{su22definition}). }.   In either cases, $X^{AB}$ satisfy the $SU(2,2)(\simeq SO(4,2)\simeq SO(2,4))$ algebra.   Thus, the total algebras that correspond to $H_F^{2,2}$ and $H^{4,0}_F$ are identically given by $SU(2,2)$. 
This is a non-compact counterpart \cite{arXiv:0902.2523} of the enhanced $SU(4)$ algebra of fuzzy four-sphere \cite{HoRamgoolam2002,Kimura2002,Kimura2003}:  
\begin{center}
\begin{tabular}{cccccc}
 Fuzzy manifold & ~~~~ $S_F^4$ & ~~~~ &   $H_F^{4,0}$     &  ~~~~& $H_F^{2,2}$  
\\ \\
 \vspace{0.3cm} 
Original algebra  & ~~~~~ $SO(5)$ &~~~~  & $SO(1,4)$  & ~~~~ & $SO(3,2)$  \\
  \vspace{0.3cm}
  Enhanced algebra & ~~~~~ $SO(6)\simeq SU(4)$ &~~~~ & $SO(2,4)\simeq SU(2,2)$  & ~~~~&   $SO(4,2)\simeq SU(2,2)$   \\ 
\end{tabular}
\end{center}

Thus, the total fuzzy coordinates of fuzzy four-hyperboloid may be considered as $X^a$ and $X^{ab}$.  Existence of $X^{ab}$ suggests ``internal structure'' of fuzzy four-hyperboloid.  
With a fixed latitude of $X^5$, the remaining $X^{\mu}$  ($\mu=1,2,3,4$) construct   $X^{\mu\nu}=-i\frac{1}{4}[X^{\mu},X^{\nu}]$ ($\mu,\nu=1,2,3,4$) that satisfy  $SO(2,2)$ or  $SO(4)$ algebra for $H^{2,2}_F$ or  $H^{4,0}_F$. Notice that both of these are semi-simple algebras, $i.e.$ $so(2,2)\simeq su(1,1)\oplus su(1,1)$ and $so(4)\simeq su(2)\oplus su(2)$. Since $SU(1,1)$ and $SU(2)$ are the algebras  to define  $H^{2,0}_F$ and $S_F^2$ respectively, such algebraic decomposition implies that at a fixed latitude on fuzzy four hyperboloid, there exists  ``fuzzy bundle'' made of  two fuzzy two-hyperboloids for $H_F^{2,2}$, and two fuzzy two-spheres for $H_{F}^{4,0}$. Indeed, the  fuzzy four-hyperboloids  have the following geometries   
\begin{equation}
H^{2,2}_F(n)|_{X^5}~\sim~ H^{2,0}_F\biggl(\frac{n+X^5}{2}\biggr)\oplus H^{2,0}_F\biggl(\frac{n-X^5}{2}\biggr), 
\label{intuitivepich22}
\end{equation}
and 
\begin{equation}
H^{4,0}_F(n)|_{X^5}~\sim ~S_F^{2}\biggl(\frac{n+X^5}{2}\biggr)\oplus S_F^{2}\biggl(\frac{n-X^5}{2}\biggr), 
\label{intuitivepich22h}
\end{equation}
where $H^{2,0}_F(n)$ and $S_F^{2}(n)$ denote the fuzzy $H^{2,0}$ and fuzzy $S^2$ with radius $n$, respectively. 
According to the similar arguments in fuzzy four-sphere \cite{Hasebe2010},  it can be shown that the radii of fuzzy hyperboloid- and  fuzzy sphere-bundles are specified by $n$ and  $X^5$ as given by (\ref{intuitivepich22}) and (\ref{intuitivepich22h}). 
Eq.(\ref{intuitivepich22}) indicates that the Hilbert space of the $SO(3,2)$ representation (\ref{so32eigen}) which lives on the latitude  $X^5$ of $H_F^{2,2}$ with radius $n$  is given by the direct-sum of the Hilbert spaces of two $SU(1,1)$ representations (\ref{n1n2ket}) of  fuzzy two-hyperboloids with radii $(n+X^5)/2$  and  $(n-X^5)/2$. Similarly, Eq.(\ref{intuitivepich22h}) represents that the Hilbert space of the $SO(1,4)$ representation (\ref{irrso41re}) which lives on the latitude $X^5$ of $H_F^{4,0}$ with radius $n$  is given by the direct-sum of the Hilbert spaces of two $SU(2)$ representations of  fuzzy two-hyperboloids with radii $(n+X^5)/2$  and  $(n-X^5)/2$ \footnote{
Non-unitary representations of $H^{2,2}_F$ (\ref{so32eigen}) and $H^{4,0}_F$ (\ref{irrso41re}) carry  four quantum numbers, $n_1$, $n_2$, $n_3$ and $n_4$. Two of them  specify the radius of fuzzy four-hyperboloid $n(=n_1+n_2+n_3+n_4)$ and the spectrum of 
$X^5$.  
The other two  specify two latitudes of the two fuzzy hyperboloid- or two fuzzy sphere-bundles. }.   
Around the ``north pole'' $X^5\sim n$, (\ref{intuitivepich22}) and (\ref{intuitivepich22h}) are respectively reduced to  
$H^{2,2}_F(n)|_{X^5\sim n}~\sim~ H^{2,0}_F(n)$ 
and 
$H^{4,0}_F(n)|_{X^5\sim 0}~\sim ~S_F^{2}(n )$, 
and hence  $H_F^{2,2}$ and $H^{4,0}_F$ are locally expressed as 
\begin{equation}
H^{2,2}_F(n)~\sim~H^{2,2}(n)\otimes H^{2,0}_F(n) 
\label{localhprodh22}
\end{equation}
and 
\begin{equation}
H^{4,0}_F(n)~\sim ~H^{4,0}(n)\otimes S_F^{2}(n).  
\label{localhprodh40}
\end{equation}
 Thus,   $H_F^{2,2}$ is locally equivalent to  fibration of fibre $H^{2,0}_F$ over the basemanifold $H^{2,2}$ and  similarly,  $H_F^{4,0}$ is equivalent to fibration of $S^{2}_F$ over $H^{4,0}$.   
With use of the original symmetries of the hyperbolic basemanifolds, (\ref{localhprodh22}) and (\ref{localhprodh40}) can be expressed as  
\begin{equation}
H_F^{2,2}~\simeq~ SO(3,2)/U(1,1)
 \label{cosetrepfuzzyfour1}
\end{equation}
and 
\begin{equation} 
H_F^{4,0}~\simeq~ SO(1,4)/U(2), 
\label{cosetrepfuzzyfour2}
\end{equation}
where we utilized $SO(3,2)/U(1,1) \sim H^{2,2}\otimes SO(2,2)/U(1,1) \sim H^{2,2}\otimes SU(1,1)/U(1)$ for (\ref{cosetrepfuzzyfour1}), and $SO(1,4)/U(2) \sim H^{4,0}\otimes SO(4)/U(2) \sim H^{4}\otimes SU(2)/U(1)$ for (\ref{cosetrepfuzzyfour2}).

The non-compact Hopf maps are a ``classical'' counterpart of the fuzzy hyperboloids, and then the corresponding structures can also be observed in the geometry of the non-compact Hopf maps.  
To see this, consider  the total manifold of the  split 2nd Hopf map, $H^{4,3}$, and its corresponding  symplectic manifold $\mathbb{C}P^{1,2}$ \cite{arXiv:0902.2523}\footnote{ Eq.(\ref{coset1}) implies that $H^{4,3}$ can be expressed as two distinct fibrations,  $H^{4,3}\sim \mathbb{C}P^{1,2}\otimes S^1$ and $H^{4,3}\sim H^{2,2}\otimes H^{2,1}$. We reconsider this in the context of lowest Landau level physics in Section \ref{subsec:lllfourhy}.  } 
\begin{equation}
\mathbb{C}P^{1,2}\simeq H^{4,3}/S^1\sim H^{2,2}\otimes H^{2,1}/S^1\simeq H^{2,2}\otimes H^{2,0}.
\label{coset1}
\end{equation}
One may find that the last expression on the right-hand side of (\ref{coset1}) corresponds to (\ref{localhprodh22}). 
Similarly for the hybrid 2nd Hopf map, we have 
\begin{equation}
\mathbb{C}P^{1,2}\simeq H^{4,3}/S^1 \sim H^{4,0}\otimes S^{3}/S^1\simeq H^{4,0}\otimes S^{2}. 
\label{coset2}
\end{equation}
Again, the last expression on the right-hand side of (\ref{coset2}) corresponds to (\ref{localhprodh40}). 
Thus, the indefinite complex projective space $\mathbb{C}P^{1,2}$ is considered as the classical counterpart of the $H_F^{2,2}$ and $H_F^{4,0}$ (see Appendix \ref{sec:appeninfcom} for  $\mathbb{C}P^{p,q}$).  
Meanwhile,  $\mathbb{C}P^{1,2}$ is represented by the coset:  
\begin{equation} 
\mathbb{C}P^{1,2}\simeq SU(2,2)/U(1,2).  
\end{equation}
The $SU(2,2)$ symmetry naturally appears as the isometry of $\mathbb{C}P^{1,2}$.  This is another way of understanding the appearance of the $SU(2,2)$ in the geometry of fuzzy $H_F^{2,2}$ and fuzzy $H^{4,0}$. 

The relevant mathematical structures of 4D fuzzy sphere and fuzzy hyperboloids are summarized in Table \ref{5Dspacetable}.  
\begin{table}
\renewcommand{\arraystretch}{1}
\vspace{0.5cm}
\hspace{-1cm}
\begin{tabular}{|c||c|c|c|c|c|
}
\hline   Symmetry     & Original manifolds  &  Hopf Maps & $k$ matrix  
 & Enhanced Algebra & Fuzzy Manifolds \\ 
\hline
\hline  $SO(5)$ 
     & $S^{4}= H^{0,4}$  & Compact  & Hermite 
   & $SO(6)\simeq SU(4)$   & $S_F^{4}\simeq \mathbb{C}P^{3,0}$  \\  
\hline $SO(4,1)$ & $dS^4= H^{1,3}$  &  \slash & Anti-hermite 
 &   \slash  &   \slash \\ 
\hline $SO(3,2)$ & $H^{2,2}$ & Split & Hermite 
  & $SO(2,4)\simeq SU(2,2)$    &$H_F^{2,2}\simeq \mathbb{C}P^{1,2}$  \\
\hline 
$SO(2,3)$ & $AdS^4= H^{3,1}$ &   \slash  &  Anti-hermite 
 &   \slash  &   \slash\\ 
\hline $SO(1,4)$ & $EAdS^4=H^{4,0}$ & Hybrid & Hermite   
 &$SO(4,2)\simeq SU(2,2)$
    & $H_F^{4,0}\simeq \mathbb{C}P^{1,2}$  \\ 
\hline
\end{tabular}
 \caption{Fuzzyfication of four-hyperboloids and related properties. } 
 \label{5Dspacetable}
\end{table} 

\subsubsection{Quantum fluctuations of geometry}

Physically, the symmetry enhancement is brought by quantum fluctuations of the geometry on fuzzy hyperboloid \cite{balachandran2001,Hasebe2011}.   
To see this, we first introduce the coherent state  $|\omega\rangle$ to satisfy 
\begin{equation}
\eta_{ab}x^a X^b|\omega\rangle ={n}|\omega\rangle.  
\end{equation}
Here, $X^a$ and $x^a$ are coordinates on fuzzy and classical four-hyperboloid respectively, and $\eta_{ab}$ is the corresponding indefinite metric, $i.e.$ the $SO(1,4)$ metric for $H_F^{4,0}$ or the $SO(3,2)$ metric for $H_F^{2,2}$.    
The coherent state is derived as 
\begin{equation}
|\omega\rangle =\frac{1}{\sqrt{n!}}(\Phi^{\dagger}k\phi)^n|0\rangle, 
\end{equation}
where $k$ is the hermitianizing matrix, $\Phi$ is the Schwinger operator whose components satisfy $[\Phi_{\alpha},\Phi^{\dagger}_{\beta}]=k_{\alpha\beta}$, $[\Phi_{\alpha},\Phi_{\beta}]=[\Phi_{\alpha}^{\dagger},\Phi_{\beta}^{\dagger}]=0$, and $\phi$ is the non-compact Hopf spinor.   
The dual state of $|\omega\rangle$ is given by 
\begin{equation}
\langle \omega| =\frac{1}{\sqrt{n!}}\langle 0| (\phi^{\dagger}k\Phi)^n,  
\end{equation}
which satisfies $\langle \omega|\omega\rangle=1$.   
With the coherent state, one-point functions are derived as 
\begin{equation}
\langle \omega|X^a|\omega\rangle=n x^a.  
\end{equation}
Thus, one-point functions reduce to coordinates on the corresponding classical hyperboloid. 
Meanwhile, two-point functions are calculated as 
\begin{equation}
\langle \omega|X^a X^b|\omega \rangle= n^2 x^a x^b +n(-x^ax^b+2ix^{ab}+\eta^{ab}), 
\label{twopointfun}
\end{equation}
where $x^{ab}$ are defined by 
\begin{equation}
x^{ab}\equiv \frac{1}{n}\langle \omega|X^{ab}|\omega\rangle. 
\end{equation}
$x^{ab}$ do not have any counterpart in coordinates on the classical hyperboloid.  
Since $x^a$ and $x^{ab}$ are the expectation values of the $SU(2,2)$ operators,  they amount to  coordinates on $\mathbb{C}P^{1,2}$. 
The sub-dominant second  term  of the order $n$ on the right-hand side of (\ref{twopointfun}) signifies quantum fluctuations, and it includes  $x^{ab}$: the $\mathbb{C}P^{1,2}$geometry.   In the ``classical limit'' $n\rightarrow \infty$, we see that (\ref{twopointfun})   indeed reduces to its classical counterpart $\langle \omega|X^a X^b|\omega \rangle\rightarrow n^2 x^ax^b$, while in the quantum limit $n\simeq O(1)$, the particular $\mathbb{C}P^{1,2}$ coordinates $x^{ab}$ become comparable to the classical term $x^ax^b$ of the order $n^2$,  
 indicating that quantum fluctuations on fuzzy hyperboloid smear the original hyperbolic geometry to generate a new enhanced geometry of  $\mathbb{C}P^{1,2}$.

\section{Even Higher Dimensional Generalization of Fuzzy Hyperboloid}\label{sec:fuzh}

For the previous realization of fuzzy hyperboloids, 
we utilized gamma matrices of low dimensional indefinite orthogonal groups.   
In this section, we extend the previous analysis  to construct even higher dimensional fuzzy hyperboloids  based on general indefinite gamma matrices  (Section \ref{subsec:algebra} and \ref{subsec:classicalcou}). We also investigate  fuzzy-bundle structure of  fuzzy hyperboloids (Section \ref{subsec:fibrestruhyper}).

\subsection{Hierarchical structure of fuzzy ultra-hyperboloid}\label{subsec:algebra}

Remember, in the case of fuzzy $H^{2,2}$, its coordinates are essentially given by the $SO(3,2)$ gamma matrices. Therefore, 
 it  may be natural to define general fuzzy $H^{2p,2q}$ for their coordinates to satisfy the algebra of  $SO(2q+1,2p)$ gamma matrices. Such fuzzy coordinates do not close algebra by themselves without  introducing  $SO(2q+1,2p)$ generators.  In total, they  amount to $SO(2q+2,2p)$ algebra, and in this sense the $SO(2q+2,2p)$ algebra can be considered as the total algebra of fuzzy $H^{2p,2q}$.  For fuzzy four-hyperboloids, $H_F^{2,2}$ and $H_F^{4,0}$, their total algebras were identically given by  $SO(2,4)\simeq SO(4,2)\simeq SU(2,2)$.    
Thus, besides the original fuzzy coordinates that reduce to the classical coordinates of the original hyperboloid $H^{2p,2q}$, the closure of the algebra ``requires'' an extra fuzzy space spanned by the $SO(2q+1,2p)$ algebra. Such newly introduced fuzzy space is considered as   $H^{2p,2q-2}_F$ as we shall see below.  
On a fixed latitude of the basemanifold of $H^{2p,2q}$, which is simply realized by taking an eigenvalue of the fuzzy coordinate $X^{2p+2q+1}$,   the commutators between the remaining fuzzy coordinates $X^{\mu}$ ($\mu=1,2,\cdots,2p+2q$) yield the $SO(2q,2p)$ generators.  Thus at each latitude of the original hyperboloid, the $SO(2q,2p)$ algebra is defined. This can be regarded as a sort of ``fibration''.    
The $SO(2q,2p)$ algebra is the defining algebra of fuzzy $H^{2p,2q-2}$ as discussed above, and then 
``$SO(2q,2p)$''-fiber geometrically indicates $H_F^{2p,2q-2}$. 
In this sense,  $H_F^{2p,2q}$ is understood as such a fuzzy manifold whose  basemanifold is $H^{2p,2q}$ and  fibre is $H_F^{2p,2q-2}$: 
\begin{equation}
H_F^{2p,2q}~\sim~ H^{2p,2q}\otimes H_F^{2p,2q-2},  
\label{geometryofindeffuzzy}
\end{equation}
for $q\neq 0$. Consequently,  fuzzy hyperboloid is considered as a fibration of  lower dimensional fuzzy hyperboloid-bundle  on a hyperbolic basemanifold.  We designate such hierarchical geometry (\ref{geometryofindeffuzzy}) as the hyperbolic hierarchy. 
This is a generalization of the split Hopf map, since (\ref{geometryofindeffuzzy}) for $p=q=1$  reduces to (\ref{coset1}). 

For $q=0$, a special care is needed. In the case $q=0$, the ``fibre'' is given by $SO(2p)$ algebra that defines  $S_F^{2p-2}$. Then, the relation (\ref{geometryofindeffuzzy}) is modified to give   
\begin{equation}
H_F^{2p,0}~\sim~ H^{2p}\otimes S_F^{2p-2}.  
\label{geometryofindeffuzzy2}
\end{equation}
We designate the structure (\ref{geometryofindeffuzzy2}) as the hybrid hierarchy, since (\ref{geometryofindeffuzzy2}) is a generalization of the  hybrid Hopf map, in the sense that (\ref{geometryofindeffuzzy2}) for $p=2$ reduces  to the hybrid Hopf map (\ref{coset2}).  

The ultra-hyperboloids thus ``contain'' lower dimensional fuzzy hyperboloids (or spheres) as their fuzzy-fibre.  
Inversely, higher dimensional hyperboloids can be constructed by low dimensional fuzzy manifolds. 
Such geometrical structure is called  the dimensional hierarchy \cite{hep-th/0310274,Kimura2003}.  

\subsubsection{Hyperbolic hierarchy: Construction I 
}\label{subsub:stI}

We argue the dimensional  hierarchy in view of the structure of indefinite gamma matrices. 
The gamma matrices of $SO(2q+1,2p)$  are ``constructed'' by those of  $SO(2q-1,2p)$.  
We first introduce the 
 $SO(2q-1,2p)$ gamma matrices $\gamma^i$  $(i=1,2,\cdots,2p+2q-1)$ that satisfy 
\begin{equation}
\{\gamma^i, \gamma^j\}=2\eta^{ij},  
\end{equation}
with $\eta_{ij}=diag(\overbrace{+,+,\cdots,+}^{2q-1},\overbrace{-,-,\cdots,-}^{2p})$.  
The hermitianizing matrix $k$ is  hermitian, and $k\gamma^i$ are hermitianized gamma matrices. 
The $SO(2q+1,2p)$ gamma matrices are constructed as 
\begin{equation}
\Gamma^i=-\gamma^i\otimes \sigma^2=
\begin{pmatrix}
0 & i\gamma^i \\
-i\gamma^i & 0 
\end{pmatrix},~~
\Gamma^{2p+2q}=
\begin{pmatrix}
0 & \bold{1} \\
\bold{1} & 0 
\end{pmatrix},~~\Gamma^{2p+2q+1}=
\begin{pmatrix}
\bold{1} & 0 \\
0 & -\bold{1}
\end{pmatrix}, 
\label{formula2+gamma} 
\end{equation}
which satisfy 
\begin{equation}
\{\Gamma^a,\Gamma^b\}=2\eta^{ab}, 
\end{equation}
with $\eta^{ab}=(\eta^{ij},+,+)$. 
The construction of $SO(3,2)$ gamma matrices (\ref{explicitso32constgam}) is the simplest demonstration of the formula (\ref{formula2+gamma}) . 
The hermitianizing matrix $K$ is also given by 
\begin{equation}
K=k\otimes 1_2=
\begin{pmatrix}
k & 0 \\
0 & k 
\end{pmatrix}, 
\label{hermitanizingso2q+12p}
\end{equation}
and indeed the $SO(2q+1,2p)$ gamma matrices become hermitian:  
\begin{equation}
K\Gamma^i= 
\begin{pmatrix}
0 & ik\gamma^i \\
-ik\gamma^i & 0 
\end{pmatrix},~~K\Gamma^{2p+2q}=
\begin{pmatrix}
0 & k \\
k & 0 
\end{pmatrix},~~K\Gamma^{2p+2q+1}=
\begin{pmatrix}
k & 0 \\
0 & -k
\end{pmatrix}. 
\label{so2q+12pgammamatricesexp}
\end{equation}
In this way, the $SO(2q+1,2p)$ gamma matrices are constructed by the $SO(2q-1,2p)$ gamma matrices. 
Since the gamma matrices correspond to coordinates of fuzzy hyperboloid, in the language of geometry,  such hierarchical structure suggests that $H^{2p,2q}_F$ contains $H^{2p,2q-2}_F$ as its internal fuzzy space. This agrees with the observation (\ref{geometryofindeffuzzy}).

\subsubsection{Hybrid hierarchy: Construction II  
}\label{subsub:stII}

The $SO(2p+1,2q)$ gamma matrices can be constructed from the $SO(2q-1,2p)$ gamma matrices $\gamma^i$ as  
\begin{equation}
\Gamma^i=\gamma^i\otimes i\sigma^1=
\begin{pmatrix}
0 & i\gamma^i \\
i\gamma^i & 0 
\end{pmatrix},~~\Gamma^{2p+2q}=  
\begin{pmatrix}
0 & \bold{1} \\
-\bold{1} & 0 
\end{pmatrix},~~\Gamma^{2p+2q+1}=
\begin{pmatrix}
\bold{1} & 0 \\
0 & -\bold{1}
\end{pmatrix}, 
\label{formula2+gammadash}
\end{equation}
which satisfy 
\begin{equation}
\{\Gamma^a,\Gamma^b\}=2\eta^{ab}, 
\end{equation}
with $\eta^{ab}=(-\eta^{ij},-,+)$. 
The construction of $SO(1,4)$ gamma matrices (\ref{so14gammaexp})  is the simplest demonstration of the formula (\ref{formula2+gammadash}).  
The hermitianizing matrix $K$ is given by 
\begin{equation}
K=k\otimes \sigma^3=
\begin{pmatrix}
k & 0 \\
0 & -k 
\end{pmatrix}, 
\label{prehermiII}
\end{equation}
and the $SO(2p+1,2q)$ gamma matrices are hermitianized as  
\begin{equation}
K\Gamma^i= 
\begin{pmatrix}
0 & ik\gamma^i \\
-ik\gamma^i & 0 
\end{pmatrix},~~K\Gamma^{2p+2q}=
\begin{pmatrix}
0 & k \\
k & 0 
\end{pmatrix},~~K\Gamma^{2p+2q+1}=
\begin{pmatrix}
k & 0 \\
0 & k
\end{pmatrix}. 
\end{equation}
In particular $p=0$,  the $SO(1,2q)$ gamma matrices are constructed by the $SO(2q-1)$ gamma matrices:  With $SO(2q-1)$ gamma matrices $\gamma^i$  satisfying $\{\gamma^i,\gamma^j\}=2\delta^{ij}$,  we have  
\begin{equation}
K=\bold{1}\otimes \sigma^3=
\begin{pmatrix}
\bold{1} & 0 \\
0 & -\bold{1} 
\end{pmatrix},  
\label{hermiII}
\end{equation}
and 
\begin{equation}
K\Gamma^i= 
\begin{pmatrix}
0 & i\gamma^i \\
-i\gamma^i & 0 
\end{pmatrix},~~K\Gamma^{2p+2q}=
\begin{pmatrix}
0 & \bold{1} \\
\bold{1} & 0 
\end{pmatrix},~~K\Gamma^{2p+2q+1}=
\begin{pmatrix}
\bold{1} & 0 \\
0 & \bold{1}
\end{pmatrix}. 
\label{expso12qgamma}
\end{equation}
Such $SO(2q-1)$ gamma matrix structure is consistent with the above observation that 
 $H^{2p,0}_F$ contains $S^{2p-2}_F$ as its internal fuzzy space (\ref{geometryofindeffuzzy2}).

\subsubsection{Examples in low dimension}\label{subsec:concrete}

One can construct  gamma matrices of $SO$(odd,even) in any higher dimensions by following the method in Section \ref{subsub:stI} (Construction I) and the one in Section \ref{subsub:stII} (Construction II) from  either $SO(3)$ or $SO(1,2)$ gamma matrices \footnote{The hermitianizing matrix of $SO(2q+1,2p)$ is also constructed by  repeatedly applying the procedure (\ref{hermitanizingso2q+12p}) or (\ref{hermiII}) to the hermitianizing matrix  of $SO(1,2)$ or $SO(3)$, $i.e.$ $\sigma^3$ or $1_2$.  Then, one finds that the hermitianizing matrix of $SO(2q+1,2p)$ is unitarily equivalent to the split-signature diagonal matrix,   $K=diag(\overbrace{+,\cdots,+}^{2^{p+q-1}},\overbrace{-,\cdots,-}^{2^{p+q-1}})$.  }.  
In low dimensions, we have 

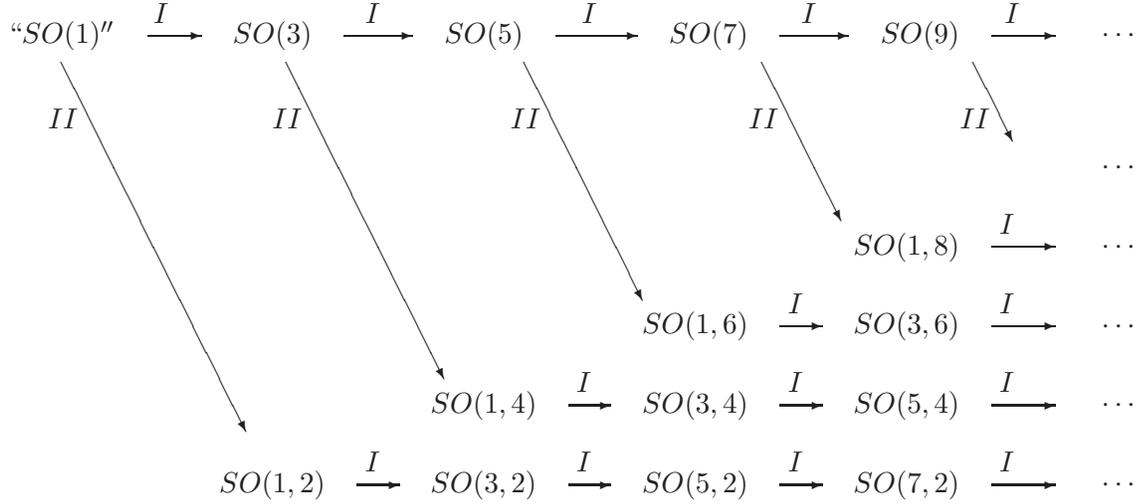
\begin{picture}(400,210)
\point{H1}(  0,  170){$~~``SO(1)''~~$}
\point{H2}(  80,  170){$~~SO(3)~~$}
\point{H3}(  160,  170){$~~SO(5)~~$}
\point{H4}(  245,  170){$~~SO(7)~~$}
\point{H5}(  325,  170){$~~SO(9)~~$}
\point{H6}(  400,  170){$~~~\cdots~~~$}
\point{I2}(  80,   0){$~~SO(1,2)~~$}
\point{I3}(  160,  0){$~~SO(3,2)~~$}
\point{I4}(  240,  0){$~~SO(5,2)~~$}
\point{I5}(  320,  0){$~~SO(7,2)~~$}
\point{I6}(  400,  0){$~~~\cdots~~~$}
\point{J3}(  160,  30){$~~SO(1,4)~~$}
\point{J4}(  240,  30){$~~SO(3,4)~~$}
\point{J5}(  320,  30){$~~SO(5,4)~~$}
\point{J6}(  400,  30){$~~~\cdots~~~$}
\point{K4}(  240,  60){$~~SO(1,6)~~$}
\point{K5}(  320,  60){$~~SO(3,6)~~$}
\point{K6}(  400,  60){$~~~\cdots~~~$}
\point{L5}(  320,  90){$~~SO(1,8)~~$}
\point{L6}(  400,  90){$~~~\cdots~~~$}
\point{M6}(  400,  120){$~~~\cdots~~~$}
\arrow{H1}{H2} \midput( -5,  5){${I}$}
\arrow{H2}{H3} \midput( -5,  5){${I}$}
\arrow{H3}{H4} \midput( -5,  5){${I}$}
\arrow{H4}{H5} \midput( -5,  5){${I}$}
\arrow{H5}{H6} \midput( -5,  5){${I}$}
\arrow{I2}{I3} \midput( -5,  5){${I}$}
\arrow{I3}{I4} \midput( -5,  5){${I}$}
\arrow{I4}{I5} \midput( -5,  5){${I}$}
\arrow{I5}{I6} \midput( -5,  5){${I}$}
\arrow{J3}{J4} \midput( -5,  5){${I}$}
\arrow{J4}{J5} \midput( -5,  5){${I}$}
\arrow{J5}{J6} \midput( -5,  5){${I}$}
\arrow{K4}{K5} \midput( -5,  5){${I}$}
\arrow{K5}{K6} \midput( -5,  5){${I}$}
\arrow{L5}{L6} \midput( -5,  5){${I}$}
\put(0,160){\vector(1,-2){70}}
\put(-5,135){$II$}
\put(85,160){\vector(1,-2){60}}
\put(80,135){$II$}
\put(175,160){\vector(1,-2){45}}
\put(170,135){$II$}
\put(265,160){\vector(1,-2){30}}
\put(260,135){$II$}
\put(345,160){\vector(1,-2){15}}
\put(340,135){$II$}
\end{picture}  
\vspace{0.5cm}\\
where $I$ and $II$ denote ``Construction I and II'', and the gamma matrix of  $``SO(1)''$ is defined as $-1$ to construct $SO(3)$ gamma matrix by (\ref{formula2+gamma}).   For the construction of fuzzy ultra-hyperboloid $H_F^{2p,2q}$, $SO(2q+1,2p)$  gamma matrices are utilized, and hence 
the above hierarchical structure of indefinite gamma matrices suggests the dimensional hierarchy of  fuzzy ultra-hyperboloids:

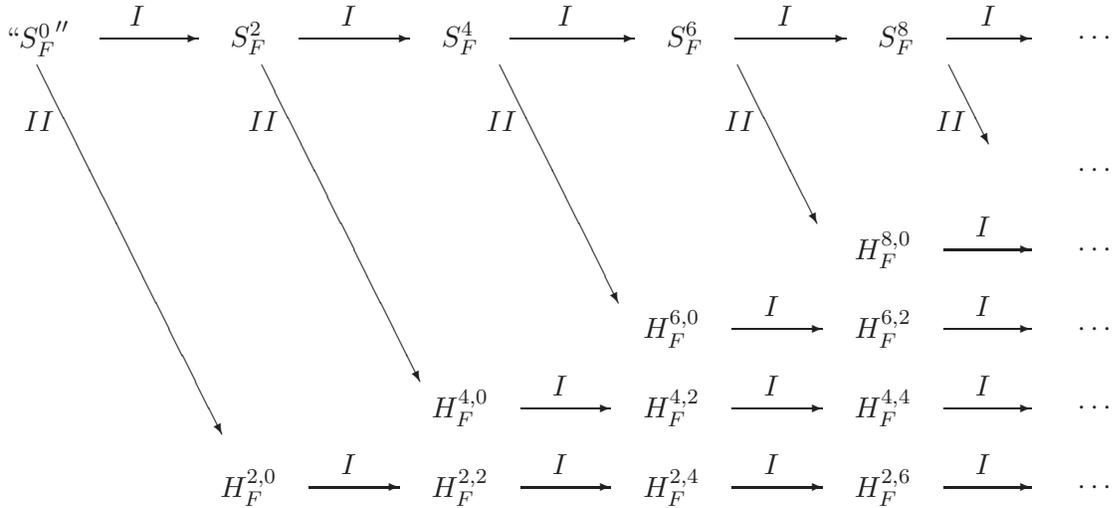
\begin{picture}(400,210)
\point{H1}(  0,  170){$~~``{S_F^0}''~~$}
\point{H2}(  80,  170){$~~S_F^2~~$}
\point{H3}(  160,  170){$~~S_F^4~~$}
\point{H4}(  245,  170){$~~S_F^6~~$}
\point{H5}(  325,  170){$~~S_F^8~~$}
\point{H6}(  400,  170){$~~~\cdots~~~$}
\point{I2}(  80,   0){$~~H^{2,0}_F~~$}
\point{I3}(  160,  0){$~~H^{2,2}_F~~$}
\point{I4}(  240,  0){$~~H^{2,4}_F~~$}
\point{I5}(  320,  0){$~~H^{2,6}_F~~$}
\point{I6}(  400,  0){$~~~\cdots~~~$}
\point{J3}(  160,  30){$~~H^{4,0}_F~~$}
\point{J4}(  240,  30){$~~H^{4,2}_F~~$}
\point{J5}(  320,  30){$~~H^{4,4}_F~~$}
\point{J6}(  400,  30){$~~~\cdots~~~$}
\point{K4}(  240,  60){$~~H^{6,0}_F~~$}
\point{K5}(  320,  60){$~~H^{6,2}_F~~$}
\point{K6}(  400,  60){$~~~\cdots~~~$}
\point{L5}(  320,  90){$~~H^{8,0}_F~~$}
\point{L6}(  400,  90){$~~~\cdots~~~$}
\point{M6}(  400,  120){$~~~\cdots~~~$}
\arrow{H1}{H2} \midput( -5,  5){${I}$}
\arrow{H2}{H3} \midput( -5,  5){${I}$}
\arrow{H3}{H4} \midput( -5,  5){${I}$}
\arrow{H4}{H5} \midput( -5,  5){${I}$}
\arrow{H5}{H6} \midput( -5,  5){${I}$}
\arrow{I2}{I3} \midput( -5,  5){${I}$}
\arrow{I3}{I4} \midput( -5,  5){${I}$}
\arrow{I4}{I5} \midput( -5,  5){${I}$}
\arrow{I5}{I6} \midput( -5,  5){${I}$}
\arrow{J3}{J4} \midput( -5,  5){${I}$}
\arrow{J4}{J5} \midput( -5,  5){${I}$}
\arrow{J5}{J6} \midput( -5,  5){${I}$}
\arrow{K4}{K5} \midput( -5,  5){${I}$}
\arrow{K5}{K6} \midput( -5,  5){${I}$}
\arrow{L5}{L6} \midput( -5,  5){${I}$}
\put(0,160){\vector(1,-2){70}}
\put(-5,135){$II$}
\put(85,160){\vector(1,-2){60}}
\put(80,135){$II$}
\put(175,160){\vector(1,-2){45}}
\put(170,135){$II$}
\put(265,160){\vector(1,-2){30}}
\put(260,135){$II$}
\put(345,160){\vector(1,-2){15}}
\put(340,135){$II$}
\end{picture}  
\vspace{0.5cm}\\

We have seen that the non-unitary  representation of $H_F^{2p,2q}$ $(p+q\le 2)$ is given by the  finite dimensional representation that is equal to the fully symmetric representation of the fuzzy sphere $S^{2p+2q+1}_F$.  Therefore, non-unitary finite dimensional  representation for arbitrary fuzzy ultra-hyperboloids can be readily constructed by applying a similar technique developed in the analysis of fuzzy spheres \cite{azuma2003,AzumaThesis}. For unitary infinite dimensional representation in low dimensions, we have seen that the unitary representation made by the Schwinger operators  corresponds to  discrete series of the corresponding indefinite orthogonal groups. According to Harish-Chandra's equal-rank condition \cite{HarishChadra2},   arbitrary $SO(2q+1,2p)$ group generally accommodates discrete series\footnote{For instance, see Refs.\cite{bookknapp,bookbarutraczka,bookgilmore} for more about non-compact groups and also Refs.\cite{barsgunaydin1983,gunaydinsaclioglu1982} for unitary representation  of the $SU(p,q)$ group made by Schwinger operator.}. 
Then, we can adopt the $SO(2q+1,2p)$ discrete series for unitary construction of $H_F^{2p,2q}$.

\subsection{Classical counterpart}\label{subsec:classicalcou}

Here, we introduce a coset for the ``classical limit'' of  fuzzy hyperboloid. 
The classical counterpart of fuzzy two-hyperboloid, $H^{2,0}_F$, is given by 
$H^{2,0}_F~\simeq ~ SO(1,2)/U(1)$,  
while those of fuzzy four-hyperboloids are   
$H^{2,2}_F~\simeq ~SO(3,2)/U(1,1)$ and $H^{4,0}_F~\simeq ~SO(1,4)/U(2).$  
Furthermore, from \cite{HoRamgoolam2002}, we know that the classical counterpart of fuzzy sphere is  
$H^{0,2p}_F= S^{2p}_F ~\simeq ~ SO(2p+1)/U(p).$ 
Therefore, one may infer the coset of fuzzy hyperboloid as 
\begin{equation}
H_{F}^{2p,2q} \simeq SO(2q+1,2p)/U(q,p) \simeq SO(2p,2q+1)/U(p,q). 
\label{generalcosethyperb}
\end{equation}
For $q\neq 0$,  (\ref{generalcosethyperb})  yields  
\begin{equation}
H_F^{2p,2q}\sim H^{2p,2q}\otimes SO(2p,2q)/U(p,q), 
\label{localexcoset1}
\end{equation}
where $H^{2p,2q}\simeq SO(2p,2q+1)/SO(2p,2q)$ was used. Thus, $H_{F}^{2p,2q}$ is regarded as a manifold whose basemanifold  is $H^{2p,2q}$ and its fibre is represented by the coset $SO(2p,2q)/U(p,q)$.  
From the relation  
\begin{equation}
SO(2p,2q)/SO(2p,2q-1)\simeq U(p,q)/U(p,q-1) ~(\simeq H^{2p,2q-1}),   
\end{equation}
one may find that (\ref{localexcoset1}) implies     
\begin{equation}
H_F^{2p,2q}\sim H^{2p,2q}\otimes SO(2p,2q-1)/U(p,q-1)\simeq H^{2p,2q}\otimes H_F^{2p,2q-2}.  
\label{fuzzyhigherhyperboloidcoset}
\end{equation}
As a special case of (\ref{generalcosethyperb}), for $q=0$, we have 
\begin{equation}
H_F^{2p,0}\sim    H^{2p,0}\otimes SO(2p)/U(p).   
\label{localexcoset2}
\end{equation}
Thus, $H_{F}^{2p,0}$ is locally given by the fibre-bundle of the fibre $SO(2p)/U(p)$ over the basemanifold  $H^{2p,0}$.  
From  
\begin{equation}
SO(2p)/SO(2p-1)\simeq U(p)/U(p-1)~(\simeq S^{2p-1}),  
\end{equation}
we find that (\ref{localexcoset2}) further suggests     
\begin{equation}
H_F^{2p,0}\sim    H^{2p,0}\otimes SO(2p-1)/U(p-1) \simeq H^{2p,0}\otimes S_F^{2p-2}. 
\end{equation}
Thus, we confirmed that (\ref{generalcosethyperb}) reproduces the geometrical structure of fuzzy hyperboloid discussed in Section \ref{subsec:algebra}, and these results support the validity of  (\ref{generalcosethyperb}). Due to the existence of internal fuzzy manifold, the dimension of $H_F^{2p,2q}$ is $\it{not}$ $2(p+q)$  but larger than that of $H^{2p,2q}$:  
\begin{equation}
\text{dim}[SO(2q,2p+1)/U(q,p)]=(p+q+1)(p+q).  
\end{equation}
The dimensions of $H_F^{2p,2q}$ and $H^{2p,2q}$ coincide only in the cases for $(p,q)=(1,0)$ and  $(0,1)$, $i.e.$ $H^{2,0}_F$ and $S^{2}_F$. 

In the following, we give several concrete examples of the dimensional hierarchy of fuzzy hyperboloids and their corresponding gamma matrices.  We use the definition,  
$S^{2p}_F\equiv H_F^{0,2p}$  and  $H^{2p}_F\equiv H_F^{2p,0}$. 
For $p+q=1$, we have 
\begin{align}
&SO(3)~\text{gamma~mat.}~~~:~~ S_F^2\sim H^{0,2}= S^2,\nonumber\\
&SO(1,2)~\text{gamma~mat.}:~H^{2}_F\sim H^{2,0}= H^2. 
\end{align}
For $p+q=2$,  
\begin{align}
&SO(5)~\text{gamma~mat.}~~~:~~S^{4}_F~\sim ~H^{0,4}\otimes H^{0,2}_F \sim S^{4}\otimes S^{2},\nonumber\\
&SO(3,2)~\text{gamma~mat.}~:~H^{2,2}_F~\sim H^{2,2}\otimes H^{2,0}_F  \sim H^{2,2}\otimes H^{2},\nonumber\\
&SO(1,4)~\text{gamma~mat.}~:~H^{4}_F~\sim ~ H^{4,0}\otimes S^2_F\sim H^{4}\otimes S^2. 
\end{align}
Note that these three respectively correspond to three kinds of the 2nd Hopf map.  
For $p+q=3$, 
\begin{align}
&SO(7)~\text{gamma~mat.}~~~:~~S^{6}_F~\sim ~ H^{0,6}\otimes H^{0,4}_F\sim S^6\otimes S^4\otimes S^2 ,\nonumber\\
&SO(5,2)~\text{gamma~mat.}~:~H^{2,4}_F\sim H^{2,4}\otimes H^{2,2}_F\sim H^{2,4}\otimes H^{2,2}\otimes H^{2} ,\nonumber\\
&SO(3,4)~\text{gamma~mat.}~:~H^{4,2}_F\sim H^{4,2}\otimes H_{F}^{4,0}\sim H^{4,2}\otimes H^{4}\otimes S^{2},\nonumber\\
&SO(1,6)~\text{gamma~mat.}~:~H^{6}_F~\sim ~ H^{6,0}\otimes S_{F}^{4}\sim H^{6}\otimes S^{4}\otimes S^{2}. 
\end{align}
For $p+q=4$,   
\begin{align}
&SO(9)~\text{gamma~mat.}~~~:~~S^{8}_F~\sim ~ H^{0,8}\otimes H^{0,6}_F \sim S^8\otimes S^6\otimes S^{4}\otimes S^{2},\nonumber\\
&SO(7,2)~\text{gamma~mat.}~:~H^{2,6}_F\sim H^{2,6}\otimes H^{2,4}_F  \sim H^{2,6}\otimes H^{2,4}\otimes H^{2,2}\otimes H^{2},\nonumber\\
&SO(5,4)~\text{gamma~mat.}~:~H^{4,4}_F\sim H^{4,4}\otimes H^{4,2}_F\sim H^{4,4}\otimes H^{4,2} \otimes H^{4}\otimes S^2,\nonumber\\
&SO(3,6)~\text{gamma~mat.}~:~H^{6,2}_F\sim H^{6,2}\otimes H^{6}_F\sim H^{6,2}\otimes H^6\otimes S^4\otimes S^2,\nonumber\\
&SO(1,8)~\text{gamma~mat.}~:~H^{8}_F~\sim ~ H^{8}\otimes S_F^6 \sim H^8\otimes S^6\otimes S^4\otimes S^2.  
\end{align}

\subsection{Non-compact monopoles}\label{subsec:fibrestruhyper}

It is known that 
monopole gauge fields are  realized as canonical connection of fibre-bundles of fuzzy spheres (see \cite{Hasebe2010} as a review).   Similarly, non-compact monopoles appear as associated connection of bundles of fuzzy hyperboloids. 

\subsection{ $U(1)$ monopole on $H^{2,0}$}

As a warm-up, we introduce $U(1)$ monopole on $H^{2,0}$  
from the 1st non-compact Hopf map. From the non-compact 1st Hopf map (\ref{differentnoncompacthopfmap}), the total manifold $H^{2,1}$ can be expressed as    
\begin{equation}
H^{2,1}\simeq H^{2,0}\otimes S^1.     
\end{equation}
Since $S^1\simeq U(1)$, 
the non-compact 1st Hopf map is physically related to  $H^{2,0}$ in the $U(1)$ monopole background.  
Indeed, by inverting (\ref{1stHopfmapexplicit}),   the total manifold $H^{2,1}$ denoted by 
 $\phi$ is  represented as  
\begin{equation}
\phi=\frac{1}{\sqrt{2(1+x^3)}}
\begin{pmatrix}
1+x^3\\
x^1+ix^2
\end{pmatrix}e^{i\chi},
\end{equation}
where $x^i$  are coordinates on $H^{2,0}$ and $e^{i\chi}$ stands for $U(1)$ phase factor.  
The associated fibre-connection is derived as 
\begin{equation}
A=-i\phi^{\dagger} \sigma^3 d\phi=dx^i A_i
\label{su11fibreconne}
\end{equation}
where 
\begin{equation}
A_i=\epsilon_{ij3}\frac{x^j}{2(1+x^3)}.
\end{equation}
The corresponding curvature is 
\begin{equation}
F_{ij}=\partial_i A_j-\partial_j A_i=-\frac{1}{2}\epsilon_{ijk}{x^k}.
\end{equation}

\subsubsection{$SO(2p)$ monopole on $H^{2p,0}$}\label{subsecsp2pmonopole}

Similarly,  from the hybrid 2nd Hopf map, $H^{4,3}$ is  expressed as   
\begin{equation}
H^{4,3}~\simeq ~  H^{4,0}\otimes S^3. 
\label{localexpreh43}
\end{equation}
Since $S^3$ is the group manifold corresponding to $SU(2)$, the hybrid 2nd Hopf map is closely related to the $SU(2)$ monopole gauge field on $H^{4,0}$. 
The total manifold  $H^{4,3}$ denoted by $\psi$  in (\ref{so41map2nd}) is parameterized as 
\begin{equation}
\psi=M\phi, 
\label{fromso3toso14}
\end{equation}
where $\phi$ is the $SO(3)$ Hopf spinor  subject to the constraint 
$\phi^{\dagger}\phi=1$, representing $S^3$, and $M$ is $4\times 2$ matrix  given by 
\begin{equation}
M=\frac{1}{\sqrt{2(1+x^5)}}
\begin{pmatrix}
(1+x^5){1}_2  \\
(x^4 {1}_2-ix^i\sigma^i)
\end{pmatrix}.
\end{equation}
Here, $x^a$  denote the coordinates on $H^{4,0}$ that satisfy  $\eta_{ab}x^ax^b=-\sum_{\mu,\nu=1}^4\delta_{\mu\nu}x^{\mu}x^{\nu}+x^5x^5=1$.  
The connection of $S^3$-bundle is evaluated to give   $SU(2)$ monopole gauge field on $H^{4,0}$: 
\begin{equation}
A=-iM^{\dagger} k dM=\frac{1}{2(1+x^5)}{\eta}_{\mu\nu i}x^{\mu}dx^{\nu}\sigma^i,
\label{su2gaugeh4}
\end{equation}
where  $\eta_{\mu\nu i}$ is  't Hooft tensor   ${\eta}_{\mu\nu i}\equiv \epsilon_{\mu\nu i 4}-{\delta}_{\mu i}{\delta}_{\nu 4}+{\delta}_{\mu 4}{\delta}_{\nu i}$ ($\epsilon_{1234}=1$,  $\mu=1,2,3,4$) \cite{'tHooft1976}, and $k$ is given by (\ref{hermitiso14}). 
The fuzzy ultra-hyperboloid $H_F^{2p,0}$ is a generalization of the hybrid Hopf maps and is locally given by   
$H^{2p,0}_F~\sim ~ H^{2p,0}\otimes {S_F^{2p-2}}$. 
Following the above procedure, we can derive the connection of fuzzy fibre $S_F^{2p-2}$  over  $H^{2p,0}$. 
We introduce the ``$SO(1,2p)$ Hopf spinor'' $\Psi$  as a classical counterpart of the fuzzy hyperboloid $H_F^{2p,0}$,   
\begin{equation}
\Psi=M\Phi, 
\label{itterapsiphi}
\end{equation}
where $M$ denotes $2^{p+1}\times 2^p$ matrix given by 
\begin{equation}
M=\frac{1}{\sqrt{2(1+x^{2p+1})}}
\begin{pmatrix}
(1+x^{2p+1}) \bold{1}\\
x^{2p}\bold{1}-ix^i\gamma^i 
\end{pmatrix}. \label{noncompacthopfspinormat}
\end{equation}
Thus, the $SO(1,2p)$ Hopf spinor $\Psi$ is generally constructed by (\ref{itterapsiphi}) from the $SO(2p-1)$ Hopf spinor $\Phi$. The simplest example of this construction is (\ref{fromso3toso14}): from the $SO(3)$ Hopf spinor to the $SO(1,4)$ Hopf spinor.   
Here, $x^a$ signify coordinates on $H^{2p,0}$ and satisfy  $\eta_{ab}x^ax^b=1$ with $SO(1,2p)$ metric $\eta_{ab}=diag(-,-,\cdots,-,+)$ ($a,b=1,2,\cdots,2p+1$), $\bold{1}$ denotes $2^p\times 2^p$ unit matrix and  $\gamma^i$ $(i=1,2,\cdots,2p-1)$ are $SO(2p-1)$ gamma matrices that satisfy 
\begin{equation}
\{\gamma^i,\gamma^j\}=2\delta^{ij}.   
\end{equation}
$\Phi$ in (\ref{itterapsiphi}) stands for the coordinates on $S_F^{2p-2}$, which satisfies $\Phi^{\dagger}\Phi=1$ [see Ref.\cite{hep-th/0310274} for realization of $\Phi$].  
From the following properties    
\begin{equation}
M^{\dagger}KM=\bold{1}, ~~~~M^{\dagger}K\Gamma^a M=x^a \bold{1}, 
\label{noncompacthopfmap}
\end{equation}
with $K$ (\ref{hermiII}) and $\Gamma^a$ (\ref{expso12qgamma}),  one may readily see that $\Psi$  satisfies $\Psi^{\dagger}K\Psi=\Phi^{\dagger}\Phi=1$  and the ``generalized'' hybrid Hopf map
\begin{equation}
\Psi^{\dagger}K^a\Psi=x^a. 
\end{equation}
The connection of fuzzy bundle $S_F^{2p-2}$ is calculated as   
\begin{equation}
A=-iM^{\dagger}K dM=dx^a k A_a,   
\end{equation}
where 
\begin{equation}
A_{\mu}=-\frac{1}{1+x^{2p+1}}\sigma_{\mu\nu}x^{\nu},~~~~A_{2p+1}=0. 
\end{equation}
Here, $\sigma_{\mu\nu}$ are the $SO(2p)$ generators\footnote{
$\sigma_{\mu\nu}$ satisfy the $SO(2p)$ algebra
\begin{equation}
[\sigma_{\mu\nu},\sigma_{\rho\sigma}]=i(\delta_{\mu\rho}\sigma_{\nu\sigma}-\delta_{\mu\sigma}\sigma_{\nu\rho}-\delta_{\nu\rho}\sigma_{\mu\sigma}+\delta_{\nu\sigma}\sigma_{\mu\rho}). 
\end{equation}
}, 
\begin{equation}
\sigma_{ij}=-i\frac{1}{4}[\gamma_i,\gamma_j],~~~~\sigma_{i, 2n}=-\sigma_{2p, i}=\frac{1}{2}\gamma_i, 
\end{equation}
with $\gamma_i\equiv -\gamma^i$. 
The field strength $F_{ab}=\partial_a A_b-\partial_b A_a-i[A_a,A_b]$ is derived as  
\begin{equation}
F_{\mu\nu}=-x_{\mu} A_{\nu}+x_{\nu}A_{\mu} +\sigma_{\mu\nu},~~~F_{\mu,2p+1}=-F_{2p+1,\mu}=(1+x^{2p+1})A_{\mu}.
\end{equation}
Thus, the connection of  $S^{2p-2}_F$-fibre represents the $SO(2p)$ gauge field on $H^{2p,0}$.

\subsubsection{$SO(2p,2q)$ monopole on $H^{2p,2q}$}
 
In the case of the split 2nd  Hopf map, the total manifold $H^{4,3}$ is locally expressed as    
\begin{equation}
H^{4,3}~\sim~H^{2,2}\otimes H^{2,1}.  
\label{localh43h22}
\end{equation}
Since $H^{2,1}\simeq SU(1,1)$,  the  split 2nd Hopf map is closely related to   $H^{2,2}$ in the $SU(1,1)$ monopole background. 
The normalized $SO(3,2)$ spinor $\psi$ representing the total manifold $H^{4,3}$  (\ref{normaso32hopfh22}) can be expressed as 
\begin{equation}
\psi=M\phi, 
\label{2ndsplithopfcon}
\end{equation}
where $M$ denotes  $4\times 2$ matrix of the form 
\begin{equation}
M=\frac{1}{\sqrt{2(1+x^5)}}
\begin{pmatrix}
(1+x^5)1_2\\
x^41_2-ix^i\tau_i
\end{pmatrix}. 
\label{explicit2ndspinor}
\end{equation}
Here, $x^a$ are coordinates on $H^{2,2}$: $\eta_{ab}x^a x^b=
\sum_{\mu,\nu=1}^4\eta_{\mu\nu}x^{\mu}x^{\nu}+x^5x^5=1$ ($\eta_{\mu\nu}=\text{diag}(-,-,+,+)$), and    
$\phi$  is the $SO(1,2)$ Hopf spinor that satisfies the condition of $H^{2,1}$-fibre, $\phi^{\dagger}\sigma^3\phi=1$.  
The associated connection of $H^{2,1}$-fibre is derived as 
\begin{equation}
A=-iM^{\dagger} k dM  =- \frac{1}{2(1+x^5)} \eta'_{\mu\nu i } \sigma^3  \tau^i x^{\mu} dx^{\nu},   
\label{su11gaugefieldexpli}
\end{equation}
where   $\eta'_{\mu\nu i}$ ($\mu,\nu=1,2,3,4$) is  ``split''-'t Hooft symbol:     
\begin{equation}
\eta'_{\mu\nu i }=\epsilon_{\mu\nu i 4}- \eta_{\mu i}\eta_{\nu 4}+\eta_{\nu i}\eta_{\mu 4}, 
\label{``split''-'t Hooftdef}
\end{equation}
with $\epsilon_{ijk4}\equiv \epsilon_{ijk}$. $\tau^i$ in (\ref{su11gaugefieldexpli}) are the $SU(1,1)$ Pauli matrices (\ref{defoftaus}), and 
 (\ref{su11gaugefieldexpli}) denotes  $SU(1,1)$ monopole gauge field. 
The commutators of the $SU(1,1)$ gauge field (\ref{su11gaugefieldexpli}) yield 
\begin{equation}
[A_{\mu},A_{\nu}]=i\frac{1}{1+x^5}(x_{\mu}A_{\nu}-x_{\nu}A_{\mu})+i\frac{1}{2}\frac{1-x^5}{1+x^5}\eta'_{\mu\nu i}\tau^i, 
\label{su11gaugecommu}
\end{equation}
and the field strength, $F_{ab}=\partial_a A_b-\partial_b A_a-i[A_a,A_b]$, satisfies  
\begin{equation}
F_{ac}F^{c}_{~~b}=-(\eta_{ac}-x_ax_b)(\frac{1}{2}\tau_i)^2-iF_{ab}.
\label{relationfsu11gauge}
\end{equation}
Eqs.(\ref{su11gaugecommu}) and (\ref{relationfsu11gauge}) will be useful in Section \ref{subsec:LLLSU(1,1)}. 

The above analysis can readily be applied to higher dimensional fuzzy hyperboloid $H^{2p,2q}$ ($q\neq 0$). 
Corresponding to $H_F^{2p,2q}$ $(q\neq 0)$ which is locally given by  $H^{2p,2q}_F~\sim ~ H^{2p,2q}\otimes H_F^{2p,2q-2}$, 
 we introduce the ``$SO(2q+1,2p)$ Hopf spinor''   
\begin{equation}
\Psi=M\psi,
\label{splitspinorconst}
\end{equation}
where $M$ denotes $2^{p+q+1}\times 2^{p+q}$ matrix of the form 
\begin{equation}
M=\frac{1}{\sqrt{2(1+x^{2p+2q+1})}}
\begin{pmatrix}
(1+x^{2p+2q+1}) \bold{1}\\
x^{2p+2q}\bold{1}-ix^i\gamma_i 
\end{pmatrix}. \label{noncompacthopfspinormat}
\end{equation}
Here,  $\gamma^i$ ($i=1,2,\cdots,2p+2q-1$) are the $SO(2q-1,2p)$ gamma matrices, $x^a$ ($a=1,2,\cdots,2p+2q+1$) are coordinates on the basemanifold $H^{2p,2q}$ that satisfy $\eta_{ab}x^ax^b=1$ with $SO(2q+1,2p)$ metric $\eta_{ab}$, and $\bold{1}$ denotes $2^{p+q}\times 2^{p+q}$ unit matrix. 
$\psi$ signifies  the fuzzy-fibre $H_F^{2p,2q-2}$. 
From the $SO(1,2p)$ Hopf spinor $\psi$, we repeat the procedure (\ref{splitspinorconst}) ``$q$ times'' to construct the $SO(2q+1,2p)$ Hopf spinor.  
(The $SO(1,2p)$ Hopf spinor itself is constructed by the procedure discussed in Section \ref{subsecsp2pmonopole}.)     
From $\Psi$, a projection matrix $P$ is constructed as  
\begin{equation}
P=\Psi\Psi^{\dagger}K, 
\end{equation}
which satisfies $P^2=P$ and $P\Psi=\Psi$.     
 $M$ (\ref{noncompacthopfspinormat}) satisfies 
\begin{equation}
M^{\dagger}KM=k, ~~~~M^{\dagger}K\Gamma^a M=x^a k, 
\label{noncompacthopfmap}
\end{equation}
where $K$ and $k$ are the hermitianizing matrices (\ref{hermitanizingso2q+12p}) and $\Gamma^a$ are $SO(2q+1,2p)$ gamma matrices 
(\ref{so2q+12pgammamatricesexp}). Here,     
  $\Gamma^a$ and $\gamma^i$ respectively signify $SO(2q+1,2p)$ and $SO(2q-1,2p)$ gamma matrices  related by (\ref{formula2+gamma}).  
We obtain  
$\Psi^{\dagger}K\Psi=\psi^{\dagger}k\psi=1$ and  the ``generalized'' split-Hopf map  
\begin{equation}
\Psi^{\dagger}K^a\Psi=x^a. 
\end{equation}

The connection of fuzzy bundle $H_F^{2p,2q-2}$ over $H^{2p,2q}$ is derived as 
\begin{equation}
A=-iM^{\dagger}K dM= dx^a k A_a ,   
\end{equation}
where $A_a$ $(a=1,2,\cdots,2p+2q+1)$ are 
\begin{equation}
A_{\mu}=-\frac{1}{1+x^{2p+2q+1}}\sigma_{\mu\nu}x^{\nu},~~~~A_{2p+2q+1}=0. 
\end{equation}
Here $\mu,\nu=1,2,\cdots,2p+2q$, and $\sigma_{\mu\nu}$ are the $SO(2p,2q)$ generators\footnote{
$\sigma_{\mu\nu}$  satisfy 
\begin{equation}
[\sigma_{\mu\nu},\sigma_{\rho\sigma}]=i(\eta_{\mu\rho}\sigma_{\nu\sigma}-\eta_{\mu\sigma}\sigma_{\nu\rho}-\eta_{\nu\rho}\sigma_{\mu\sigma}+\eta_{\nu\sigma}\sigma_{\mu\rho}), 
\end{equation}
with the $SO(2q,2p)$ metric, $\eta_{\mu\nu}$.} given by   
\begin{equation}
\sigma_{ij}=-i\frac{1}{4}[\gamma_i,\gamma_j],~~~~\sigma_{i, 2p+2q}=-\sigma_{2p+2q, i}=\frac{1}{2}\gamma_i,   
\end{equation}
with $i,j=1,2,\cdots,2p+2q-1$. 
The field strength $F_{ab}=\partial_a A_b-\partial_b A_a-i[A_a,A_b]$ is calculated as 
\begin{equation}
F_{\mu\nu}=-x_{\mu} A_{\nu}+x_{\nu}A_{\mu} +\sigma_{\mu\nu},~~~~F_{\mu,2p+2q+1}=-F_{2p+2q+1,\mu}=(1+x^{2p+2q+1})A_{\mu}.
\end{equation}
The $SO(2p,2q)$ monopole gauge field is thus  induced as the connection of fuzzy bundle  $H^{2p,2q-2}_F$ over the basemanifold $H^{2p,2q}$.  

To summarize, the connection of fuzzy bundle of  $H_F^{2p,2q}$ physically corresponds to $SO(2p,2q)$ monopole gauge field  over the basemanifold $H^{2q,2p}$.  
In particular for $(p,q)=(2,0)$, the gauge group of monopole is $SO(4,0)\simeq SU(2)\otimes  SU(2)$ or $SU(2)$.  
 Meanwhile, for $(p,q)=(1,1)$, the gauge group of monopole is $SO(2,2)\simeq SU(1,1)\otimes SU(1,1)$ or $SU(1,1)$.

\section{Fuzzy Hyperboloid realized as Lowest Landau Level}\label{sec:fuz}

The lowest Landau level physics illustrates close relations between monopoles and fuzzy geometry  \cite{hep-th/0310274}. Here, we explore such relations for fuzzy two- (Section \ref{subsec:llltwohy}) and four-hyperboloids (Section \ref{subsec:lllfourhy}).

\subsection{Lowest Landau level on two-hyperboloid}\label{subsec:llltwohy}

First, we introduce one-particle mechanics on two-hyperboloid in $U(1)$ monopole background \cite{arXiv:0809.4885}.  The Lagrangian is given by  
\begin{equation}
L=\frac{M}{2}\eta_{ij}\dot{x}^i\dot{x}^j+\dot{x}^iA_i, 
\end{equation}
with $\eta_{ij}=\text{diag}(-,-,+)$\footnote{The corresponding Landau problem is investigated in Ref.\cite{comtet1987}. See also \cite{arXiv:0809.4885,Beluccietal2011} for related works about supersymmetrization. }. 
Here, $A_i$ denotes the $U(1)$ gauge field 
\begin{equation}
A_{i}=\frac{I}{2}\epsilon_{ij3}\frac{x^j}{1+x^3}
\end{equation}
with monopole charge $I/2$ ($I$ is an integer). 
In the lowest Landau level $M\rightarrow 0$, the kinetic term is quenched and the  Lagrangian  reduces to  
\begin{equation}
L_{LLL}=\dot{x}^iA_i=-i{I}\phi^{\dagger}\sigma^3\frac{d}{dt}\phi,
\end{equation}
where   $\phi$ stands for the $SO(1,2)$ Hopf spinor with the normalization condition,     
\begin{equation}
\phi^{\dagger}\sigma^3\phi=1. 
\label{normasu11hopf}
\end{equation}
The canonical conjugate of $\phi$ is derived as 
\begin{equation}
\pi=\frac{\partial L_{LLL}}{\partial \dot{\phi}}=-iI\phi^{\dagger}\sigma^3.
\end{equation}
Then, $\pi$ is $\it{not}$ the time derivative of $\phi^{\dagger}$, but $\phi^{\dagger}$ itself.   
Therefore, when  we impose the canonical quantization condition between $\phi$ and $\pi$,
\begin{equation}
[\phi_{\alpha},\pi_{\beta}]=i\delta_{\alpha\beta},
\end{equation}
$\phi^*$ is expressed as 
\begin{equation}
\phi^*=\frac{1}{I}\sigma^3\frac{\partial}{\partial\phi}.
\end{equation}
The normalization (\ref{normasu11hopf}) becomes a constraint imposed on Hilbert space 
\begin{equation}
\phi^t\frac{\partial}{\partial\phi} f_{LLL}=If_{LLL},
\end{equation}
which determines the lowest Landau level bases, $f_{LLL}$. $f_{LLL}$ is given by the homogeneous polynomials of the Hopf spinor  constructed by replacing the Schwinger operator in (\ref{n1n2ket}) with the Hopf spinor. 
Furthermore, in the lowest Landau level,  $x^i$  (\ref{1stHopfmapexplicit}) 
are effectively represented as  
\begin{equation}
X_i=\frac{1}{I}\phi^t(\tau_i)^t\frac{\partial}{\partial\phi}, 
\end{equation}
which satisfy  
\begin{equation}
[X^i,X^j]=\frac{1}{I}i\epsilon^{ijk}X_k. 
\end{equation}
The equations of motion are derived as 
\begin{equation}
\dot{X}^i=-i[X^i,V]=\frac{1}{I}\epsilon^{ijk}X_jE_k,
\label{twohyperequaionofmotion}
\end{equation}
where $E_i=-\partial_i V$.  (\ref{twohyperequaionofmotion}) indicates a hyperbolic version of the cyclotron motion of the center of mass coordinates $X^i$. The Hall law   
\begin{equation}
E^{i}\dot{X}_i=0,
\end{equation}
and the cyclotron motion
\begin{equation}
F_{ij}\dot{X}^j=-(\eta_{ij}-X_iX_j)E^j,
\label{cyclohyper2}
\end{equation}
follow from (\ref{twohyperequaionofmotion}). Here, $F_{ij}$ is the field strength of the monopole: 
\begin{equation}
F_{ij}=\partial_i A_j-\partial_j A_i=  - \frac{I}{2}\epsilon_{iik}x^k  .
\end{equation}

\subsection{Lowest Landau level on four-hyperboloid}\label{subsec:lllfourhy}

Eq.(\ref{coset1}) suggests two superficially different expressions for $H^{4,3}$:  
\begin{equation}
H^{4,3}~\sim~ H^{2,2}\otimes H^{2,1}~\sim~ \mathbb{C}P^{1,2}\otimes S^1.
\end{equation}
In either case, the bundles are principal bundles : 
\begin{equation}
H^{2,1}\simeq SU(1,1),~~~~~~~S^1\simeq U(1). 
\end{equation}
This observation implies that  $H^{2,2}$ in $SU(1,1)$ monopole background is equivalent to $\mathbb{C}P^{1,2}$ in $U(1)$ monopole background. 
Here, we demonstrate this speculation in the context of lowest Landau level physics. 
 The following analysis is  a non-compact extension of the analysis in Ref.\cite{cond-mat/0206164}. 
 
\subsubsection{Lowest Landau level on $H^{2,2}$ in $SU(1,1)$ monopole background}\label{subsec:LLLSU(1,1)}
 
In $SU(1,1)$ monopole background,
 one-particle Lagrangian is given by 
\begin{equation}
L=\frac{M}{2}\eta_{ab}\dot{x}^a\dot{x}^b+ \dot{x}^a A_a,
\label{lagraso32one}
\end{equation}
where $x^a$ $(a=1,2,3,4,5)$ are coordinates on $H^{2,2}$ subject to 
\begin{equation}
\eta_{ab}x^ax^b=1,  
\label{so32condnor}
\end{equation}
with $SO(3,2)$ metric $\eta_{ab}=diag(-,-,+,+,+)$, and $A_a$ represents the $SU(1,1)$ monopole gauge field (\ref{su11gaugefieldexpli})\footnote{The corresponding Landau problem is investigated in Ref.\cite{arXiv:0902.2523}. The Landau problem on $H^{4,0}$ in $SU(2)$ monopole background (which corresponds to the hybrid 2nd Hopf map) is also argued in Ref.\cite{BellucciPLB2006}.}. 
The Lagrangian (\ref{lagraso32one}) apparently respects  the $SO(3,2)$ symmetry. 
Meanwhile in the lowest Landau level, the mass term drops and the gauge interaction term only survives to yield 
\begin{equation}
L_{LLL}=\dot{x}^a A_a= -iI\psi^{\dagger}k \frac{d\psi}{dt},
\label{LLLoneparticle}
\end{equation}
with the constraint for the $SO(3,2)$ Hopf spinor 
\begin{equation}
\psi^{\dagger}k\psi=1.   
\label{noramlizationso32spc}
\end{equation}
One may see that both (\ref{LLLoneparticle}) and (\ref{noramlizationso32spc}) enjoy the enhanced $SU(2,2)\simeq SO(4,2)$ symmetry as the rotational symmetry of $\psi$.  
From the lowest Landau level Lagrangian (\ref{LLLoneparticle}), the canonical conjugate of $\psi$ is derived as 
\begin{equation}
\pi=-iI\psi^{\dagger}k.
\end{equation}
With the canonical quantization 
\begin{equation}
[\psi_{\alpha},\pi_{\beta}]=i\delta_{\alpha\beta},
\end{equation}
we have 
\begin{equation}
\psi^*=\frac{1}{I}k\frac{\partial}{\partial\psi}.
\end{equation}
 In quantum mechanics, the normalization (\ref{noramlizationso32spc}) is transformed to a constraint on the Hilbert space 
\begin{equation}
\psi^t\frac{\partial}{\partial\psi} f_{LLL}=If_{LLL}. 
\end{equation}
Then, the lowest Landau level bases are given by the homogeneous polynomials of the components of the $SO(3,2)$ Hopf spinor. 
Furthermore,   $x^a$ (\ref{2ndnoncompactHopf}) reduce to   
\begin{equation}
X_a=\frac{1}{I}\psi^t\gamma_a^t\frac{\partial}{\partial\psi}.
\end{equation}
Also, the total angular momentum $L_{ab}$ reduces to the field strength, $L_{ab}\rightarrow -F_{ab}$, and hence the non-commutative relations  
\begin{equation}
[X_a,X_b]={4}iL_{ab}
\end{equation}
 are rewritten as 
\begin{equation}
[X_a,X_b]=-{4}iF_{ab}.
\end{equation}
Consequently, in the lowest Landau level, we derive the equations of motion as 
\begin{equation}
\dot{X}_a=-i[X_a,V]={4}F_{ab}E^b,  
\label{equationofmotion4d}
\end{equation}
where $E_a=-\partial_a V$. From the equations of motion (\ref{equationofmotion4d}), we obtain the Hall law 
\begin{equation}
E^a\dot{X}_a=0,
\end{equation}
and a generalized cyclotron motion 
\begin{equation}
F^{ab}\dot{X}_b=-(\eta^{ab}-x^a x^b)E_b \cdot (\tau_i)^2-i F^{ab}E_b, 
\label{equationofmotionoflaugex}
\end{equation}
where (\ref{relationfsu11gauge}) was used.

\subsubsection{ $U(1)$ monopole gauge fields on $\mathbb{C}P^{1,2}$ }

The coordinates on $H^{4,3}$ are parameterized as  (\ref{2ndsplithopfcon}) with  
(\ref{explicit2ndspinor}) and the $H^{2,1}$-fibre is represented as 
\begin{equation}
\phi=\frac{1}{\sqrt{2(1+n^3)}}
\begin{pmatrix}
1+n^3\\
n^1+in^2
\end{pmatrix}e^{i\chi}
\end{equation}
with 
\begin{equation}
\eta_{ij}n^in^j=-n^1n^1-n^2n^2+n^3n^3=1. 
\label{constraintofns}
\end{equation}
The $U(1)$ gauge field on $\mathbb{C}P^{1,2}$ is derived as   
\begin{equation}
\mathcal{A}=-i\psi^{\dagger}kd\psi=
dx^a  \mathcal{A}_a(x,n)|_{I=1} +dn^i \mathcal{A}_i(n)|_{I=1}, 
\end{equation}
where $\mathcal{A}_a$ $(a=1,2,3,4,5)$ and $\mathcal{A}_i$ $(i=1,2,3)$ are respectively given by  
\begin{subequations}
\begin{align}
&\mathcal{A}_{\mu}(x,n)=\frac{I}{2}\eta'_{\mu\nu i}\frac{x^{\nu}}{1+x^5}n^{i},~~~\mathcal{A}_5=0,
\label{U1mathcalAs}\\
&\mathcal{A}_i(n)=\frac{I}{2}\epsilon_{ij3}\frac{n^j}{1+n^3}.  
\end{align}
\end{subequations}
Here, $\eta'_{\mu\nu i}$ denotes the split-'t Hooft tensor (\ref{``split''-'t Hooftdef}) and $I/2$  ($I$: integer) stands for $U(1)$ monopole charge. 
$\mathcal{A}_a$ and $\mathcal{A}_i$  are tangent to the surface of $H^{2,2}$ and $H^{2,0}$, respectively:  
\begin{equation}
\eta_{ab}\mathcal{A}_a x^b=\eta_{ij}\mathcal{A}^i n^j=0.
\end{equation}
The $U(1)$ field strength 
\begin{equation}
\mathcal{F}_{ab}=\partial_a \mathcal{A}_b-\partial_b \mathcal{A}_a,~~~\mathcal{F}_{ai}=\partial_a \mathcal{A}_i-\partial_i \mathcal{A}_a,~~~\mathcal{F}_{ij}=\partial_i \mathcal{A}_j-\partial_j \mathcal{A}_i,  
\end{equation}
is calculated as 
\begin{subequations}
\begin{align}
&\mathcal{F}_{\mu\nu}=-\frac{2+x^5}{1+x^5}(x_{\mu}\mathcal{A}_{\nu}-x_{\nu}\mathcal{A}_{\mu})-I \eta'_{\mu\nu i }\frac{n^i}{1+x^5},~~~\mathcal{F}_{\mu 5}=(1+x^5)\mathcal{A}_{\mu},\\
&\mathcal{F}_{\mu i}=-\mathcal{F}_{i\mu}=-\frac{I}{2}\eta'_{\mu\nu i }\frac{x^{\nu}}{1+x^5}-\mathcal{A}_{\mu}n_i,~~~\mathcal{F}_{5i}=-\mathcal{F}_{i 5}=0,\\
&\mathcal{F}_{ij}=-\frac{I}{2}\epsilon_{ijk}n^k, 
\label{mathcalfijexp}
\end{align}
\end{subequations}
which is orthogonal to the surface of $\mathbb{C}P^{1,2}$ in the following sense:  
\begin{subequations}
\begin{align}
&\eta_{ab}x^a\mathcal{F}^{bc}=\eta_{ab}x^a \mathcal{F}^{bi}=0, \label{orthoxandf}\\
&\eta_{ij}n^i\mathcal{F}^{jk}=\eta_{ij}n^i\mathcal{F}^{ja}=0. \label{orthonandf}
\end{align}
\end{subequations}
Furthermore, they satisfy 
\begin{subequations}
\begin{align}
&\mathcal{F}_{ij}\mathcal{F}^{jk}=-\frac{I^2}{4}(\delta_i^{~k}-n_in^k), \label{ffrela1}\\
&\mathcal{F}^{ij}\mathcal{F}_{j a}=-\frac{I}{2}\epsilon^{ijk}\mathcal{A}_{a j}n_k, \label{ffrela2}\\
&\mathcal{F}_{\mu i}\mathcal{F}^{ij}\mathcal{F}_{j\nu}
= \frac{I^2}{4}\frac{1}{1+x^5}(x_{\mu}\mathcal{A}_{\nu}-x_{\nu}\mathcal{A}_{\mu})+
\frac{I^3}{8}\frac{1-x^5}{1+x^5}\eta'_{\mu\nu i }n^i,
\label{3fffrela}
\end{align}
\end{subequations}
where $\mathcal{A}_{ai}$ in (\ref{ffrela2}) is defined as $\mathcal{A}_a\equiv \mathcal{A}_{ai}n^i$ (\ref{U1mathcalAs}), $i.e.$  $\mathcal{A}_{\mu i}=\frac{I}{2}\eta'_{\mu\nu i}\frac{x^{\nu}}{1+x^5}$, $\mathcal{A}_{5i}=0$, and the properties of the split- `t Hooft tensor,  
$\epsilon^{ij}_{~~k}\eta'_{\mu\nu i} \eta'_{\rho\sigma j}=\eta_{\mu\rho}\eta'_{\nu\sigma k}-\eta_{\mu\sigma}\eta'_{\nu\rho k }-\eta_{\nu\rho}\eta'_{\mu\sigma k}+\eta_{\nu\sigma}\eta'_{\mu\rho k}$,  
was used to derive (\ref{3fffrela}). It should be noted that 
the right-hand side of  (\ref{su11gaugecommu}) is ``equal'' to that of (\ref{3fffrela})  by replacing the $SU(1,1)$ matrix $\tau^i$  with its  corresponding $c$-number $n^i$. Hence, we have the correspondence\footnote{For the case $(a,b)=(\mu,5)$ or $(a,b)=(5,\nu)$, the validity of (\ref{corresaaa}) is apparent since $A_5=0$ and $\mathcal{F}_{5i}=0$.}:    
\begin{equation}
 \sigma^3[A_{a},A_{b}] ~~\leftrightarrow ~~ 
i\frac{4}{I^2}\mathcal{F}_{ai}\mathcal{F}^{ij}\mathcal{F}_{jb}.   
\label{corresaaa}
\end{equation}
Eventually, one may find that the $SU(1,1)$ and $U(1)$ field strengths are related as 
\begin{equation}
\sigma^3 F_{ab}~~ \leftrightarrow ~~\mathcal{F}_{ab}+\frac{4}{I^2}
\mathcal{F}_{ai}\mathcal{F}^{ij}\mathcal{F}_{jb}.  
\label{corressu11andu1cp12}
\end{equation}

\subsubsection{Lowest Landau level on $\mathbb{C}P^{1,2}$ in $U(1)$ monopole background}

We consider the one-particle motion on $\mathbb{C}P^{1,2}$ in $U(1)$ monopole background. 
The one-particle Lagrangian is given by 
\begin{equation}
S=\int dt \biggl[\frac{M}{2}(\frac{dx^a}{dt})^2+\frac{M}{2}(\frac{dn^i}{dt})^2+\mathcal{A}_a(x,n)\frac{dx^a}{dt}+\mathcal{A}_i(n)\frac{dn^i}{dt}-V(x,n)\biggr]. 
\end{equation}
The Lagrange multipliers,  $\lambda_1$ and $\lambda_2$, are introduced to incorporate the  conditions (\ref{constraintofns}) and (\ref{constraintsofh22}).  With the Lagrange multipliers, the equations of motion are derived as 
\begin{align}
&M\ddot{x}_a=-\dot{x}^b\mathcal{F}_{ba}-\dot{n}^i\mathcal{F}_{ia}+E_a+2\lambda_1 x_a,\nonumber\\
&M\ddot{n}_i=-\dot{n}^j\mathcal{F}_{ji}-\dot{x}^a\mathcal{F}_{ai}+E_i+2\lambda_2 n_i,
\label{equation2xn}
\end{align}
where $E_a=-\partial_a V$, $E_i=-\partial_i V$. With (\ref{constraintofns}) and (\ref{constraintsofh22}),   the Lagrange multipliers are obtained as 
\begin{equation}
\lambda_1= \frac{M}{2} x^a\ddot{x}_a -\frac{1}{2}x^a E_a,~~~~\lambda_2=\frac{M}{2} n^i\ddot{n}_i -\frac{1}{2}n^i E_i. 
\label{lagrange2}
\end{equation}
Substituting (\ref{lagrange2}) to (\ref{equation2xn}), we have  
\begin{subequations}
\begin{align}
&M(\eta_{ab}-x_a x_b)\ddot{x}^b=\mathcal{F}_{ab}\dot{x}^b+\mathcal{F}_{ai}\dot{n}^i+
(\eta_{ab}-x_ax_b)E^b,\label{generahall1}\\
&M(\eta_{ij}-n_i n_j)\ddot{n}^j=\mathcal{F}_{ij}\dot{n}^j+\mathcal{F}_{ia}\dot{x}^a+
(\eta_{ij}-n_i n_j)E^j. \label{generahall2}
\end{align}\label{fulequcp12}
\end{subequations}
By multiplying $\dot{x}^a$ and $\dot{n}^i$ to (\ref{generahall1}) and (\ref{generahall2}) respectively,  their sum yields    
\begin{equation}
M(\eta_{ab}\dot{x}^a\ddot{x}^b+\eta_{ij}\dot{n}^i\ddot{n}^j)=\dot{x}_aE^a+\dot{n}_iE^i, 
\label{Mxxab}
\end{equation}
where the ``normalization'' conditions (\ref{so32condnor}) and (\ref{constraintofns}) were used.  
In the lowest Landau level ($M\rightarrow 0$), (\ref{Mxxab})  reduces to a generalized Hall law on $\mathbb{C}P^{1,2}$: 
\begin{equation}
\dot{x}_aE^a+\dot{n}_iE^i=0,
\end{equation}
and (\ref{fulequcp12}) also reduces to  
\begin{subequations}
\begin{align}
&0=\mathcal{F}_{ab}\dot{x}^b+\mathcal{F}_{ai}\dot{n}^i+
(\eta_{ab}-x_a x_b)E^b,\label{lllequationofmotion1}\\
&0=\mathcal{F}_{ij}\dot{n}^j+\mathcal{F}_{ia}\dot{x}^a+(\eta_{ij}-n_i n_j)E^j.
\label{lllequationofmotion2}
\end{align}
\end{subequations}
We arrange these first derivative equations to derive the equation of motion for $x^a$. 
From (\ref{lllequationofmotion2}), one finds that $\dot{n}^i$ are related to $\dot{x}^a$ as 
\begin{equation}
\dot{n}^i=\frac{4}{I^2}\mathcal{F}^{ij}E_j+\frac{4}{I^2}\mathcal{F}^{ij}\mathcal{F}_{ja}\dot{x}^a, 
\label{equationforxnapp}
\end{equation}
where (\ref{orthonandf}) and  (\ref{ffrela1})  were used.  
From (\ref{mathcalfijexp}) and (\ref{ffrela2}), (\ref{equationforxnapp}) can be rewritten as 
\begin{equation}
\dot{n}^i=-\frac{2}{I}\epsilon^{ijk}(E_j+\mathcal{A}_{aj}\dot{x}^a) n_k, 
\label{equationforxnrew}
\end{equation}
which is a natural generalization of the cyclotron motion (\ref{twohyperequaionofmotion}).   
By inserting (\ref{equationforxnapp})  to (\ref{lllequationofmotion1}),  we eventually obtain the first derivative equation only for $x^a$: 
\begin{equation}
(\mathcal{F}_{ab}+\frac{4}{I^2}\mathcal{F}_{ai}\mathcal{F}^{ij}\mathcal{F}_{jb})\dot{x}^b=-
\frac{4}{I^2}\mathcal{F}_{ai}\mathcal{F}^{ij}E_j-(\eta_{ab}-x_a x_b)E^b. \label{equationforx}
\end{equation}
Unlike the two-hyperboloid (\ref{cyclohyper2}), (\ref{equationforx}) contains  higher orders of $U(1)$ field strengths. 
When $E_i=0$, (\ref{equationforx}) reduces to 
\begin{equation}
(\mathcal{F}_{ab}+\frac{4}{I^2}\mathcal{F}_{ai}\mathcal{F}^{ij}\mathcal{F}_{jb})\dot{x}^b=
-(\eta_{ab}-x_a x_b)E^b.\label{equationofmotionaboutu1x}
\end{equation}
With use of the correspondence (\ref{corressu11andu1cp12}), one may find equivalence between the $SU(1,1)$ case (\ref{equationofmotionoflaugex})
and  the $U(1)$ case (\ref{equationofmotionaboutu1x}). Thus in the lowest Landau level, one-particle mechanics on  
 $H^{2,2}$ in the $SU(1,1)$ monopole background and that on $\mathbb{C}P^{1,2}$ in $U(1)$ monopole background are equivalent at the level of classical  equations of motion. 
 In this way, we confirmed  equivalence between fuzzy $H^{2,2}$ and fuzzy $\mathbb{C}P^{1,2}$ in the context of lowest Landau level. 
 
\section{Summary and Discussions}\label{sec:sum}

 We developed a systematic construction of fuzzy ultra-hyperboloids based on  gamma matrices of indefinite orthogonal groups.  
 With the cousins of quaternions, the split and hybrid Hopf maps were introduced.       
We realized  
fuzzy two- and four-hyperboloids as  the Schwinger operator version of such non-compact Hopf maps. 
We also performed a study of fuzzy hyperboloids in higher dimensional space-times with use  of indefinite  
 gamma matrices. There are two ways to describe fuzzy hyperboloids; one is to utilize non-unitary finite dimensional representation, while the other is to utilize unitary infinite dimensional representation.    
With the appropriate choice of vacuum of Schwinger operators, we showed that the Schwinger operator formalism yields infinite dimensional representation of the  discrete series of non-compact groups. 
The geometry of fuzzy ultra-hyperboloids reflects the generalized structures of  the split and hybrid Hopf maps.  
We illuminated such  
generalized enhanced symmetry and dimensional hierarchy 
 in the geometry of fuzzy ultra-hyperboloids. 
 Non-compact monopole gauge field is  naturally induced as connection of the fibration of  fuzzy hyperboloid.   
We also argued the identification between  
 the fuzzy four-hyperboloid  and the six-dimensional fuzzy indefinite  complex projective space  in the context of  the lowest Landau level physics.   
 We believe  that 
the present study may be useful not only for  fuzzy physics itself but also for further understanding of brane geometry, twistors, and higher spin theory. 
 
Finally, we mention the limitation of the present work. 
Irreducible representation of non-compact group generally contains discrete and principal series.  
We focused on the discrete series and the corresponding fuzzy hyperboloids of the type $H_F^{\text{even},\text{even}}$.  
The fuzzy hyperboloids  by the principal series are of the type $H_F^{\text{odd},\text{odd}}$, which includes for instance,  $H^{1,1}_F(=dS^2_F=AdS^2_F)$ and  $H^{1,3}_F(=dS^4_F)$.    Such fuzzy hyperboloids are beyond the scope of the present study\footnote{Still, we partially discuss a construction of fuzzy $H^{odd,odd}$ in Appendix \ref{sec:split-al} with use of the split-complex number. }, and systematic construction of such fuzzy hyperboloids should  be addressed in future works.

\section*{Acknowledgments}\label{secacknowledg}

I would like to thank Koichi Murakami and Satoshi Watamura for helpful discussions. 
I am also grateful to Taichiro Kugo for telling a useful reference in his homepage.  
This work was supported in part by a 
Grant-in-Aid for Scientific Research from the Ministry of Education, Science, Sports and Culture of Japan (Grant No.23740212). 

\appendix

\section{Hybrid 3rd Hopf map}

The 1st and 2nd Hopf maps were realized by sandwiching the Pauli and $SO(5)$ gamma matrices by  Hopf spinors. One may expect that such realization can be readily applied to the 3rd Hopf map. However, it is not so straightforward, since the octonions cannot be realized by matrices due to their non-associative property. Instead of using the octonions themselves, 
 the octonion structure constants are utilized to derive the following $8\times 8$ matrices\cite{Hasebe2010,Bernevig2003} 
\begin{align} 
&\lambda^{1}= -i\left(
 \begin{array}{@{\,}cccc@{\,}}
 \sigma_2    &   0       &     0         &   0
\\   0        & \sigma_2 &     0         &   0
\\   0        &   0       &   \sigma_2    &   0
\\   0        &   0       &     0         & -\sigma_2
 \end{array}\right),~~~ 
\lambda^{2} = \left(
 \begin{array}{@{\,}cccc@{\,}}
        0     & -\sigma_3 &      0        & 0
\\  \sigma_3   &   0       &      0        & 0
\\      0     &   0       &      0        & -1_2 
\\      0     &   0       &  1_2    & 0
 \end{array}
 \right),\nonumber\\
&\lambda^{3}= \left(
 \begin{array}{@{\,}cccc@{\,}}
          0   & -\sigma_1 &    0        &  0
\\  \sigma_1 &    0      &    0        &  0  
\\        0   &    0      &    0        & -i\sigma_2
\\        0   &    0      & {-i\sigma_2}   &  0
 \end{array}\right),~~~~\lambda^{4} = \left(
 \begin{array}{@{\,}cccc@{\,}}
         0    &     0      &    -\sigma_3    &   0
\\       0    &     0      &    0      &    1_2 
\\     \sigma_3    &     0      &    0      &   0
\\        0   &    -1_2     &    0      &   0
 \end{array}
 \right),\nonumber\\
&\lambda^{5}= \left(
 \begin{array}{@{\,}cccc@{\,}}
        0    &      0      &    -\sigma_1   &    0
\\      0    &      0      &      0        &  i\sigma_2  
\\ \sigma_1 &      0      &      0        &    0
\\      0    &    {i\sigma_2} &      0        &    0
 \end{array}\right)\!,~~~~\lambda^{6} = \left(
 \begin{array}{@{\,}cccc@{\,}}
        0     &      0      &    0        &   -1_2
\\      0     &      0      &   -\sigma_3      &    0
\\      0     &    \sigma_3     &    0        &    0
\\     1_2    &      0      &    0        &    0
 \end{array}
 \right),~~~~~~\nonumber\\
&\lambda^{7}= \left(
 \begin{array}{@{\,}cccc@{\,}}
       0      &      0      &     0       &  -i\sigma_2 
\\     0      &      0      &  -\sigma_1   &    0 
\\     0      & \sigma_1   &     0       &    0
\\  -i\sigma_2 &      0      &     0       &    0
\end{array} \right). 
\end{align}
They are real antisymmetric matrices that satisfy 
\begin{equation}
\{\lambda^I,\lambda^J\}=-2\delta^{IJ}.
\end{equation}
With $\lambda^0\equiv 1_8$, $\lambda^0$ and $\lambda^I$ $(I=1,2, \cdots,7)$ are regarded as the $SO(8)$ ``Weyl $+$'' gamma matrices. 
From $\lambda^0$ and $\lambda^I$, the $SO(1,8)$ gamma matrices $\Gamma^A$  are constructed as  
\begin{equation}
\Gamma^I=\lambda^{I}\otimes \sigma^1,~~\Gamma^8=i1_8\otimes \sigma^2,~~\Gamma^9=1_8\otimes \sigma^3,
\end{equation}
or 
\begin{equation}
\Gamma^I=\begin{pmatrix}
0 & \lambda^{I} \\
\lambda^{I} & 0
\end{pmatrix},~~
\Gamma^8=
\begin{pmatrix}
0 & 1_8 \\
-1_8 & 0 
\end{pmatrix},~~
\Gamma^9=
\begin{pmatrix}
1_8 & 0 \\
0 & -1_8 
\end{pmatrix}, 
\end{equation}
which satisfy 
\begin{equation}
\{\Gamma^A,\Gamma^B\}=2{\eta}^{AB},
\end{equation}
where $A,B=1,2,\cdots,9$ and ${\eta}_{AB}={\eta}^{AB}=diag(-,-,-,-,-,-,-,-,+)$. 
$\Gamma^I$ and $\Gamma^8$ are real antisymmetric matrices: 
\begin{equation}
(\Gamma^I)^t=\Gamma_I=-\Gamma^I,~~~(\Gamma^8)^t=\Gamma_8=-\Gamma^8. 
\end{equation}
Since $\Gamma_A$ are real matrices, the $SO(1,8)$ generators, $\Sigma_{AB}=-i\frac{1}{4}[\Gamma_A,\Gamma_B]$, are purely imaginary: $\Sigma_{AB}^*=-\Sigma_{AB}$. Thus, the present representation is Majorana representation, in which the charge conjugation matrix is given by an unit matrix, and the $SO(1,8)$ Majorana spinor 
is simply represented by (16-component) real spinor.  From (\ref{so12nhermiti}), the hermitianizing matrix  $K$ is constructed as 
\begin{equation}
K=\Gamma^1\Gamma^2\cdots \Gamma^8=\Gamma^9=\begin{pmatrix}
1_8 & 0 \\
0 & -1_8 
\end{pmatrix}, 
\end{equation}
and the  gamma matrices are hermitianized as 
\begin{equation}
K^A=K\Gamma^A.   
\end{equation}
In detail, 
\begin{equation}
K^I=\begin{pmatrix}
0 & \lambda^{I} \\
-\lambda^{I} & 0
\end{pmatrix},~~
K^8=
\begin{pmatrix}
0 & 1_8 \\
1_8 & 0 
\end{pmatrix},~~
K^9=
\begin{pmatrix}
1_8 & 0 \\
0 & 1_8 
\end{pmatrix}.
\end{equation}

The $SO(1,8)$ Hopf spinor is an $SO(1,8)$ Majorana spinor\footnote{
The $SO(p,q)$ with $p+q=9$ accommodate Majorana spinor only when $(p,q)=(9,0)$, $(5,4)$ and $(1,8)$.   
The former two cases correspond to the compact and split 3rd Hopf maps, and the last corresponds to the hybrid 3rd Hopf map.} subject to the normalization condition 
\begin{equation}
\Psi^t K \Psi  ={\Psi_1}^2+{\Psi_2}^2+\cdots+{\Psi_8}^2 -{\Psi_9}^2-{\Psi_{10}}^2-\cdots -{\Psi_{16}}^2 =1, 
\label{normalization3rdhopf}
\end{equation}
and hence $\Psi$ is regarded as coordinates on $H^{8,7}$. 
 By sandwiching $\Gamma_A$  between the 3rd Hopf spinors, 
we realize the hybrid 3rd  Hopf map, $H^{8,7}\overset{S^7}\longrightarrow H^{8,0}$, as 
\begin{equation}
\Psi\rightarrow x^A=\Psi^t K^A \Psi.
\label{map3rdhopfII}
\end{equation}
Here, $x_A$ are coordinates on $H^{8,0}$, since    
\begin{equation}
\sum_{A,B=1,2, \cdots,9}{\eta}_{AB}x^A x^B=(\Psi^t K\Psi)^2=1.
\end{equation}
The $SO(1,8)$ Hopf spinor $\Psi$ is represented as  
\begin{equation}
\Psi= \frac{1}{\sqrt{2(1+x^9)}}
\begin{pmatrix}
(1+x^9) 
\Phi \\
(x^8-\lambda^{I} x^I) \Phi
\end{pmatrix}, 
\end{equation}
where $\Phi$ is an $SO(7)$ real 8-component spinor subject to the constraint 
\begin{equation}
\Phi^t \Phi     =1, 
\end{equation}
representing the $S^7$-bundle. 
The connection of $S^7$-bundle is evaluated as 
\begin{equation}
A=-i\Psi^{t}K d\Psi=-\frac{1}{2(1+x^9)}\sigma_{MN}x^Nd x^M,
\end{equation}
where 
\begin{equation}
\sigma_{IJ}=-i\frac{1}{4}[\lambda_I,\lambda_J],~~~~\sigma_{I8}=-\sigma_{8I}=i\frac{1}{2}\lambda_I,  
\end{equation}
with $\lambda_I\equiv -\lambda^I$. 
These represent the $SO(8)$ monopole gauge field on $H^{8,0}$.  The corresponding field strength 
\begin{equation}
F_{AB}=\partial_A A_B -\partial_B A_A -i[A_A,A_B]
\end{equation}
is derived as 
\begin{equation}
F_{MN}=x_{M}A_{N}-x_{N}A_{M}-\sigma_{MN},~~~~F_{M 9}=-F_{9M}=(1+x^9)A_{M}. 
\end{equation}

\section{Split Algebra and Fuzzy Split-Hyperboloid}\label{sec:split-al}

The split-imaginary unit $j$ is introduced so as to satisfy  
\begin{equation}
j^2=1, ~~~~~~j^*=-j,
\label{propertyofj}
\end{equation}
where $*$ denotes complex conjugation. 
With two real numbers $x$ and $y$, the split-complex number is defined as    
\begin{equation}
z=x+jy.
\end{equation}
Its complex conjugation is  given by 
\begin{equation}
z^*=x-jy, 
\end{equation}
and then 
\begin{equation}
z^*z=zz^*=x^2-y^2.
\end{equation}
The split Hopf maps are naturally introduced by adopting the split-complex number \cite{arXiv:0905.2792}. 
 Similarly, the unitary and special unitary groups of split-complex numbers are introduced as 
\begin{subequations}
\begin{align} 
&\mathcal{SU}(p)\equiv SU(p;\mathbb{C}')=SL(p,\mathbb{R}), \\
&\mathcal{U}(p)\equiv U(p;\mathbb{C}')=\mathcal{U}(1)\otimes \mathcal{SU}(p).  
\end{align}
\end{subequations}
Note that the (quasi-)split hyperboloid $H^{p+1,p}$ is represented by the coset 
\begin{equation}
H^{p+1,p}\simeq \mathcal{SU}(p+1)/\mathcal{SU}(p).
\label{expresplithyp}
\end{equation}
  
\subsection{Fuzzy two-hyperboloid: $\mathcal{H}_F^{1,1}$}

 The coordinates on $S_F^2$ satisfy the $SU(2)$ algebra and square of the radius of $S_F^2$ is specified by the eigenvalues of the $SU(2)$ Casimir.   
In a similar manner, we introduce fuzzy split-hyperboloid, $\mathcal{H}_{F}^{1,1}$, based on the split imaginary unit. 
The coordinates on $\mathcal{H}_F^{1,1}$ are constructed as 
\begin{equation}
X^i=\Phi^{\dagger}\sigma^i\Phi,  
\label{splixsu11}
\end{equation}
where $\sigma^i$ $(i=1,2,3)$ are ``Pauli matrices'' with split imaginary unit:   
\begin{equation}
\sigma^1=
\begin{pmatrix}
1 & 0 \\
0 & -1
\end{pmatrix},~~~~
\sigma^2=
\begin{pmatrix}
0 & 1 \\
1 & 0
\end{pmatrix},~~~~
\sigma^3=
\begin{pmatrix}
0 & -j \\
j & 0
\end{pmatrix},
\label{splitpaulimatex}
\end{equation}
and $\Phi$ denotes a two-component Schwinger operator whose components satisfy  
\begin{equation}
[\Phi_{\alpha},\Phi_{\beta}^{\dagger}]=\delta_{\alpha\beta}.  
\end{equation}
(\ref{splitpaulimatex}) gives $SO(2,1)$ gamma matrices in the sense that the  ``Pauli matrices'' satisfy  
\begin{equation}
\{\sigma^i,\sigma^j\}=2\eta^{ij},
\end{equation}
with $\eta^{ij}=diag(+,+,-)$, and then 
 $X^i$ (\ref{splixsu11})  satisfy the $SU(1,1)$ algebra 
\begin{equation}
[X^i,X^j]=j\epsilon^{ijk}X_k. 
\end{equation}
Square of the radius of $\mathcal{H}_{F}^{1,1}$ is given by the $SU(1,1)$ Casimir  
\begin{equation}
\eta_{ij}X^iX^j=X^2+Y^2-Z^2=(\Phi^{\dagger}\Phi)(\Phi^{\dagger}\Phi+2).  
\label{splitsauradi}
\end{equation}
Note that the right-hand side of (\ref{splitsauradi}) is invariant under  ``$SU(1,1)$ rotations'' generated by (\ref{splitpaulimatex}).    
Thus, the spectra of  square of the radius of $H_F^{1,1}$ are given by $l(l+1)$ with $l=0,1/2,1,3/2,\cdots$. 
The commutative counterpart of $H_F^{1,1}$ is one-leaf hyperboloid 
\begin{equation}
H^{1,1}\simeq SU(1,1)/SO(1,1)\simeq AdS^2\simeq \mathbb{C}'P^1.
\end{equation}
 $\mathbb{C}'P^1$ will be introduced in Appendix \ref{subsec:appeninfcom2}.

\subsection{Fuzzy four-hyperboloid: $\mathcal{H}_F^{2,2}$}

 Next, we discuss the fuzzy four-hyperboloid based on the split-imaginary unit.  
With the Pauli matrices $\sigma^i$  made of the split-imaginary unit  (\ref{splitpaulimatex}),   ``$SO(3,2)$'' gamma matrices are constructed as 
\begin{align}
\gamma^i=
\begin{pmatrix}
0 & j\sigma^i \\
-j \sigma^i & 0
\end{pmatrix},~~\gamma^4=
\begin{pmatrix}
0 & 1_2 \\
1_2 & 0
\end{pmatrix},~~\gamma^i=
\begin{pmatrix}
1_2 &  0  \\
 0 & -1_2
\end{pmatrix},~~
\label{so32splitgamma}
\end{align}
which satisfy 
\begin{equation}
\{\gamma^a,\gamma^b\}=2\eta^{ab},
\end{equation}
where  $\eta_{ab}=(-,-,+,+,+)$.  
The ``$SO(3,2)$ generators'' are also given by 
\begin{equation}
\sigma^{ab}=-j\frac{1}{4}[\gamma^a,\gamma^b], 
\end{equation}
which satisfy 
\begin{align}
&[\gamma_a,\sigma_{bc}]=-j(\eta_{ab}\gamma_c-\eta_{ac}\gamma_b),\nonumber\\
&[\sigma_{ab},\sigma_{cd}]=j(\eta_{ac}\sigma_{bd}-\eta_{ad}\sigma_{bc}+\eta_{bd}\sigma_{ac}-\eta_{bc}\sigma_{ad}). 
\end{align}
With four-component Schwinger operator $\Phi$, 
$[\Phi_{\alpha},\Phi_{\beta}^{\dagger}]=\delta_{\alpha\beta}$ 
 $(\alpha,\beta=1,2,3,4)$, we introduce the fuzzy coordinates on $\mathcal{H}_F^{2,2}$ as 
\begin{equation}
X^a=\Phi^{\dagger}\gamma^a\Phi,  
\end{equation}
and square of the radius of $\mathcal{H}_F^{2,2}$ is derived as   
\begin{equation}
\eta_{ab}X^a X^b=(\Phi^{\dagger}\Phi)(\Phi^{\dagger}\Phi+4).  
\label{h22invsplit} 
\end{equation}
With  $X_{AB}$ $(A,B=1,2,3,4,5,6)$;  $X_{a6}\equiv -\frac{1}{2}X_a$ and $X_{ab}\equiv -i\frac{1}{4}[X_a,X_b]$, $X_{AB}$ satisfy the closed algebra:  
\begin{equation}
[X_{AB},X_{CD}]=j(\eta_{AC}X_{BD}-\eta_{AD}X_{BC}+\eta_{BD}X_{AC}-\eta_{BC}X_{AD}),
\end{equation}
where $\eta_{AB}=\text{diag}(-,-,+,+,+,-)$, $i.e.$ the $SO(3,3)$ metric. 
(Remember, in the case of $H^{2,2}_F$, $X_{AB}$ satisfy the $SO(4,2)$ algebra (\ref{so24algebraandso42}).) 
Note  that  $SO(3,3)$ is isomorphic to the split-imaginary special unitary group: $SO(3,3)\simeq \mathcal{SU}(4)$. 
Then,  $\mathcal{H}^{2,2}_F$ is represented as the coset  
\begin{equation}
{\mathcal{H}}_F^{2,2} \simeq SO(3,3)/\mathcal{U}(3) 
\simeq \mathcal{SU}(4)/\mathcal{U}(3)\simeq \mathbb{C}'P^{3}, 
\label{commusf4lim}
\end{equation}
where  $\mathbb{C}'P^3$ denotes  split-complex projective space (see Appendix \ref{subsec:appeninfcom2}). With the original symmetry $SO(3,2)$, $\mathcal{H}_F^{2,2}$ can be expressed as 
 \begin{equation}
 \mathcal{H}_F^{2,2}\simeq SO(3,2)/\mathcal{U}(2), 
 \end{equation}
 since  
 $\mathbb{C}'P^{3}\sim H^{2,2}\otimes H^{1,1}\simeq SO(3,2)/SO(2,2)\otimes SO(2,1)/SO(1,1)\simeq SO(3,2)/(SO(2,1)\otimes SO(1,1))$\footnote{Here, we used  
\begin{equation}
SO(2,2)\simeq SU(1,1)\otimes SU(1,1),~~~~~SO(2,1)\simeq SU(1,1)\simeq \mathcal{SU}(2),~~~~~SO(1,1)\simeq \mathcal{U}(1).
\end{equation}
}.
This result is a natural split-complex number version of fuzzy four-sphere: $S_F^{4}\simeq  SO(5)/U(2)$.   Also note that $\mathcal{H}_F^{2,2}$ is different from $H^{2,2}_F$ (\ref{cosetrepfuzzyfour1}).

\subsection{Fuzzy split-hyperboloids:  $\mathcal{H}_F^{p,p}$}

From the above discussions, 
it may be natural to expect that  
  fuzzy split-hyperboloids are generally given by the coset:  
\begin{equation}
\mathcal{H}_{F}^{p,p}\simeq SO(p+1,p)/\mathcal{U}(p).  
\label{defofcomulimhyperI}
\end{equation}
This is a natural split signature counterpart of the fuzzy sphere, $S_F^{2p}\simeq SO(2p+1)/U(p)$ \cite{HoRamgoolam2002}.  
 $\mathcal{H}_F^{p,p}$ is locally given by 
\begin{equation}
\mathcal{H}^{p,p}~\sim~ H^{p,p}\otimes SO(p,p)/\mathcal{U}(p),  
\end{equation}
where $H^{p,p}$  and $SO(p,p)/\mathcal{U}(p)$ respectively represent the basemanifold and the fibre on it. With 
\begin{equation}
SO(p,p)/SO(p,p-1)~~\simeq~~\mathcal{U}(p)/\mathcal{U}(p-1)~(\simeq ~H^{p,p-1}),   
\end{equation}
the fuzzy split-hyperboloid may be expressed as 
\begin{align}
&\mathcal{H}_{F}^{p,p} 
\sim H^{p,p}\otimes  SO(p,p-1)/\mathcal{U}(p-1)\nonumber\\
&~~~~~~\simeq~ H^{p,p}\otimes \mathcal{H}^{p-1,p-1}_F\nonumber\\
&~~~~~~\sim ~H^{p,p}\otimes H^{p-1,p-1}\otimes H^{p-2,p-2}\otimes\cdots \otimes H^{2,2}\otimes H^{1,1}.   
\label{mathfuzzyhyperhie}
\end{align}
Then in low dimensions, we have 
\begin{align}
&\mathcal{H}^{1,1}_F\simeq SO(2,1)/\mathcal{U}(1)\simeq H^{1,1},\nonumber\\
&\mathcal{H}^{2,2}_F\simeq SO(3,2)/\mathcal{U}(2)\sim H^{2,2}\otimes H^{1,1},\nonumber\\
&\mathcal{H}^{3,3}_F\simeq SO(4,3)/\mathcal{U}(3)\sim H^{3,3}\otimes H^{2,2}\otimes H^{1,1},\nonumber\\
&\mathcal{H}^{4,4}_F\simeq SO(5,4)/\mathcal{U}(4)\sim H^{4,4}\otimes H^{3,3}\otimes H^{2,2}\otimes H^{1,1}. 
\label{commulimitsplithyper}
\end{align}
The dimension of the fuzzy split-hyperboloid $H^{p,p}_F$ is given by  
$\sum_{k=1}^p 2k=p(p+1)$.  

The coordinates of $\mathcal{H}_F^{p,p}$ are regarded as the gamma matrices of the (quasi-)split orthogonal groups $SO(p+1,p)$. Then, 
the hierarchical geometry (\ref{commulimitsplithyper}) can also be observed in the structure of the $SO(p+1,p)$ gamma matrices. 
The $SO(p+1,p)$ gamma matrices, $\Gamma^a$ $(a=1,2,\cdots,2p+1)$, are given by 
\begin{equation}
\Gamma^i=\begin{pmatrix}
0 & j\gamma^{2p-i} \\
-j\gamma^{2p-i} & 0
\end{pmatrix},~~~\Gamma^{2p}=\begin{pmatrix}
0 & 1\\
1 & 0
\end{pmatrix},~~~\Gamma^{2p+1}=\begin{pmatrix}
1 & 0 \\
0 & -1
\end{pmatrix}, 
\label{construsplitgam}
\end{equation}
where $i=1,2,\cdots,2p-1$. 
The anti-commutation relations of $\Gamma^a$ read as 
\begin{equation}
\{\Gamma^a,\Gamma^b\}=2\xi^{ab}, 
\end{equation}
where $\xi^{ab}=(-\eta^{2p-i, 2p-j}, +, + )$.  
Thus, from $SO(p+1,p)$ gamma matrices, we can construct $SO(p+2,p+1)$ gamma matrices. 
With iterative use of (\ref{construsplitgam}), we have the gamma matrices of the following groups 
\begin{equation}
SO(2,1)\rightarrow SO(3,2) \rightarrow SO(4,3) \rightarrow SO(5,4) \rightarrow \cdots. 
\end{equation}
The split Hopf spinor for $\mathcal{H}_F^{p,p}$ is constructed as 
\begin{equation}
\Psi=M\psi, 
\end{equation}
where $M$ denotes $2^{p}\times 2^{p-1}$ matrix of the form\footnote{ $M$ satisfies 
\begin{align}
&M^{\dagger}M=\bold{1},\nonumber\\
&MM^{\dagger}=  \frac{1}{2}
\begin{pmatrix} 
1_{2^{p-1}}+x^{2p+1} & x^{2p}1_{2^{p-1}}+jx^i\gamma_{2p-i} \\
x^{2p}1_{2^{p-1}}-jx^i\gamma_{2p-i} & 1_{2^{p-1}}-x^{2p+1} 
\end{pmatrix}=\frac{1}{2}(1+x_a\Gamma^a)\equiv P,   
\end{align}
where $P$ is a  projection operator:  
\begin{equation}
P^2=P,~~~~~ P\Psi=\Psi. 
\end{equation}
} 
\begin{equation}
M=
\frac{1}{\sqrt{2(1+x^{2p+1})}} 
\begin{pmatrix}
(1+x^{2p+1}) 1_{2^{p-1}} \\
x^{2p}1_{2^{p-1}}-jx^i\gamma_{2p-i}
\end{pmatrix}, 
\end{equation}
with $x^i\gamma_{2p-i}\equiv \xi_{ij} x^i\gamma^{2p-j}$.  
As found in (\ref{mathfuzzyhyperhie}),  $\mathcal{H}_F^{p,p}$ is regarded as  fuzzy fibre-bundle of the fibre $\mathcal{H}_F^{p-1,p-1}$ over the basemanifold  $H^{2,2}$.   
The connection of fuzzy bundle $\mathcal{H}_F^{p-1,p-1}$ is derived as 
\begin{equation}
A=dx^a A_a=-jM^{\dagger}dM, 
\end{equation}
where 
\begin{equation}
A_{\mu}=-\frac{1}{1+x^{2p+1}}\sigma_{\mu\nu}x^{\nu},~~~~A_{2p+1}=0.   
\end{equation}
Here, $\sigma_{\mu\nu}$  ($\mu,\nu=1,2,\cdots,2n$) are $SO(p,p)$ generators given by  
\begin{equation}
\sigma_{ij}=-j\frac{1}{4}[\gamma_{2p-i},\gamma_{2p-j}],~~~\sigma_{i,2p}=-\sigma_{2p,i}=\frac{1}{2}\gamma_{2p-1}, 
\end{equation}
with $i=1,2,\cdots,n-1$.  
$A_a$ represent $SO(p,p)$ monopole gauge field.  
The field strength, $F_{ab}=\partial_a A_b-\partial_b A_a -j[A_a,A_b]$, is derived as 
\begin{equation}
F_{\mu\nu}=-x_{\mu}A_{\nu}+x_{\nu}A_{\mu}+\sigma_{\mu\nu},~~~~F_{\mu ,2p+1}=-F_{2p+1 ,\mu} =(1+x^{2p+1})A_{\mu}. 
\end{equation}

\section{Indefinite Complex Projective Spaces}\label{sec:appeninfcom}

We briefly introduce indefinite complex and split-complex projective spaces.  
 
\subsection{Indefinite complex projective spaces}\label{subsec:appeninfcom}

Indefinite complex projective space signifies complex projective space in indefinite complex space. $\mathbb{C}P^{p,q}$ is defined so as to  satisfy  the condition of $H^{2q,2p+1}$
\begin{equation}
\sum_{i=1}^{p+1} {z_i}^* z_i-\sum_{j=1}^{q} {z'}_j^*z'_j =1,
\end{equation}
 modulo $U(1)$: 
\begin{equation}
(z_1, z_2,\cdots,z_{p+1},z_1',z_2',\cdots,z_{q}')\sim e^{i\theta}(z_1,z_2,\cdots,z_{p+1},z_1',z_2',\cdots,z_{q}'), 
\end{equation}
and hence $\mathbb{C}P^{p,q}$ is expressed as the coset:  
\begin{equation}
\mathbb{C}P^{p,q} \simeq H^{2q,2p+1}/S^{1},
\label{s1devidecp}
\end{equation}
or 
\begin{equation}
\mathbb{C}P^{p,q}  \simeq SU(p+1,q)/U(p,q) \simeq  SU(q,p+1)/U(q,p). 
\end{equation}
(\ref{s1devidecp}) can be regarded as a higher dimensional generalization of the non-compact 1st Hopf map. 
Indeed, for $(p,q)=(0,1)$, we reproduce the non-compact 1st Hopf map, 
\begin{equation}
\mathbb{C}P^{0,1}\simeq SU(1,1)/U(1)\simeq H^{2,1}/S^{1}\sim H^{2,0}. 
\end{equation}
Also, $\mathbb{C}P^{p,q}$ are related to the 2nd and 3rd split Hopf maps:  
\begin{subequations}
\begin{align}
&\mathbb{C}P^{1,2}\simeq SU(2,2)/U(1,2) \simeq H^{4,3}/S^{1}\sim H^{2,2}\otimes H^{2,0},\label{2ndsplith}\\
&\mathbb{C}P^{3,4}\simeq SU(4,4)/U(3,4)\simeq H^{8,7}/S^{1}\sim H^{4,4}\otimes H^{2,2}\otimes H^{2,0}. 
\label{noncompactCPhieII}
\end{align}
\end{subequations}
(\ref{2ndsplith}) is the basic relation of the discussion in Section \ref{subsec:lllfourhy}.  

\subsection{Split-complex projective spaces}\label{subsec:appeninfcom2}

The split-complex projective spaces ${\mathbb{C}}' P^{p}$ are introduced
 by replacing the usual imaginary unit  with the split imaginary unit in the definition of $\mathbb{C}P^p$: With $z_{i}=x_i+jy_i ~(i=1,2,\cdots,p+1)\in {\mathbb{C}'}^{p+1}$, $\mathbb{C}'P^{p}$ is defined so as to satisfy the condition of ${H}^{p+1,p}$ 
\begin{equation}
\sum_{i=1}^{p+1} z_i^*z_i=   
\sum_{i=1}^{p+1} {x}_i {x}_i   -\sum_{i=1}^{p+1}{y}_i {y}_i    =1,
\label{firstdefcdashp}
\end{equation}
modulo $\mathcal{U}(1)$ 
\begin{equation}
(z_1, z_2,\cdots,z_{p+1})\sim e^{j\theta}(z_1,z_2,\cdots,z_{p+1}).   
\end{equation}
Therefore, $\mathbb{C}'P^p$ can be expressed by the coset: 
\begin{equation}
\mathbb{C}'P^{p} \simeq H^{p+1,p}/{H}^1, 
\end{equation}
or 
\begin{equation}
\mathbb{C}'P^{p}\simeq \mathcal{SU}(p+1)/\mathcal{U}(p),   
\label{cosetCPI2}
\end{equation}
where we used $H^{1}\simeq \mathcal{U}(1)$ and (\ref{expresplithyp}). 
In particular, related to the split Hopf maps, we have\footnote{Eq.(\ref{cosetCPI2}) is a non-compact version of the expression $\mathbb{C}P^p \simeq SU(p+1)/U(p) \simeq S^{2p+1}/S^1$.  Eq.(\ref{noncompactCPhieI}) is the split-signature version of 
\begin{align}
&\mathbb{C}P^1\simeq SU(2)/U(1)\simeq S^3/S^1\sim S^2,\nonumber\\
&\mathbb{C}P^3\simeq SU(4)/U(3)\simeq S^7/S^1\sim S^4\otimes S^2\nonumber\\
&\mathbb{C}P^7\simeq SU(8)/U(7)\simeq S^{15}/S^1\sim S^8\otimes S^4\otimes S^2. 
\end{align}
}
\begin{align}
&\mathbb{C}'P^1\simeq \mathcal{SU}(2)/\mathcal{U}(1)\simeq H^{2,1}/H^{1,0}\sim H^{1,1},\nonumber\\
&\mathbb{C}'P^3\simeq \mathcal{SU}(4)/\mathcal{U}(3)\simeq H^{4,3}/H^{1,0}\sim H^{2,2} \otimes H^{1,1}.\nonumber\\
&\mathbb{C}' P^7\simeq \mathcal{SU}(8)/\mathcal{U}(7)
\simeq H^{8,7}/H^{1,0}
\sim H^{4,4}\otimes H^{2,2}\otimes H^{1,1}. 
\label{noncompactCPhieI}
\end{align}

Notice that  for split-complex projective space,   
we need not define its indefinite version.   It is because   $\mathbb{C}'P^{p,q}$ satisfies   
\begin{equation}
\sum_{i=1}^p z_i^*z_1-\sum_{j=1}^{q+1} \tilde{z}^*_j\tilde{z}_j= (\sum_{i=1}^p x_i x_i+\sum_{j=1}^{q+1}\tilde{y}_j\tilde{y}_j) - (\sum_{j=1}^{p}{y}_j {y}_j+\sum_{i=1}^{q+1} \tilde{x}_i \tilde{x}_i)     =1, 
\label{condcpmndash}
\end{equation}
modulo $\mathcal{U}(1)$, however with $z'_i\equiv x_i+ j y_i$ ($i=1,2,\cdots,p$)  and 
${z}'_{p+i} \equiv \tilde{y}_i+j \tilde{x}_i$ ($i=1,2,\cdots,q+1$),  (\ref{condcpmndash}) can be rewritten in the form of (\ref{firstdefcdashp}):  
\begin{equation}
\sum_{i=1}^{p+q+1} {z'_i}^*z'_i=1.
\end{equation}
This indicates\footnote{
(\ref{equivindefspiltcomp}) is also consistent with the cosets for $\mathbb{C}'P^{p,q}$  and $\mathbb{C}'P^{p+q}$: 
\begin{equation}
\mathbb{C}'P^{p,q}\simeq \mathcal{SU}(p+1,q)/\mathcal{U}(p,q)  
\end{equation}
and 
\begin{equation}
\mathbb{C}'P^{p+q}\simeq \mathcal{SU}(p+q+1)/\mathcal{U}(p+q), 
\end{equation}
since 
\begin{equation}
\mathcal{SU}(p,q) \simeq \mathcal{SU}(p+q),~~~\mathcal{U}(p,q) 
\simeq \mathcal{U}(p+q). 
\end{equation}
}
\begin{equation}
\mathbb{C}'P^{p,q} = \mathbb{C}'P^{p+q}. 
\label{equivindefspiltcomp}
\end{equation}

\newpage 


\end{document}